\documentclass[12pt]{article}

\usepackage{epsfig,epsf}
\usepackage{amsmath}
\usepackage{amsthm}
\usepackage{amsfonts}
\usepackage{amssymb}
\usepackage{dsfont}

\usepackage{multirow}

\usepackage{slashed}

\usepackage[active]{srcltx}
\usepackage{wick}
\usepackage{psfrag}

\renewcommand{\thefootnote}{\fnsymbol{footnote}}

\newcommand{\be}{\begin{equation}}
\newcommand{\ee}{\end{equation}}
\newcommand{\ba}{\begin{eqnarray}}
\newcommand{\ea}{\end{eqnarray}}
\newcommand{\baa}{\begin{eqnarray*}}
\newcommand{\btab}{\begin{tabular}}
\newcommand{\etab}{\end{tabular}}
\newcommand{\eaa}{\end{eqnarray*}}

\def \e {\mbox{e}}

\def\inbar{\,\vrule height1.5ex width.4pt depth0pt}
\def\IC{\relax\hbox{$\inbar\kern-.3em{\rm C}$}}
\def\IZ{\relax{\hbox{\cmss Z\kern-.4em Z}}}
\def\IR{{\hbox{{\rm I}\kern-.2em\hbox{\rm R}}}}

\def\IP{{\hbox{{\rm I}\kern-.2em\hbox{\rm P}}}}
\def\II{\hbox{{1}\kern-.25em\hbox{l}}}

\newcommand{\twist}{ t}

\numberwithin{equation}{section}

\begin{document}

\begin{titlepage}

\vskip1cm
\begin{center}
  {\large \bf
     Renormalization of Twist-Four Operators in QCD
  \\}

\vspace{1cm}
{\sc V.M.~Braun\,$^{a}$},
{\sc A.N.~Manashov\,$^{a,b}$}
and {\sc J. Rohrwild\,$^{a}$}
\\[0.5cm]
\vspace*{0.1cm} $^{a}${\it
   Institut f\"ur Theoretische Physik, Universit\"at
   Regensburg, \\ D-93040 Regensburg, Germany}  \\[5mm]
\vspace*{0.1cm} $^{b}${\it Department of Theoretical Physics,  St.-Petersburg State
University\\
199034, St.-Petersburg, Russia}\\[10mm]

Version of \today\\[1cm]

\def\thefootnote{\arabic{footnote}}
\setcounter{footnote} 0

\vskip0.8cm
{\bf Abstract:\\[10pt]} \parbox[t]{\textwidth}{
Extending the work by Bukhvostov, Frolov, Lipatov and Kuraev (BFLK) on the
renormalization of quasipartonic operators
we derive a complete set of two-particle renormalization group kernels
that enter QCD evolution equations to twist-four accuracy.
It is shown that the $2\to 2$ evolution kernels which involve ``non-partonic'' components of
field operators, and, most remarkably, also $2\to3$ kernels
do not require independent calculation and can be restored
from the known results for quasipartonic operators using
conformal symmetry and Lorentz transformations.
The kernels are presented for the renormalization of light-ray
operators built of chiral fields in a particular basis such that the conformal
symmetry is manifest. The results can easily be recast in momentum space, in the form
of evolution equations for generalized parton distributions.
}
\vskip1cm

\end{center}

\end{titlepage}

{\small \tableofcontents}

\newpage

%
\section{Introduction}
%

Higher-twist effects generically correspond to corrections to hadronic observables that are
suppressed by a power of the (large) momentum transfer or the heavy quark mass.
One application where such effects are phenomenologically relevant are high precision
studies of the total cross section of deep-inelastic lepton-hadron scattering (DIS).
In this case the leading-twist calculations within the standard DGLAP formalism
are advanced to the next-to-next-to-leading order. Also lattice calculations of the moments
of parton distributions with an accuracy at a percent level are becoming feasible.
In this situation taking into account twist-four corrections that are suppressed by a power
of the photon virtuality $Q^2$ proves to be increasingly important for the analysis
of modern data, see e.g.
\cite{Leader:2006xc,Alekhin:2007zz,Blumlein:2008kz,Alekhin:2008ua,Fiore:2008nj,Leader:2009va}.
Twist-three effects are actively discussed in the context of exclusive and semi-inclusive reactions,
e.g. deeply-virtual Compton scattering \cite{Belitsky:2000vx,Kivel:2000fg}
 and diffractive electroproduction of vector mesons \cite{Anikin:2009hk},
 single spin asymmetry in various reactions
\cite{Efremov:1984ip,Efremov:1994dg,Eguchi:2006mc,Koike:2007rq,Kang:2008ih,Kang:2008ey,Vogelsang:2009pj}, etc.
One can expect that with the increasing accuracy of the experimental data the twist-four effects
will start playing a role here as well. Another large field of applications are the studies
of higher-twist hadron distribution amplitudes
(e.g. \cite{Braun:1989iv,Braun:2000kw,Ball:2006wn,Ball:2007zt}) that provide one with the
important input to the so-called light-cone sum rules \cite{Balitsky:1989ry,Chernyak:1990ag}.

The theoretical description of higher-twist corrections is based on the
Wilson Operator Product Expansion (OPE) and involves contributions of a
large number of local operators.
The corresponding leading-order coefficient functions are usually easy to calculate
(e.g. twist-four contributions to DIS are known since many years~\cite{Jaffe:1982pm,Ellis:1982cd})
but the operator renormalization for the operators of twist-four and higher
has not been studied systematically.
Up to now, twist-four anomalous dimensions are only known for a subset of
four-quark operators~\cite{Okawa:1980ei} and
for a few quark-gluon operators of lowest dimension
(e.g. \cite{Okawa:1981uw,Shuryak:1981kj,Shuryak:1981pi,Morozov:1983qr}).  In
addition, the structure of  the most singular parts of the mixing kernels for small values
of the  Bjorken  variable that are relevant for the contribution of two-pomeron  cuts in
high-energy scattering processes was considered in \cite{Levin:1992mu,Bartels:1993it}.

\begin{figure}[t]
\centerline{\epsfxsize12cm\epsfbox{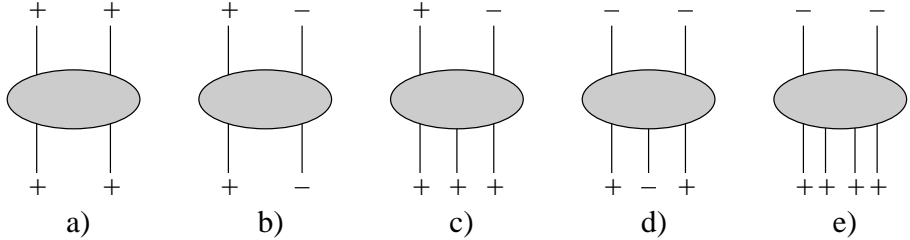}}
\caption{\small{Schematic structure of one-loop renormalization group kernels in QCD}}
\label{fig:schema}
\end{figure}

A general formalism  was  developed  by Bukhvostov, Frolov, Lipatov and Kuraev (BFLK)
\cite{Bukhvostov:1985rn} for the special class of so-called quasipartonic operators
that are built of ``plus'' components of quark and gluon fields. For each  twist, the set
of quasipartonic operators is closed under renormalization and the renormalization group
(RG) equation can be written in a Hamiltonian form that involves two-particle
``interaction'' kernels, cf. Fig.~\ref{fig:schema}a, that can be expressed in terms of
two-particle Casimir operators of the collinear subgroup
$SL(2,{\mathbb{R}})$  of the conformal group. In this formulation symmetries of the RG
equations  become explicit.  Moreover, the corresponding
three-particle  quantum-mechanical problem turns out to be completely integrable
for a few important cases, and in
fact  reduces to a Heisenberg spin chain \cite{BDM}. An almost complete
understanding  achieved at present of the renormalization of twist-three operators  is due
to all  these formal developments, see \cite{Braun:2003rp,Belitsky:2004cz} for a review
and further  references.

The goal of our study is to generalize the BFLK approach to the situation
where not all contributing operators are quasipartonic, as it proves to be the case
starting with  twist four.  On this way, there are two complications.

{}First, the number of fields (``particles'') is not conserved.
To one-loop accuracy, the mixing matrix
of operators with a given twist has a block-triangular structure as the operators with less
fields can mix with ones containing more fields but not vice versa. Operators with the maximum
possible number of fields are quasipartonic.

Second, operators involving  ``minus'' and ``transverse'' derivatives and/or field components
must be included. The problem is that transverse derivatives generally do not have good
transformation properties with respect to the $SL(2,\mathbb{R})$ group.
In concrete applications
it may be possible to get rid of such operators using equations of motion (EOM)
and exploiting specific structure of the matrix elements of interest,  e.g. if
there is no transverse momentum transfer between the initial and the final state.
The main problem as far as the operator renormalization
is concerned is that after this reduction the conformal symmetry becomes obscured.
In the work~\cite{Braun:2008ia} we have suggested a different, general approach
based on the construction of a complete conformal operator basis for all twists. In this basis,
the $SL(2,{\mathbb R})$ symmetry of the RG equations is manifest.

Another important observation was made in \cite{BFKS04} where it was shown
that the diagonal part of
one-loop QCD  RG equations (for arbitrary twist) can be written in a Hamiltonian form in
terms of  quadratic Casimir operators of the full conformal group $SO(4,2)$.
This implies that the $2\to 2$ kernels of the type shown in Fig.~\ref{fig:schema}b
can be obtained from the BFLK  kernels in Fig.~\ref{fig:schema}a by the corresponding
replacement. Our first goal will be  to work out the corresponding relations using the
conformal basis of Ref.~\cite{Braun:2008ia} for generic twist-four operators.

The main new contribution of this work is the calculation of 16 independent
$2\to 3$ kernels corresponding to operator mixing of one ``partonic'' and
one ``non-partonic'' field in three-particle quasipartonic operators, shown schematically
in Fig.~\ref{fig:schema}c. To this end we suggest a new technique based
on application of Lorentz transformations
(translations and rotations in the transverse plane) to the ``diagonal'' $2\to2$ kernels,
which bypasses calculation of Feynman diagrams.

Combining the kernels in Fig.~\ref{fig:schema}a,b,c one obtains a complete set
of building blocks for the renormalization of twist-4 operators that involve at most one
``non-partonic'' field. Strictly speaking, (two-particle) twist-4 operators
containing two ``non-partonic'' fields also exist and have to be included.
They can mix into three-particle and
also four-particle operators as in Fig.~\ref{fig:schema}d,e. However, contributions
of such operators can always be dispensed off using equations of motion.
Therefore, the results presented in our work are in fact sufficient for
writing down arbitrary QCD evolution equations to the twist-four accuracy
and e.g. calculation of the spectrum of anomalous dimensions of arbitrary
twist-four operators.
The kernels are written for the renormalization of coordinate-space light-ray
operators~\cite{Balitsky:1987bk} built of chiral fields and are manifestly $SL(2)$
invariant. We believe that this form is most suitable in practical applications.
The results can easily be recast in momentum space, in the form of evolution
equations for generalized parton distributions~\cite{Mueller:1998fv} .
The application  of our formalism to DIS will be presented elsewhere.

The presentation is organized as follows. Sect.~2 is introductory; we explain
some basic ideas and the coordinate-space formalism.
In Sect.~3 formal definitions are given, and we specify
the conformal operator basis that is used throughout this work.
Sect.~4 contains a summary of the evolution kernels for quasipartonic operators from
Ref.~\cite{Bukhvostov:1985rn} which we rewrite  in our language. The generalization
of these results to the $2\to2$ and $2\to3$ mixing kernels for
non-quasipartonic operators is considered in Sect.~4 and Sect.~5, respectively,
where we explain our method on simple examples.
The complete results for the $2\to 3$ mixing kernels are collected in Sect.~7.
The final Sect.~8 is reserved for conclusions.

\section{Background}
In this work we will use
the light-ray operator formalism in the spinor representation.
The basic elements  and the notation are explained in what follows.

\subsection{Light-ray operators}
We refer to non-local gauge-invariant operators with all the fields lying on a light-like
line $n^2=0$ as light-ray operators \cite{Anikin:1978tj}. The simplest example is
\begin{align}
\mathcal{O}(z_1,z_2)=\bar q(z_1n)\gamma_+ [z_1,z_2]q(z_2n)\,,
\label{2:LRO}
\end{align}
where $q(x)$ is the quark field and $[z_1,z_2]$ is the (light-like) Wilson line
\begin{align}
[z_1,z_2]=P\exp\left\{ig\, z_{12}\! \int_0^1 du\, A_+(z^u_{12} n )\right\}.
\label{2:PEXP}
\end{align}
Here and below we use ``plus'' for the projection on the light-cone direction
\begin{align}
              a_+ \equiv n_\mu a^\mu
\label{2:plus}
\end{align}
and also a shorthand notation
\begin{align}
 z_{12} = z_1-z_2\,, & & z_{12}^u = \bar u  z_1 + u z_2\,, & & \bar u = 1-u \,.
\label{2:baru}
\end{align}
Because of the light-like separation, the operator in (\ref{2:LRO}) contains
additional (ultraviolet) divergences apart from usual field renormalization, which
have to be renormalized. For practical purposes it is sufficient to {\it define}
the renormalized light-ray operator $[\mathcal{O}]_R$
as the generating function for the renormalized local operators
\begin{align}
\big[\mathcal{O}\big]_R(z_1,z_2)=
\sum_{N=0}^\infty\sum_{k=0}^N\frac{z_1^k\, z_2^{N-k}}{k!(N-k)!}
\big[\bar q \stackrel{\leftarrow}{D}_+^{k} \gamma_+ \stackrel{\rightarrow}{D}_+^{N-k}\!\! q\big]_R\,.
\label{2:GF}
\end{align}
Note that all local operators appearing on the r.h.s. of (\ref{2:GF}) have the
same (geometric) {\it twist}~$t=2$, defined as $t=$dimension-spin~\cite{Gross:1971wn}.

The scale-dependence of the renormalized light-ray operator is governed by the
re\-norma\-li\-zation-group (RG) equation
\begin{align}
\Big(\mu\frac{\partial}{\partial\mu}+\beta(g)\frac{\partial}{\partial
g}+ \frac{\alpha_s}{2\pi}\mathbb{H}\Big)[\mathcal{O}(z_1,z_2)]_R=0\,,
\label{2:RGE}
\end{align}
where $\mathbb{H}$ is the integral operator~\cite{Balitsky:1987bk}
\begin{eqnarray}\label{2:Hqq}
[\mathbb{H}\cdot
\mathcal{O}](z_1,z_2)&=&2C_F\biggl\{\int_0^1\frac{d\alpha}{\alpha}
\bigl[2\mathcal{O}(z_1,z_2)-
\bar\alpha \mathcal{O}(z_{12}^\alpha,z_2)
-\bar\alpha \mathcal{O}(z_1,z_{21}^\alpha)\bigr]
\nonumber\\
&& \hspace*{0.5cm}{}-\int_0^1d\alpha\int_0^{\bar\alpha}d\beta\,
\mathcal{O}(z_{12}^\alpha,z_{21}^\beta)-\frac32\mathcal{O}(z_1,z_2)\biggr\}\,.
\end{eqnarray}
Expansion of Eq.~(\ref{2:Hqq}) in powers of $z_1,z_2$ generates the mixing
matrix for local operators (\ref{2:GF}). Staying with the generating function
(light-ray operator) offers, however, several technical advantages.
One of them is that $\mathbb{H}$ and hence the RG equation in this form
are manifestly covariant under the $SL(2,\mathbb{R})$ transformations
of the light-cone coordinates
which correspond to the collinear subgroup of the conformal group~\cite{Braun:2003rp},
see Sec.~2.3.
Another advantage is that the RG equation~(\ref{2:RGE}), (\ref{2:Hqq}) is completely general.
It can be rewritten as the evolution equation for the generalized (quark) parton
distribution~\cite{Mueller:1998fv} and reduces to DGLAP and ERBL evolution equations
in the appropriate kinematic limits. For example, in the case of the
total cross section of deep-inelastic scattering there is no momentum transfer
between the initial and the final state, so that light-ray
operators that differ by a total translation can be identified:
$\langle P|\mathcal{O}(z_1+y,z_2+y)|P\rangle = \langle P|\mathcal{O}(z_1,z_2)|P\rangle$.
In this case Eq.~(\ref{2:Hqq}) simplifies to~\cite{Balitsky:1987bk}
\begin{eqnarray}
 [\mathbb{H}\cdot \mathcal{O}](0,z)&=& - 2 C_F \int_0^1 du\, K(u)\, \mathcal{O}(0,u z)
\end{eqnarray}
with the kernel
\begin{eqnarray}
  K(u) &=& \frac{1+u^2}{[1-u]_+} + \frac{3}{2}\delta(1-u)\,.
\end{eqnarray}
Writing the DIS matrix element as a Fourier transform of the parton
distribution~\cite{Collins:1981uw}
\begin{eqnarray}
\langle P|\mathcal{O}(0,z)|P\rangle &=& 2 P_+\int_{-1}^1dx e^{-iP_+ z x} F(x)
\end{eqnarray}
and picking up, e.g. the quark contribution, $F_q(x) \equiv F(x), x>0$,
Eq.~(\ref{2:RGE}) becomes
\begin{eqnarray}
 \mu^2\frac{d}{d\mu^2}F_q(x,\mu) &=& \frac{\alpha_s}{2\pi}C_F
 \int_x^1 \frac{dy}{y} K(y) F_q(x/y,\mu)
\end{eqnarray}
so that $C_F K(x)$ is nothing but the familiar DGLAP quark splitting function $P_{q\leftarrow q}$.
On the other hand, moments of $K(x)$ correspond to anomalous dimensions of
(flavor-nonsinglet) local operators:
\begin{eqnarray}
\gamma_n \,= \, - 2 C_F \int_0^1 dx\, x^{n-1} K(x)
&=& C_F\left[ 4\big(\psi(n+1)+\gamma_E\big)-\frac{2}{n(n+1)}-3\right].
\end{eqnarray}
The expressions in Eqs.~(\ref{2:RGE}), (\ref{2:Hqq}) give an example of what we are aiming at
for the description of higher-twist operators. The advantage of this, coordinate space
formulation is that conformal symmetry is manifest and also there is an immediate relation
both to renormalization of local operators and DGLAP-type equations for the corresponding
(multi) parton distributions.

\subsection{Spinor Representation}

We use spinor formalism and follow the conventions
adopted in Ref.~\cite{Braun:2008ia}. To this end, each covariant four-vector
$x_\mu$ is mapped to a hermitian matrix $x$:
 $$
x_{\alpha\dot\alpha}=x_\mu (\sigma^{\mu})_{\alpha\dot\alpha}\,,
\qquad
\bar x^{\dot\alpha\alpha}=x_\mu (\bar\sigma^{\mu})^{\dot\alpha\alpha}\,,
$$
where $\sigma^\mu=(\II,\vec{\sigma})$, $\bar\sigma^\mu=(\II,-\vec{\sigma})$ and
$\vec{\sigma}$
are the usual Pauli matrices.
The Dirac (quark) spinor $q$ is written as
\begin{align}
q=\begin{pmatrix}\psi_\alpha\\ \bar\chi^{\dot\beta}\end{pmatrix},
&& \bar q=(\chi^\beta,\bar\psi_{\dot\alpha})\,,
\end{align}
where $\psi_{\alpha}$, $\bar \chi^{\dot\beta}$ are two-component Weyl spinors,
$\bar\psi_{\dot\alpha}=(\psi_{\alpha})^\dagger$,
$\chi^{\alpha}=(\bar\chi^{\dot\alpha})^\dagger$.
The gluon strength tensor $F_{\mu\nu}$ can be decomposed as
\begin{align}
F_{\alpha\beta,\dot\alpha\dot\beta}=\sigma^\mu_{\alpha\dot\alpha}\sigma^\nu_{\beta\dot\beta}
F_{\mu\nu}=
2\left(\epsilon_{\dot\alpha\dot\beta} f_{\alpha\beta}-
\epsilon_{\alpha\beta} \bar f_{\dot\alpha\dot\beta}
\right).
\end{align}
Here $f_{\alpha\beta}$ and $\bar f_{\dot\alpha\dot\beta}$ are chiral and antichiral
symmetric tensors, $f^*=\bar f$, which belong to $(1,0)$ and $(0,1)$ representations
of the Lorenz group, respectively.  For the dual strength tensor
$\widetilde F^{\mu\nu}=\frac12 \epsilon^{\mu\nu\rho\sigma} F_{\rho\sigma}$ one obtains
\begin{align}	
i {\widetilde F}_{\alpha\beta,\dot\alpha\dot\beta}=2(\epsilon_{\dot\alpha\dot\beta}
f_{\alpha\beta}+
\epsilon_{\alpha\beta}\bar f_{\dot\alpha\dot\beta})\,,
\end{align}	
so that  $f_{\alpha\beta}$ ($\bar f_{\dot\alpha\dot\beta}$) can also be identified as
the selfdual (anti-selfdual) component of the field strength.
The corresponding explicit expressions are
\begin{align}
f_{\alpha\beta}=\frac14\left({D_{\alpha}}^{\dot\alpha} \bar A_{\dot\alpha\beta}+
 {D_{\beta}}^{\dot\alpha} \bar A_{\dot\alpha\alpha}\right), &&
\bar f_{\dot\alpha\dot\beta}=\frac14\left({\bar D_{\dot \alpha}}^{\phantom{\beta}\alpha}
  A_{\alpha\dot\beta}+
{\bar D_{\dot \beta}}^{\phantom{\beta}\alpha}  A_{\alpha\dot\alpha} \right),
\end{align}
where the covariant derivative is defined as $D_\mu=\partial_\mu-ig A_\mu$.
{}For convenience, we present the expressions for Dirac matrices
in the spinor basis:
\begin{align}
\gamma^{\mu}=\begin{pmatrix}0&[\sigma^\mu]_{\alpha\dot\beta}\\
                            [\bar\sigma^{\mu}]^{\dot\alpha\beta}&0 \end{pmatrix},&&
\sigma^{\mu\nu}=\begin{pmatrix}{[\sigma^{\mu\nu}]_{\alpha}}^{\beta}&0\\
                           0& {[\bar\sigma^{\mu\nu}]^{\dot\alpha}}_{\dot\beta} \end{pmatrix},
&&
\gamma_5=\begin{pmatrix}-\delta_{\alpha}^\beta&0\\
                           0&\delta^{\dot\alpha}_{\dot\beta}  \end{pmatrix}\,.
\end{align}
Here $\sigma^{\mu\nu}=\frac{i}2[\gamma^\mu,\gamma^\nu]$,
$\gamma_5=i\gamma^0\gamma^1\gamma^2\gamma^3$
and
\begin{align}
{(\sigma^{\mu\nu})_{\alpha}}^{\beta}=
\frac{i}2{\left[\sigma^{\mu}\bar\sigma^\nu-\sigma^{\nu}\bar\sigma^\mu\right
]_{\alpha}}^\beta\,, &&
{(\bar\sigma^{\mu\nu})^{\dot\alpha}}_{\dot\beta}=
\frac{i}2{\left[\bar\sigma^{\mu}\sigma^\nu-\bar\sigma^{\nu}\sigma^\mu\right
]^{\dot\alpha}}_{\dot\beta}\,.
\end{align}
%
Any
light-like vector can be represented as a product
of two spinors. We introduce two independent light-like vectors as
\begin{align}
  n_{\alpha\dot\alpha}=\lambda_{\alpha}\bar\lambda_{\dot\alpha}\,, \qquad n^2=0\,,
\nonumber  \\
  \tilde n_{\alpha\dot\alpha}=\mu_\alpha\bar\mu_{\dot\alpha}\,, \qquad \tilde n^2 =0\,,
\label{2:nlambda}
\end{align}
where $\bar\lambda=\lambda^\dagger$,  $\bar\mu=\mu^\dagger$. The basis vectors in the
plane transverse to $n,\tilde n$ can be chosen as $\mu_{\alpha}\bar\lambda_{\dot\alpha}$
and $\lambda_{\alpha}\bar\mu_{\dot\alpha}$. An arbitrary four-vector
$x_{\alpha\dot\alpha}$ can be represented as
\begin{align}
x_{\alpha\dot\alpha}=
z\,\lambda_{\alpha}\bar\lambda_{\dot\alpha}+\tilde z\,\mu_{\alpha} \bar\mu_{\dot\alpha}
+w\,\lambda_{\alpha}\bar\mu_{\dot\alpha}+\bar w\,\mu_{\alpha}\bar\lambda_{\dot\alpha}\,, &&
x^2=(\mu\lambda)(\bar\lambda\bar\mu)[z\tilde z-w\bar w]\,,
\end{align}
where $z$ and $\tilde z$ are real and $w$, $\bar w =w^*$ complex coordinates in the 
two light-like directions and the transverse plane, respectively.

The ``$+$'' and ``$-$'' fields are defined as the projections onto $\lambda$ and $\mu$ spinors,
respectively:
\begin{align}\label{2:plusf}
\psi_+=\lambda^\alpha\psi_\alpha\,,&&\chi_+=\lambda^\alpha\chi_\alpha\,, &&
f_{++}=\lambda^\alpha\lambda^\beta f_{\alpha\beta}\,,
\nonumber\\
\bar\psi_+=\bar\lambda^{\dot\alpha}\bar\psi_{\dot\alpha}\,,&&
\bar \chi_+=\bar\lambda^{\dot\alpha}\chi_{\dot\alpha}\,, &&
\bar f_{++}=\bar\lambda^{\dot\alpha}\bar\lambda^{\dot\beta} \bar f_{\dot\alpha\dot\beta}\,,
\nonumber\\
 \psi_-=\mu^\alpha \psi_\alpha\,,&&
 \bar \psi_-=\bar\mu^{\dot\alpha} \bar\psi_{\dot\alpha}&&f_{+-}=\lambda^\alpha\mu^\beta
f_{\alpha\beta}\,,
\end{align}
etc. Fields with free spinor indices can be written in terms of the 
``$+$'' and ``$-$'' components, e.g.
\begin{align}
 (\mu\lambda)\,\psi_\alpha(z)&=\lambda_\alpha \,\psi_-(z)-\mu_\alpha \,\psi_+(z)\,,
 \notag\\
 (\mu\lambda)^2 f_{\alpha\beta}(z)&=\lambda_\alpha\lambda_\beta\, f_{--}(z)
 -(\lambda_\alpha\mu_{\beta}+\lambda_\beta\mu_{\alpha})\,f_{+-}(z)
 +\mu_{\alpha}\mu_{\beta}\, f_{++}(z)\,.
\label{2:psidecomp}
\end{align}
Note that in difference to Ref.~\cite{Braun:2008ia} we do not impose any particular
normalization condition on the auxiliary spinors $\lambda$, $\mu$. Without loss of generality
one can put e.g. $(\mu\lambda)=1$. However, keeping this factor proves to be
convenient as it allows one to keep track of the balance of
$\lambda,\mu,\bar\lambda,\bar\mu$ in the equations.

The light-ray operator (\ref{2:LRO}) is decomposed as
\begin{eqnarray}
  \bar q(z_1)\gamma_+q(z_2) &=&
 \bar \psi_+(z_1)\psi_+(z_2) +  \chi_+(z_1)\bar\chi_+(z_2)\,,
\nonumber\\
 \bar q(z_1)\gamma_5\gamma_+q(z_2) &=&
 \bar \psi_+(z_1)\psi_+(z_2) -  \chi_+(z_1)\bar\chi_+(z_2)\,,
\end{eqnarray}
where $q(z_2)\equiv q(z_2 n)$ etc., and the Wilson lines are implied.

As a less trivial example, consider the three-particle light-ray operators
$S^\pm_\mu$~\cite{Balitsky:1987bk}
\begin{align}
S_{\mu}^\pm(z_1,z_2,z_3)=g\bar q(z_1)\left[
F_{\mu+}(z_2)\pm \widetilde F_{\mu+}(z_2)i\gamma_5\right]\gamma_+ q(z_3)
\end{align}
which contribute to polarized deep-inelastic scattering to twist-three
accuracy (to the structure function $g_2(x,Q^2)$).
Going over to $S_{\alpha\dot\alpha}^{\pm}=\sigma_{\alpha\dot\alpha}^\mu S^\pm_\mu$ one finds easily
\begin{align}
S_{\alpha\dot\alpha}^{+}=&2g\left[\bar\lambda_{\dot\alpha}\bar\psi_+(z_1)
f_{+\alpha}(z_2)\psi_+(z_3)
+\lambda_{\alpha}\chi_+(z_1)\bar f_{+\dot\alpha}(z_2)\bar\chi_+(z_3)
\right],
\\
S_{\alpha\dot\alpha}^{-}=&2g\left[\bar\lambda_{\dot\alpha}\chi_+(z_1)
f_{+\alpha}(z_2)\bar\chi_+(z_3)
+\lambda_{\alpha}\bar\psi_+(z_1)\bar f_{+\dot\alpha}(z_2)\psi_+(z_3)
\right].
\end{align}
The operators $S^{\pm}_{\alpha\dot\alpha}$ contain both twist-three and twist-four
contributions. Indeed, expansion of
$\bar\psi_+(z_1)f_{+\alpha}(z_2) \psi_+(z_3)$ (and similarly for the other terms) at short
distances goes over local operators
$O_{\alpha,\alpha_1,\ldots\alpha_{k+1},\dot\alpha_1,\ldots,\dot\alpha_k}$
(where $k$ is the total number of covariant derivatives)
which are symmetric in all dotted indices  $\dot\alpha_1,\ldots,\dot\alpha_{k}$
and also in the subset of the undotted ones, $\alpha_1,\ldots,\alpha_{k+1}$. In order to
separate the (leading) twist-three contribution one has to symmetrize in the remaining
undotted index $\alpha$. The answer can be written as follows (cf.~\cite{Balitsky:1987bk}):
\begin{align}
[\bar\psi_+(z_1)f_{+\alpha}(z_2) \psi_+(z_3)]^{\mathrm{tw}-3}=
\frac{\partial}{\partial\lambda^\alpha}\int_0^1d\tau \tau^2
\bar\psi_+(\tau z_1)f_{++}(\tau z_2) \psi_+(\tau z_3)\,.
\end{align}
As it should be, the twist-three part of this operator involves only ``+'' fields.
{}For higher-twist operators  twist separation becomes rather cumbersome in the ``vector''
formalism and going over to the spinor basis yields considerable advantages. We will
encounter further examples in what follows.

\subsection{$SL(2,\mathbb{R})$ invariance}\label{sl2r-inv}

It is well known that conformal symmetry of the QCD Lagrangian imposes nontrivial constraints
on the structure of RG equations to the one-loop accuracy, see e.g. Ref.~\cite{Braun:2003rp}.
For fields ``living'' on the light-cone, $\Phi(zn)$, it is sufficient to consider the
so-called collinear subgroup $SL(2,\mathbb{R})$ of the conformal group $SO(4,2)$,
corresponding to projective (M\"obius)  transformations of the line~$x=zn$:
$$
z\to\frac{az+b}{cz+d}\,,\qquad ab-cd=1\,,
$$
where  $a,b,c,d$ are  real numbers.
A field with definite spin projection $s$ on the light-cone transforms according to the
irreducible representation of the $SL(2,\mathbb{R})$ group with  the conformal spin
\begin{align}
    j = \frac12(\ell^{\rm can}+s) = \ell^{\rm can} -E/2\,,
\end{align}
where $\ell^{\rm can}$ is the (canonical) dimension and $E=\ell^{\rm can}-s$ is the collinear twist.
Action of the $SL(2)$ generators on quantum fields can be traded for the
differential operators acting on field coordinates. In this representation
the generators become
\begin{align}\label{2:diff-form}
S_+=z^2\partial_z+2jz\,, && S_0=z\partial_z+j\,, && S_-=-\partial_z\,.
\end{align}
They obey the standard commutation relations
\begin{align}\label{2:algebra}
[S_+,S_-]=2S_0\,, && [S_0,S_\pm]=\pm S_\pm\,.
\end{align}
 A finite form of the group transformations is
\begin{align}\label{2:Tg}
[T^j(g^{-1})\Phi](z)=\frac1{(cz+d)^{2j}}\Phi\left(\frac{az+b}{cz+d}\right)\,,\qquad
g=\begin{pmatrix}a&b\\c&d\end{pmatrix}.
\end{align}
Note that the field decomposition in ``$+$'' and ``$-$'' components as in Eq.~(\ref{2:psidecomp})
is equivalent to the separation of different spin projections, $s=+1/2$ and $s=-1/2$,
resulting in different values of the conformal spin $j=1$ and $j=1/2$ for
$\psi_+(z)$ and $\psi_-(z)$, respectively.  Similarly for gluon fields:
$f_{++}$ corresponds to $j=3/2$, $f_{+-}$ to $j=1$, etc.

The functional form of the ``Hamiltonians'' $\mathbb{H}$ that appear in the
RG equations for light-ray operators (\ref{2:RGE}), cf. Eq.~(\ref{2:Hqq}), is constrained
by the $SL(2)$ invariance. For illustration, consider the simplest case:
A light-ray operator built of fields with conformal spins $j_1$ and $j_2$ mixing
into a light-ray operator built of the fields with the same spins.
We are looking for an invariant kernel $\mathcal{H}_{12}$
acting on functions of two variables $\varphi(z_1,z_2)$ which transform according to
the representation $T^{j_1}\otimes T^{j_2}$:
\begin{align}
\varphi(z_1,z_2)\to [T^{j_1}(g)\otimes T^{j_2}(g)\varphi](z_1,z_2)=(cz_1+d)^{-2j_1}(cz_2+d)^{-2j_2}
\varphi\left(\frac{az_1+b}{cz_1+d},\frac{az_2+b}{cz_2+d}\right).
\end{align}
The $SL(2)$ invariance means that $\mathcal{H}$ commutes with group transformations:
\begin{align}	\label{2:HT}
\mathcal{H}_{12} \,T^{j_1}(g)\otimes T^{j_2}(g)= T^{j_1}(g)\otimes T^{j_2}(g)\,\mathcal{H}_{12}\,.
\end{align}

There are different ways to define an operator.
The first one is to specify the eigenvalues of $\mathcal{H}_{12}$ on each irreducible
component in the tensor product decomposition
\begin{align}	 \label{2:TD}
T^{j_1}\otimes T^{j_2}=\sum_{n=0}^{\infty} \oplus\, T^{j_1+j_2+n},\qquad
\varphi(z_1,z_2)=\sum_{n=0}^{\infty}\varphi_n(z_1,z_2)\,.
\end{align}	
Let
\begin{align}
 \mathcal{H}_{12}\,\varphi_n(z_1,z_2)=h_n\varphi_n(z_1,z_2)\,.
\label{2:hhh}
\end{align}	
On the other hand
\begin{align}
 S_{12}^2\,\varphi_n(z_1,z_2)=j_n(j_n-1)\varphi_n(z_1,z_2)\,,\qquad j_n = n+j_1+j_2\,,
\label{2:phin}
\end{align}	
where
\begin{equation}\label{S12}
  S_{12}^2 = - \partial_1\partial_2 \,z_{12}^2 + 2(j_1-1)\partial_2 z_{21} +
2(j_2-1)\partial_1 z_{12} + (j_1+j_2-1)(j_1+j_2-2)
\end{equation}	
is the two-particle Casimir operator, $S_{12}^2 =(\vec{S}_1+\vec{S}_2)^2$. Here
$\vec{S}_1$ and $\vec{S}_2$
are the generators (\ref{2:diff-form}) acting on $z_1$ and $z_2$ coordinates, respectively,
$\partial_k \equiv \partial/\partial z_k$ and $z_{12}=z_1-z_2$, cf. (\ref{2:baru}).
It proves to be convenient to define the operator $J_{12}$ as a formal solution of the
operator equation
\begin{equation}
  S_{12}^2 = J_{12}(J_{12}-1)\,,\qquad J_{12} \,\varphi_n(z_1,z_2)=j_n \varphi_n(z_1,z_2)\,.
\label{2:sss}
\end{equation}	
Since $\mathcal{H}_{12}$ and $J_{12}$ have the same eigenfunctions, they can be diagonalized
simultaneously which means that $\mathcal{H}_{12} = h(J_{12})$ where $h(j)$ is an
ordinary function which can be found by expressing the eigenvalues
$h_n$ (\ref{2:hhh}) in terms of $j_n$ (\ref{2:sss}).
For example, the operator in Eq.~(\ref{2:Hqq}) can be written as
\begin{equation}
  {\mathbb H} = 2C_F\left[\psi(J_{12}+1) + \psi(J_{12}-1) -2\psi(1)-\frac32\right],
\label{2:kurz}
\end{equation}	
where $\psi(x) = d\ln\Gamma(x)/dx$ is the Euler $\psi$-function. This is the most
concise form. Moreover, as noticed in Ref.~\cite{BFKS04}, the
renormalization group kernels involving fields with other spin projections
can be obtained from this expression replacing the $SL(2)$ Casimir operators
by the ones of the full four-dimensional conformal group $O(4,2)$. We will use this
technique in Sect.~5.

Often it is preferable to have a more  functional definition of an operator. It can be
shown that action of any $SL(2,R)$ invariant operator $\mathcal{H}_{12}$ on a function
$\varphi(z_1,z_2)$ can be written in the form
\begin{align}
[\mathcal{H}_{12}\varphi](z_1,z_2)=\int_0^1d\alpha\int_0^1d\beta\,
\bar\alpha^{2j_1-2}\,\bar\beta^{2j_2-2}
\,\omega\left(\frac{\alpha\beta}{\bar\alpha\bar\beta}\right)\,\varphi(z_{12}^\alpha,z_{21}^\beta)\,,
\label{2:inv}
\end{align}
where $\omega(x)$ is an arbitrary function of one variable. For the same example
in Eq.~(\ref{2:Hqq})
\begin{equation}
\label{2:lang}
[\mathbb{H}\cdot
\mathcal{O}](z_1,z_2)=2C_F \int_0^1\!d\alpha\int_0^{\bar\alpha}\!d\beta\,
\Big\{\delta\left(\frac{\alpha\beta}{\bar\alpha\bar\beta}\right)
\left[\mathcal{O}(z_1,z_2)-\mathcal{O}(z^\alpha_{12},z^\beta_{21})\right]
 - \mathcal{O}(z^\alpha_{12},z^\beta_{21}) -3\Big\}.
\end{equation}
We stress that the conformal symmetry only becomes manifest in the properly chosen
operator basis. Problem is that light-ray operators involving ``$-$'' components of
the fields can mix in operators containing ``$-$'' or transverse derivatives that do not
have, in general, good properties under conformal transformations.
A general approach how to deal with such contributions in a conformally covariant way was
developed in Ref.~\cite{Braun:2008ia}.
The idea is to expand the light-ray operator basis by adding primary fields
with a particular transverse derivative, e.g. $\bar D_{-+}\psi_+$,
and at the same time eliminating the other derivative, $\bar D_{+-}\psi_+$,
using equations of motion. We will elaborate on this proposal in the following Section.

\section{Conformal Operator Basis}

\subsection{Light-ray fields }

We define one-particle quark, antiquark and gluon light-ray operators
(alias light-ray fields) including the Wilson line
\begin{align}
     [0,z]\Phi(z) \equiv ~\mbox{\rm Pexp}\left\{-ig z \int_0^1 du\, (n\cdot A)(uzn)\right\}\Phi(zn)\,.
\label{3:lf}
\end{align}
The Wilson lines are always assumed although in many cases they will not be written explicitly.
Here and below we use a shorthand notation $\Phi(z)$ for $ \Phi(nz)$ etc.

Light-ray fields can be viewed as generating functions  for local
operators with covariant derivatives that arise through the (formal) Taylor expansion
\begin{align}
  [0,z] \Phi(z) = \sum_k \frac{z^k}{k!}(n\cdot D)^k\Phi(0) =
   \sum_k \frac{z^k}{2^kk!}(\bar\lambda\bar D\lambda)^k\Phi(0) \,.
\label{3:light-ray-oper}
\end{align}
Note that all local operators on the r.h.s. of (\ref{3:light-ray-oper})
have the same collinear twist as the
field $\Phi$ itself since each $(n\cdot D)$ derivative adds one unit of dimension
and spin projection, simultaneously.

It is implied that the Wilson line in Eq.~(\ref{3:lf}) is written in the appropriate
representation of the color group. In order to unify the notation
we introduce the $SU(N)$ generators acting on quark, antiquark and gluon fields as follows:
\begin{align}
t^a \Phi=
\begin{cases}\label{3:generators}
(t^a\psi)^i=T^a_{ii'}\psi^{i'}  & \text{for the quark fields}\quad \psi,\bar\chi
\\
(t^a \bar\psi)^i=-T^a_{i'i}\bar\psi^{i'}&
  \text{for the antiquark fields}\quad \bar\psi,\chi
\\
(t^a f)^b=i\,\mathrm{f}^{bab'}\! f^{b'}
 &  \text{for the (anti)self-dual gluon fields}\quad  f,\bar f
\end{cases}
\end{align}	
where $T^a$ are the usual generators in the fundamental representation
and f$^{abc}$ are the structure constants. Then, in particular
\begin{eqnarray}
 D\psi &\equiv& (\partial -ig A^a T^a)\psi\,, \nonumber\\
 D\bar\psi &\equiv& \bar \psi (\stackrel{\leftarrow}{\partial} + ig A^a T^a)\,.
\end{eqnarray}
We tacitly assume existence of $n_F$ quark flavors; flavor indices will not be shown explicitly
in most cases.

\begin{table}[t]
\begin{center}
\begin{tabular}{|c|c|c|c|c|c|c|}
\hline\hline
       & $\psi_+$ &  $f_{++}$ & $\psi_-$ & $f_{+-}$ & $\bar D_{-+}\psi_+$ &  $\bar D_{-+}f_{++}$
\\
\hline
$j$&     $1$ &  $3/2$ & $1/2$ & $1$ & $3/2$ &  $2$
\\
\hline
$E$&     $1$ &  $1$ & $2$ & $2$ & $2$ &  $2$
\\
\hline
$H$&     $1/2$ &  $1$ & $-1/2$ & $0$ & $3/2$ &  $2$
\\
\hline
\end{tabular}
\end{center}
\caption{\small The conformal spin $j$, collinear twist $E$ and helicity $H$ of the
light-ray primary fields.
         The quantum numbers of the $\chi_+$ and $\chi_-$ fields coincide with those for
         $\psi_+$ and $\psi_-$, respectively. The anti-chiral
         fields $\bar\psi_+,\ldots$ have the same conformal spin as their chiral counterparts
          and the opposite chirality.
}
\label{tab:1}
\end{table}

We will use the notation $\Phi_+$ for the chiral ``+'' fields and $\bar \Phi_+$ for the
antichiral ones:
\begin{align}\label{defPhi}
\Phi_+=\{\psi_+,\chi_+, f_ {++}\}\,, && \bar\Phi_+=\{\bar\psi_+,\bar\chi_+,\bar f_{++}\}\,,
\nonumber\\
\Phi_-=\{\psi_-,\chi_-, f_ {+-}\}\,, && \bar\Phi_-=\{\bar\psi_-,\bar\chi_-,\bar f_{+-}\}\,.
\end{align}
The fields $\Phi_+,\bar \Phi_+$ have collinear twist $E=1$, while $\Phi_-,\bar \Phi_-$ have $E=2$.
In addition, one-particle $E=2$ operators can be constructed by adding transverse derivatives
to $E=1$ ``plus'' fields, e.g. $ \bar D_{-+}\psi_+, \bar D_{+-}\psi_+$, where
$\bar D_{-+}=(\bar \mu \bar D \lambda)$ etc.  Some of them  (but not all, in a general situation)
can be excluded from consideration with the help of equations of motion (EOM).
A systematic procedure to treat such contributions is presented in Ref.~\cite{Braun:2008ia}
where we have shown that EOM can be used to eliminate a ``half'' of the transverse derivatives
in such a way that the remaining fields (with a derivative) transform as primary fields under
the $SL(2,\mathbb{R})$ transformations. In the above example one has to
eliminate $ \bar D_{+-}\psi_+$ in favor of $\bar D_{-+}\psi_+$.

The complete basis of light-ray one-particle primary fields that is sufficient for the studies of
twist-4 operators includes the following operators:
\begin{eqnarray}
X&=& \{\Phi_+\,,~\bar \Phi_+\,,~\Phi_-\,,~\bar \Phi_-\,,~\bar D_{-+}\Phi_+\,,~D_{-+}\bar\Phi_+\}.
\label{3:X}
\end{eqnarray}
Each primary field $X$ carries two more quantum numbers in addition to the conformal spin ---
collinear twist $E$ and chirality $H$ --- which are eigenvalues of the two remaining generators
of the full conformal group that commute with the light-cone $SL(2,R)$ collinear subgroup,
see Table.~\ref{tab:1}. The fields $\psi_\pm$ and $\chi_\pm$
have the same quantum numbers, so that we display them for the $\psi$--field only. The anti-chiral
fields $\bar\Phi$ have the same conformal spin as the chiral ones $\Phi$ but opposite
chirality $H\to -H$.

\subsection{Composite light-ray operators}\label{section:l-ray}

Gauge-invariant $N$-particle light-ray operators can be defined as a product of primary
fields
\begin{eqnarray}
\mathcal{O}(z_1,\ldots,z_N)&=& S X(z_1)\otimes X(z_2)\otimes\ldots\otimes X(z_N)
\nonumber\\
&\equiv& S_{i_1\ldots i_N} \big([0,z_1]X(z_1)\big)^{i_1} \big([0,z_2]X(z_2)\big)^{i_2}
\ldots \big([0,z_N]X(z_N)\big)^{i_N}\,,
\label{3:O}
\end{eqnarray}
where $X(z_k)$ are the fields from the set in Eq.~(\ref{3:X}), $i_1,\ldots,i_N$ are
color indices and $S_{i_1\ldots i_N}$ an invariant color tensor such that
\begin{align}
\label{3:S-inv}
[(t_1)^a_{k_1i_1} + (t_2)^a_{k_2i_2} +\ldots+ (t_N)^a_{k_Ni_N}]S_{i_1,\ldots,i_N}=0\,.
\end{align}
Here it is implied that the generators $t^a$ are taken in the appropriate
representation, cf. Eq.~(\ref{3:generators}). The condition in Eq.~(\ref{3:S-inv})
ensures that $\mathcal{O}(z_1,\ldots,z_N)$ is a color singlet.

Composite light-ray operators (\ref{3:O}) transform as a product of  primary fields
under collinear conformal transformations
\begin{align}\label{3:sl2-rule}
\mathcal{O}(z_1,\ldots,z_k)\to
\prod_{i=1}^k (cz_i+d)^{-2j_i} \,\mathcal{O}(z'_1,\ldots,z'_k)\,,
\end{align}
where $z'=(az+b)/(cz+d)$ and $j_i$ is the conformal spin of the $i$-th ``constituent'' field.

The collinear twist of a multiparticle light-ray operator is equal, obviously, to the sum of
twists of the fields, $E = E_1+\ldots+E_N$. For a given $N$, the lowest possible
twist is $E=N$ and it corresponds to operators built of ``plus'' components of the fields
only, $ X = \{\Phi_+,\bar \Phi_+\} $. Such operators are known as
quasipartonic~\cite{Bukhvostov:1985rn}.
The structure of logarithmic ultraviolet singularities in the relevant
one-loop Feynman diagrams is such that the operators with less fields can mix into the
operators containing more fields,
but not the other way around. As a consequence, the set of quasipartonic operators is closed
under renormalization. The corresponding RG equations were derived in Ref.~~\cite{Bukhvostov:1985rn}.
They are sufficient for a calculation of the scale-dependence of arbitrary twist-three observables
in QCD, e.g. in polarized DIS and also leading-twist baryon distribution amplitudes.

Multiparticle operators with $E=N+1$ built of one ``minus'' field with $E=2$ and $N-1$ ``plus'' fields
with $E=1$ are the next in complexity. They are subject of this paper. These operators can mix
among themselves and also with $N+1$-particle quasipartonic operators. The RG equations for such
operators derived below are sufficient for a calculation of the scale dependence of arbitrary
twist-four observables.

As an illustration, consider a typical twist-4 operator
$i\bar q \widetilde F_{\mu\nu}n^\mu\gamma^\nu \gamma_5 q$ which contributes to the DIS
structure functions
at the level of power-suppressed $1/Q^2$ corrections in the ``longitudinal'' operator basis of
Ref.~\cite{Jaffe:1982pm}. Going over to spinor notation one obtains
\begin{align}
i\bar q \widetilde F_{\mu\nu} n^\mu\gamma^\nu \gamma_5 q=&
\bar\psi_+ (f\psi)_+-(\bar\psi\bar f)_+\psi_+-\chi_+(\bar f\bar\chi)_++(\chi  f)_+ \bar \chi_+\,.
\end{align}
Each of the four terms can now be rewritten as a combination of two ``plus'' fields
and one ``minus'' field, e.g.
\begin{align}
(\mu\lambda)\bar\psi_+ (f\psi)_+=\bar\psi_+ f_{+-}\psi_+-
\bar\psi_+f_{++}\psi_-\,.
\end{align}
Twist-four operators built of two ``minus'' fields also exist, e.g. $\bar\psi_- \psi_-$, but they
can (and should) be eliminated using EOM so that one does not need to consider them explicitly.
In this particular case (DIS) twist-four operators involving $\bar D_{-+}\Phi_+\,,~D_{-+}\bar\Phi_+$
can be eliminated using EOM as well, however, only at the cost of
loosing manifest $SL(2,\mathbb{R})$ covariance of the evolution equations,
see Ref.~\cite{Braun:2008ia}.
To avoid confusion, we stress that the operators with transverse derivatives
$\bar D_{-+}\Phi_+\,,~D_{-+}\bar\Phi_+$ are introduced here in order to maintain conformal covariance
as an extension of the ``longitudinal'' operator basis of Ref.~\cite{Jaffe:1982pm}.
They are not the same as the ``transverse'' operators of Ellis--Furmanski-Petronzio (EFP)
\cite{Ellis:1982cd}
that are advantageous in another aspect: they yield simpler coefficient functions.

\subsection{Renormalization-group equations for light-ray operators}\label{Renorm}

Operators with the same quantum numbers mix under renormalization.
Let $\mathcal{O}_i(X)$ $i=1,\ldots,L$ be the complete set of such operators.
A renormalized operator is written as
\begin{align}
[\mathcal{O}_i(X)]_R=\mathbb{Z}_{ik}\mathcal{O}_{k}(X_0)\,,
\end{align}
where $X_0=Z_X X$ is the bare field. Renormalized operators satisfy the RG equation
\begin{align}
\left(\mu\frac{\partial}{\partial\mu}+\beta(g)\frac{\partial}{\partial g}
+{\gamma}_{ik}\right)
[O_k(X)]_R=0\,.
\end{align}
Here $\beta(g)$ is the (QCD) beta function and
$$
\gamma=-\mu\dfrac{d}{d\mu} \mathbb{Z}\, \mathbb{Z}^{-1}
$$
is the matrix of anomalous dimensions.
To the one-loop accuracy in dimensional regularization $D=4-2\epsilon$ one obtains
\begin{align}
\mathbb{Z}=\mathds{I}+\frac{\alpha_s}{4\pi\epsilon} \mathbb{H} && \text{and} &&
\gamma=\frac{\alpha_s}{2\pi} \mathbb{H}\,.
\end{align}
The operator $\mathbb{H}$ (Hamiltonian) has a block-triangular form (at one loop). It
follows from the fact that the $N-$particle operators can only mix  with $M\geq N-$particle operators,
thus
\begin{align}\label{BigH}
\mathbb{H}\begin{pmatrix}\mathcal{O}^{(2)}\\
\mathcal{O}^{(3)}\\\vdots\\
\mathcal{O}^{(N)}
\end{pmatrix}=
\begin{pmatrix} \mathbb{H}^{(2\to 2)}&\mathbb{H}^{(2\to 3)}&\cdots&\cdots \\
                     0& \mathbb{H}^{(3\to 3)} &\mathbb{H}^{(3\to 4)}&\cdots\\
                     \vdots& \vdots &\ddots &\\
                     0&0&0& \mathbb{H}^{(N\to N)}\end{pmatrix}
\begin{pmatrix}\mathcal{O}^{(2)}\\
\mathcal{O}^{(3)}\\\vdots\\
\mathcal{O}^{(N)}
\end{pmatrix}\,.
\end{align}
Further, it follows from the inspection of Feynman diagrams that the diagonal blocks are
given by the sum of two-particles kernels, $\mathbb{H}_{ik}^{(2\to 2)}$
\begin{equation}
\mathbb{H}^{(n\to n) }=\sum_{i,k}^{n} \mathbb{H}_{ik}^{(2\to 2)}\,.
\end{equation}
The general structure of the kernels is
\begin{align}\label{3:2-2}
\mathbb{H}^{(2\to 2)}_{12}[X^{i_1}(z_1)\otimes X^{i_2}(z_2)] \,=\,
\sum_{q}\sum_{i'_1 i'_2} [C_q]^{i_1i_2}_{i'_1i'_2}
[\mathcal{H}^{(q)}_{12} X^{i'_1}\otimes X^{i'_2}](z_1,z_2)\,.
\end{align}
Here $ [C_q]^{i_1i_2}_{j_1j_2}$ is a color tensor, $\mathcal{H}^{(q)}_{12}$ is an $SL(2,\mathbb{R})$
invariant operator which acts on  coordinates of the fields, and $q$ enumerates
different structures.  Except for the cases when $X\otimes
X=\bar\psi\otimes\psi,~\chi\otimes\bar\chi,~f\otimes \bar f$, the operator $\mathbb{H}$
does not change the components of the primary fields, i.e.
$\mathbb{H}: \psi\otimes\psi\to \psi\otimes\psi,\psi\otimes f\to\psi\otimes f$ and so on.

Similarly, for $\mathbb{H}^{(n\to n+1) }$ one gets
\begin{equation}
\mathbb{H}^{(n\to n+1) }=\sum_{i,k}^{n} \mathbb{H}_{ik}^{(2\to 3)}\,
\end{equation}
with
\begin{align}\label{3:2-3}
\mathbb{H}^{(2\to 3)}_{12}[X^{i_1}(z_1)\otimes X^{i_2}(z_2)] \,=\,
\sum_{q}\sum_{i'_1 i'_2} [C_q]^{i_1i_2}_{i'_1i'_2i'_3}
[\mathcal{H}^{(q)}_{12} X^{i'_1}\otimes X^{i'_2} \otimes X^{i'_3}](z_1,z_2)\,,
\end{align}
where the $(2\to 3)$ kernels $\mathcal{H}^{(q)}_{12}$ are of course different from the
$(2\to2)$ ones in Eq.~(\ref{3:2-2}), and so on.

%
%
The two-particle kernels $\mathbb{H}_{ik}^{(2\to 2)}$, $\mathbb{H}_{ik}^{(2\to 3)}$ correspond to
the counterterms to the product of light-ray fields (\ref{3:lf}) with open color indices
$M = X(z_1)\otimes X(z_2)$, where $X(z)$ belongs to the set~(\ref{3:X}).
Such objects are, obviously, not gauge invariant and discussing their properties one
has to specify the gauge fixing scheme. We will use the light-cone gauge
and the background field formalism~(see~\cite{Abbott} for a review).

We recall that in this formalism one splits the fields in the quantum and classical
components, $q \to q^{cl} + q$, $A_\mu \to A^{cl}_\mu + a_\mu$.
Taking into account quantum corrections to the composite operator $M=M(q,A)$
corresponds to the calculation of the path integral over the quantum fields
\begin{align}\label{ZM}
\langle\langle{M}(q^{cl}, A^{cl})\rangle\rangle\,=\,
\int \mathcal{D}q \mathcal{D}a\, {M}(q^{cl}+q, A^{cl}+a)\,
\exp\left\{iS_R(q^{cl}+q, A^{cl}+a)+\frac{i}{2\xi}{a_+^2}\right\}\,.
\end{align}
Here $S_R(q,A)$ is the renormalized QCD action, $a_+=(a\cdot n)$, where $n^2=0$ and
$\xi$ is the gauge fixing parameter which, in the light-cone gauge, has to be sent
to zero, $\xi\to 0$.
One has a certain freedom to choose the transformation properties of quantum and
classical gauge fields under gauge transformations~\cite{Abbott}. In particular one
can assume that
\begin{align}\label{trule}
a'\,=\,U a U^\dagger\,, \qquad
A'\,=\,U A U^\dagger-\dfrac{i}g U\partial U^\dagger\,.
\end{align}
The light-ray fields (\ref{3:lf}) transform homogeneously, $X' = U(z=0)X$, so that
$M(q,A)$ transforms as the product of gauge matrices, $U(0)\otimes U(0)$,
taken at space-time point zero and in the appropriate representation of the color group.
As it is easy to see, Eq.~(\ref{trule}) guarantees that
$\langle\langle{M}(q^{cl}, A^{cl})\rangle\rangle$ transforms in exactly the same way.
This means that the counterterms to the product $X(z_1)\otimes X(z_2)$
are given in terms of products of the fields~$X$ themselves.
For example, the gluon field $A^{cl}_\mu$ can only appear as a part of the
field strength tensor, $F^{cl}_{\mu\nu}$, or inside a covariant derivative.

The advantage of using the light-cone gauge is that it makes explicit the two-particle structure
of the RG equations. The price one pays for this property is, however,
breaking of the Lorentz invariance. In particular, ``$+$'' and ``$-$'' components
of the fields are renormalized in a different
way
$$[q_{\pm}]_0=Z_{\pm} q_\pm, \qquad [A_\mu]_{(0)}={R_\mu}^\nu A_\nu\,.$$
At one loop \cite{Bassetto}
%
\begin{align}\label{Zquark}
Z_+=1+\frac{3\alpha_s}{8\pi\epsilon}C_F\,,&&
Z_-=1-\frac{\alpha_s}{8\pi\epsilon}C_F\,,
\end{align}
and
\begin{align}\label{Z-A}
R_{\mu\nu}=Z_{3}^{1/2}\left[g_{\mu\nu}-
\left(1-\widetilde Z_3^{-1}\right)\frac{n_\mu \bar n_\nu}{(n\bar n)}\right],
\end{align}
with
\begin{align}\label{Zgluon}
Z_3\,=\,1+\frac{\alpha_s}{4\pi\epsilon}\left(\frac{11}{3} N_c-\frac23 n_f\right)\,,\qquad
\widetilde Z_3\,=\,1+\frac{\alpha_s}{2\pi\epsilon} N_c\,.
\end{align}
{}The renormalized coupling constant is related to the bare one as $g_0=Z_3^{-1/2} g$.
{}In physical quantities such as the $S$-matrix or correlation functions
of gauge-invariant operators, the Lorentz invariance is, of course, restored.

For more details on the renormalization of QCD in the light-cone gauge see
Ref.~\cite{Bassetto,Bassetto-2} and references therein.

\section{Quasipartonic Operators}\label{QO}

The primary field $X$ (\ref{3:X}) has six different ``plus'' components so that one has to know $21$
$(2\to2)$ kernels to describe the renormalization of an arbitrary quasipartonic operator.
Parity and charge conjugation symmetry leave only seven independent kernels which were all
calculated in Ref.~\cite{Bukhvostov:1985rn}. These results are summarized below.

Conformal symmetry dictates that the two-particle kernels acting on the coordinates of the fields
must have the general form in Eq.~(\ref{2:inv})~\footnote{
Group theory tells us  that a nontrivial kernel $\mathcal{H}$ which satisfies~(\ref{2:HT})
exists if and only if $j_1+j_2=j'_1+j'_2+m$, where $m$ is an integer number.
For our purpose it is sufficient to consider the following cases: $j_1=j'_1,~j_2=j'_2$,
$j_1=j'_1\pm1/2,~j_2=j'_2\pm 1/2$. All of them can be reduced to the first
one using that
$\mathcal{H}^{j_1j_2\to j_1-1/2,j_2-1/2}= z_{12}\,\mathcal{H}^{j_1j_2\to j_1j_2}$
and
$\mathcal{H}^{j_1j_2\to j_1+1/2,j_2+1/2}= z_{12}^{-1}\,\mathcal{H}^{j_1j_2\to j_1j_2}$.
}.
In one-loop calculations in QCD only a limited amount
of ``standard'' functions appear:
\begin{eqnarray}\label{H-list}
{}[\widehat{\mathcal{H}}^{\phantom{v}}\varphi](z_1,z_2)&=&
\int_0^1\frac{d\alpha}{\alpha}\Big[2\varphi(z_1,z_2)-
\bar\alpha^{2j_1-1}\varphi(z_{12}^\alpha,z_2)-\bar\alpha^{2j_2-1}\varphi(z_1,z_{21}^\alpha)\Big]\,,
\\
{}[{\mathcal{H}}^d\varphi](z_1,z_2)&=&\int_0^1d\alpha\,
\bar\alpha^{2j_1-1}\alpha^{2j_2-1}\,\varphi(z_{12}^\alpha,z_{12}^\alpha)\,,
\\
{}[\mathcal{H}^+\varphi](z_1,z_2)&=&\int_0^1d\alpha\int_0^{\bar\alpha}d\beta\,
\bar\alpha^{2j_1-2}\bar\beta^{2j_2-2}\,\varphi(z_{12}^\alpha,z_{21}^\beta)\,,
\\
{}[\widetilde{\mathcal{H}}^+\varphi](z_1,z_2)&=&\int_0^1d\alpha\int_0^{\bar\alpha}d\beta\,
\bar\alpha^{2j_1-2}\bar\beta^{2j_2-2}\,\left(\frac{\alpha\beta}{\bar\alpha\bar\beta}\right)
\varphi(z_{12}^\alpha,z_{21}^\beta)\,,
\\
{}[\mathcal{H}^-\varphi](z_1,z_2)&=&\int_0^1d\alpha\int_{\bar\alpha}^1d\beta\,
\bar\alpha^{2j_1-2}\bar\beta^{2j_2-2}\,\varphi(z_{12}^\alpha,z_{21}^\beta)\,
\label{4:HH}
\end{eqnarray}
all of which correspond to a ``diagonal'' mapping of conformal spins
$T^{j_1}\otimes T^{j_2}\to T^{j_1}\otimes T^{j_2}$.
In particular the operator $B(2j_1,2j_2)^{-1}\mathcal{H}^d$, where $B(x,y)$ is Euler beta
function, defines the projector  $\mathcal{P}_0$
onto the invariant subspace $T^{j_1+j_2}$ with the lowest spin $n=0$ in the tensor product
decomposition in Eq.~(\ref{2:TD}). We will also use the notation $\Pi_0$
for the projector
\begin{align}\label{Pi0}
\Pi_0=I-\mathcal{P}_0=I-B(2j_1,2j_2)^{-1}\mathcal{H}^d\,.
\end{align}
More functions, $\mathcal{H}^{e,(k)}$ with $0 < k < 2 j_1$,
are needed  for the case that the conformal spins are reshuffled as
$T^{j_1}\otimes T^{j_2}\to T^{j_1-k/2}\otimes T^{j_2+k/2}$
\begin{eqnarray}
[\mathcal{H}^{e,(k)}_{12}\varphi](z_1,z_2)&=&\int_0^1d\alpha\,\bar\alpha^{2j_1-k-1}\,
\alpha^{k-1} \varphi(z_{12}^\alpha,z_2)\,,\qquad 0 < k < 2 j_1\,.
\end{eqnarray}
Last but not least, we need two constants
\begin{eqnarray}
  \sigma_q &=& \frac34\,,
\nonumber\\
  \sigma_g &=& b_0/4N_c\,, \qquad b_0\,=\,\dfrac{11}{3}N_c-\dfrac{2}3n_f\,,
\end{eqnarray}
which correspond to the "plus" quark field (\ref{Zquark}) and transverse gluon field renormalization
(\ref{Zgluon}) in the axial gauge
\begin{align}
Z_q=1+\frac{\alpha_s}{2\pi\epsilon}\sigma_q C_F\,, &&
Z_g=1+\frac{\alpha_s}{2\pi\epsilon}\sigma_g C_A\,.
\end{align}
Thanks to the constraint in Eq.~(\ref{3:S-inv}) the contribution due to field renormalization
for the color-singlet operators~(\ref{3:O}) can be rewritten as~\cite{Bukhvostov:1985rn}
\begin{eqnarray}
\lefteqn{
\Big(1-\frac{\alpha_s}{2\pi\epsilon}\sum_i \sigma_{X_i} C_{X_i}\Big)S(X_1\otimes\ldots\otimes X_N)
=}
\nonumber\\&&\hspace*{1.5cm}=\,
S\Big(1+\frac{\alpha_s}{2\pi\epsilon}\sum_{i<j}
(\sigma_{X_i}+\sigma_{X_j})t^a_i\otimes t^a_j \Big)
(X_1\otimes\ldots\otimes X_N)\,,
\end{eqnarray}
where $\sigma_{X_i} = \{\sigma_q,\sigma_g\}$ and $C_{X_i} = \{C_F,C_A\}$ for quarks and gluons,
respectively. This property allows one to include the field renormalization factors
in the definition of the two-particle kernels~\cite{Bukhvostov:1985rn}.

\subsection{Coordinate-space representation}
\begin{description}

\item[A)]{}\framebox{
$X\otimes X=\{\psi_+\otimes\psi_+,~\psi_+\otimes\chi_+,~\bar\psi_+\otimes\bar\psi_+,
 ~\bar\psi_+\otimes\bar\chi_+,~\chi_+\otimes\chi_+,~\bar \chi_+\otimes\bar \chi_+\}$
                }\\[2mm]
In compact notation that we will also use for the other cases
\begin{align}\label{4:Hqq}
\mathbb{H}\,\, X(z_1)\otimes X(z_2)=-2(t^a\otimes t^a)
\left[\widehat{\mathcal{H}}-2\sigma_q\right]
 X(z_1)\otimes X(z_2)\,.
\end{align}
The operator in square brackets on the r.h.s. acts on the function
$\varphi(z_1,z_2)=X(z_1)\otimes X(z_2)$ according to Eq.~(\ref{4:HH}) where one has to
substitute the values $j_1=j_2=1$ for the conformal spins of the participating fields.
Here and below a symbolic expression $X\otimes X$ implies that both fields have open color indices
and
$$
(t^b\otimes t^b)\, X(z_1)\otimes X(z_2)\equiv  t^b X(z_1)\otimes t^b X(z_2)=
t^b_{i_1i'_1} X^{i'_1}(z_1)\, t^b_{i_2i'_2} X^{i'_2}(z_2)\,.
$$

\item[B)]\framebox{
$X\otimes X=\{
\psi_+\otimes\bar\chi_+,~\bar\psi_+ \otimes\chi_+,~ \psi_+\otimes\bar\psi_+,~\bar\chi_+
\otimes\chi_+\}$ }
\\[2mm]
For quarks of different flavor:
\begin{align}\label{4:Hqbq}
\mathbb{H}\,\, X(z_1)\otimes X(z_2)=-2(t^b\otimes t^b)
\left[\widehat{\mathcal{H}}-\mathcal{H}^+-2\sigma_q\right]
 X(z_1)\otimes X(z_2)\,.
\end{align}
The generators $t^a$ acting on the quark and the antiquark fields are defined in
Eq.~(\ref{3:generators}).
After the contraction of open color indices $S_{i_1i_2}=\delta_{i_1i_2}$, cf. (\ref{3:O}),
this expression reproduces the result in Eq.~(\ref{2:Hqq}).

For the quark-antiquark pair of the same flavor, $X\otimes X=\psi\otimes \bar
\psi,\chi\otimes\bar\chi$,
there are two extra terms:
\begin{multline}\label{4:Hfs}
\mathbb{H}\, X^i(z_1)\otimes X^j(z_2)= \ldots
-4 t^a_{ij}\, \mathcal{H}^d\, {J}^a(z_1,z_2)
\\
-2iz_{12} \Big((t^at^b)_{ij}\left[\mathcal{H}^++\widetilde{\mathcal{H}}^+\right]+
2(t^bt^a)_{ij} \mathcal{H}^-
\Big)\, f^{a}_{++}(z_1)\otimes \bar f^b_{++}(z_2)\,.
\end{multline}
The ellipses stand for the contribution of Eq.~(\ref{4:Hqbq}).
Here and below
\begin{align}
P_{12}\,\varphi(z_1,z_2)=\varphi(z_2,z_1)
\end{align}
is the permutation operator acting on the field coordinates and
\begin{align}
{J}^a(z_1,z_2)=\sum_{A}
\left(\bar \psi^A_+(z_1) T^a\psi^A_+(z_2)+\chi^A_+(z_1) T^a\bar\chi^A_+(z_2)\right),
\end{align}
where the sum runs over all possible flavors.
The generators $t^a$ in Eq.~(\ref{4:Hfs}) have to be taken in the quark, $t^a_{ij}=T^a_{ij}$,
and the antiquark, $t^a_{ij}=-T^a_{ji}$, representations for the pairs
$\psi^i\otimes\bar\psi^j$ and $\chi^i\otimes\bar\chi^j$, respectively.

\item[C)]\framebox{$X^a\otimes X=\{
 f^a_{++}\otimes\psi_+,~f^a_{++}\otimes\chi_+,~\bar f^a_{++}\otimes \bar\psi_+,~\bar
f^a_{++}\otimes \bar\chi_+\}$
                }
\begin{eqnarray}\label{4:Hfq}
\mathbb{H}\,\,X^a(z_1)\otimes X^{i}(z_2)&=&-2(t^b_{aa'}\otimes t^b_{ii'})
\left[\widehat{\mathcal{H}}-\sigma_q-\sigma_g\right]X^{a'}(z_1)\otimes X^{i'}(z_2)
\nonumber\\
&& - 2(t^{a'}t^a)_{ii'}P_{12}\mathcal{H}^{e,(1)}_{12}\,X^{a'}(z_1)\otimes X^{i'}(z_2)\,.
\end{eqnarray}
Here $j_1=3/2$, $j_2=1$ and the $SU(N)$ generator acting on the first
(gluon) field (in the first line) is in the adjoint representation.

\item[D)]\framebox{$X^a\otimes X=\{
 f^a_{++}\otimes\bar\psi_+,~f^a_{++}\otimes\bar \chi_+,~\bar f^a_{++}\otimes \psi_+,~\bar
f^a_{++}\otimes \chi_+\}$
                }
\begin{eqnarray}\label{4:Hfbq}
\mathbb{H}\,X^a(z_1)\otimes X^{i}(z_2)&=&-2(t^b_{aa'}\otimes t^b_{ii'})
\left[\widehat{\mathcal{H}}-2\mathcal{H}^+-\sigma_q-\sigma_g\right]X^{a'}(z_1)\otimes X^{i'}(z_2)
\nonumber\\
&&{} + 4(t^{a'}t^a)_{ii'}\mathcal{H}^{-}\,X^{a'}(z_1)\otimes X^{i'}(z_2)\,.
\end{eqnarray}

\item[E)]\framebox{$X\otimes X=\{
 f_{++}\otimes f_{++},~\bar f_{++}\otimes \bar f_{++}\}$
                }
\begin{eqnarray}\label{4:Hff}
\mathbb{H}\,\, X(z_1)\otimes X(z_2) &=& -2(t^b\otimes t^b)
\left[\widehat{\mathcal{H}}-2\sigma_g\right] X(z_1)\otimes X(z_2)\,.
\end{eqnarray}
\item[F)]\framebox{$X\otimes X=\{
 f_{++}\otimes \bar f_{++}\}$
                }
\begin{multline}\label{4:Hgbg}
\mathbb{H}\, f^a_{++}(z_1)\otimes \bar f^c_{++}(z_2)=-2(t^b_{aa'}\otimes t^b_{cc'})
\Biggl(\widehat{\mathcal{H}}-4\mathcal{H}^+ - 2\, \widetilde{\mathcal{H}}^+ -
2\sigma_g \Biggr)\,f_{++}^{a'}(z_1)\otimes \bar f_{++}^{c'} (z_2)
\\
+12 (t^{a'}t^a)_{cc'} \mathcal{H}^{-} f^{a'}_{++}(z_1)\otimes \bar f^{c'}_{++}(z_2)+
\frac{2i}{z_{12}}\Biggl(2\mathcal{H}^{+}P_{12}-P_{ac}\Biggr)\Pi_0\,{J}^{ac}(z_2,z_1)\,,
\end{multline}
where
\begin{align}
{J}^{ac}(z_1,z_2)=\sum_{A}\bar\psi^A_+(z_1) T^a T^c\psi^A_+(z_2)-
\chi^A_+(z_2) T^cT^a\bar\chi^A_+(z_1)\,
\end{align}
and
\begin{align}
P_{ac}\, {J}^{ac}(z_1,z_2) =  {J}^{ca}(z_1,z_2)\,.
\end{align}
The operator $\Pi_0$
\begin{align}
\Pi_0\varphi(z_1,z_2)=(\mathbb{I}-6\mathcal{H}^d)\varphi(z_1,z_2)
= 6\int_0^1d\alpha\,\alpha\bar\alpha\,[\varphi(z_1,z_2)-\varphi(z_{12}^\alpha,z_{12}^\alpha)]
\end{align}
is the projector onto the subspace $J_{12}>2$ in the tensor product decomposition~(\ref{2:TD}),
 $\Pi_0=\Theta(J_{12}>2)$: The function
$\psi(z_1,z_2)=\Pi_0\varphi(z_1,z_2)$ vanishes on the line $z_2=z_1$, $\psi(z,z)=0$.

\end{description}

\subsection{Invariant representation}

As discussed in Sect.~2.3, calculating the eigenvalues of the two-particle kernels on the
set of functions that correspond to irreducible representations of the $SL(2,\mathbb{R})$ group
one can write the kernels as functions of the two-particle conformal spin operator,
 $J(J-1)=(\vec{S}_1+\vec{S}_2)^2$, cf. Eq.~(\ref{2:sss}).
This representation will be useful to restore the two-particle kernels
for non-quasipartonic operators.
\begin{description}
\item[A)]{}\framebox{
$X\otimes X=\{\psi_+\otimes\psi_+,~\psi_+\otimes\chi_+,~\bar\psi_+\otimes\bar\psi_+,
 ~\bar\psi_+\otimes\bar\chi_+,~\chi_+\otimes\chi_+,~\bar \chi_+\otimes\bar \chi_+\}$
                }\\[2mm]
\begin{align}\label{4:HqqJ}
\mathbb{H}\,\, X(z_1)\otimes X(z_2)=-4(t^a\otimes t^a)
\Big[\psi(J)-\psi(1)-\sigma_q\Big]
 X(z_1)\otimes X(z_2)\,.
\end{align}
\item[B)]{}\framebox{
$X\otimes X=\{
\psi_+\otimes\bar\chi_+,~\bar\psi_+ \otimes\chi_+,~ \psi_+\otimes\bar\psi_+,~\bar\chi_+
\otimes\chi_+\}$ }
\\[2mm]
{}For quarks of different flavor:
\begin{multline}\label{4:HqbqJ}
\mathbb{H}\,\, X(z_1)\otimes X(z_2)=\\
-2(t^a\otimes t^a)
\Big[\psi(J+1)+\psi(J-1)-2\psi(1)-2\sigma_q\Big]
 X(z_1)\otimes X(z_2)\,.
\end{multline}
{}For quarks of the same flavor ($\psi\otimes \bar
\psi$ or $\chi\otimes\bar\chi$ only) there are extra terms
\begin{multline}\label{4:Hqbq-J}
\mathbb{H}\,\, X(z_1)\otimes X(z_2)=\ldots-
\frac23\, t_{ij}^a\, \delta_{J,2}\, \mathcal{J}^a(z_1,z_2)
\\
-{2i}{z_{12}}(t^at^b)_{ij}\frac{1}{(J+1)(J-2)}
\left(1-\frac{2(-1)^{J} P_{ab}}{J(J-1)}\right)
\, f^{a}_{++}(z_1)\otimes \bar f^b_{++}(z_2)\,.
%
\end{multline}
\item[C)]\framebox{$X^a\otimes X=\{
 f^a_{++}\otimes\psi_+,~f^a_{++}\otimes\chi_+,~\bar f^a_{++}\otimes \bar\psi_+,~\bar
f^a_{++}\otimes \bar\chi_+\}$
                }
\begin{multline}\label{4:HfqJ}
\mathbb{H}\,\,X^a(z_1)\otimes X^{i}(z_2)=
-2\Biggl\{(t^b_{aa'}\otimes t^b_{ii'})
\Big[\psi(J+1/2)+\psi(J-1/2)-2\psi(1)-\sigma_q-\sigma_g\Big]
 \\
+(t^{a'}t^a)_{ii'}\frac{(-1)^{J-1/2}}{J-1/2}\,\Biggr\}X^{a'}(z_1)\otimes X^{i'}(z_2)\,.
\hspace*{2cm}\phantom{.}
\end{multline}
\item[D)]\framebox{$X^a\otimes X=\{
 f^a_{++}\otimes\bar\psi_+,~f^a_{++}\otimes\bar \chi_+,~\bar f^a_{++}\otimes \psi_+,~\bar
f^a_{++}\otimes \chi_+\}$
               }
\begin{multline}\label{4:HfbqJ}
\mathbb{H}\,\,X^a(z_1)\otimes X^{i}(z_2)=
-2\Biggl\{(t^b_{aa'}\otimes t^b_{ii'})
\Big[\psi(J+3/2)+\psi(J-3/2)-2\psi(1)-\sigma_q-\sigma_g\Big]
 \\
+(t^{a'}t^a)_{ii'}2(-1)^{J-3/2}\frac{\Gamma(J-3/2)}{\Gamma(J+3/2)}\,\Biggl\}
X^{a'}(z_1)\otimes X^{i'}(z_2)\,.
\end{multline}
\item[E)]\framebox{$X\otimes X=\{
 f_{++}\otimes f_{++},~\bar f^a_{++}\otimes \bar f^a_{++}\}$
                }
\begin{eqnarray}\label{4:HffJ}
\mathbb{H}\,\, X(z_1)\otimes X(z_2) &=& -4(t^b\otimes t^b)
\Big[\psi(J)-\psi(1)-\sigma_g\Big] X(z_1)\otimes X(z_2)\,.
\hspace*{2cm}\phantom{.}
\end{eqnarray}

\item[F)]\framebox{$X\otimes X=
 f_{++}\otimes \bar f_{++}$}
\begin{multline}\label{4:HfbfJ}
\mathbb{H}\,f^a_{++}(z_1)\otimes \bar f^c_{++}(z_2)=-2\Biggl\{
(t^b_{aa'}\otimes t^b_{cc'})
\Big[\psi(J+2)+\psi(J-2)-2\psi(1)-2\sigma_g\Big]
\\
\phantom{.}\hspace*{4.5cm}
+ (t^{a'}t^a)_{cc'} 6 (-1)^{J}
\frac{\Gamma(J-2)}{\Gamma(J+2)}
\Biggr\}\,f^{a'}_{++}(z_1)\otimes \bar f^{c'}_{++}(z_2)
\\
\phantom{.}\hspace*{1.5cm}
+\frac{2i}{z_{12}}\left(\frac{2(-1)^{J}}{J(J-1)}-P_{ac} \right)
\Theta(J>2)\,
\mathcal{J}^{ac}(z_2,z_1)\,.
\hspace*{0.35cm}\phantom{.}
\end{multline}
\end{description}

Note that with the exception of quark--gluon transitions 
(that are only possible for 
$X_1\otimes X_2 = f\otimes \bar f$ or flavor-singlet  $\psi\otimes\bar\psi,
\chi\otimes\bar\chi$ pairs), the pair-wise Hamiltonians contain two color structures only, 
$t^c\otimes t^c$ and $(t^{a'}t^a)_{ii'}$. Moreover, apart from the contributions due 
to field renormalization, the corresponding  $SL(2,\mathbb{R})$ invariant
kernels can be written in a universal form, as a function of one parameter: the difference 
of field helicities
$$ h=|h_{X_1}-h_{X_2}|\,. $$ 
The corresponding expressions are
\begin{eqnarray}
\mathcal{H}_1 &=& -2[\psi(J+h)+\psi(J-h)-2\psi(1)-\sigma_{X_1}-\sigma_{X_2}]\,,
\nonumber\\
\mathcal{H}_2 &=&-2(-1)^{J-h}\frac{\Gamma(2h)\Gamma(J-h)}{\Gamma(J+h)}\,
\end{eqnarray}
for the first, $t^c\otimes t^c$, and the second, $(t^{a'}t^a)_{ii'}$, color structures,
respectively.  

It is interesting to note that only these two Hamiltonians appear in supersymmetric
extensions of QCD, see Refs.~\cite{Belitsky:2004yg,Belitsky:2004sc}.

\section{Non-Quasipartonic Operators}\label{NQO}
As noticed in Ref.~\cite{BFKS04}, knowledge of the two-particle evolution kernels for
quasipartonic operators allows one to restore the similar kernels for
non-quasipartonic operators by ``submerging'' the $SL(2,\mathbb{R})$ subgroup into the
full conformal group $SO(4,2)$.

To understand the idea, recall that the power of conformal symmetry in the description
of renormalization of the quasipartonic operators is due to the fact that the tensor
product of any two one-particle representations of the $SL(2,\mathbb{R})$ group
is  decomposed into irreducible components without multiplicities, cf. Eq.~(\ref{2:TD}).
Different irreducible components can be labeled by the eigenvalues of the
single quadratic two-particle Casimir operator $S^2_{12}$ (\ref{2:sss}). It follows that
the evolution kernels $\mathbb{H}$ and the Casimir operator $S^2_{12}$ have
the same eigenfunctions and can be diagonalized simultaneously; hence
$\mathbb{H}$ can be written as a function of $S^2_{12}$, as exemplified in Eq.~(\ref{2:kurz}).
This is what we call an invariant representation.
It is easy to see that the non-degeneracy is of principle importance: if two operators
exist with the same transformation properties, they can mix in an arbitrary way.

The full conformal group $SO(4,2)$ is more complicated and the same property does {\it not}
hold for the tensor product of arbitrary representations: in general, there are nontrivial
multiplicities. The observation of Ref.~\cite{BFKS04} is that the degeneracy does not occur, however,
for the products of representations of the special type that we need for the construction of
the QCD operators.

It can be shown that the tensor product of two primary fields (\ref{3:X}) is
decomposed into irreducible components of $SO(4,2)$ without multiplicities. Schematically
\begin{align}
\mathbb{T}_1\otimes \mathbb{T}_2=\sum_{n}\mathbb{T}^{(n)}\,,
\end{align}
and each term (component) in the sum can be characterized uniquely by the value of the
quadratic (two-particle) Casimir operator $\mathbb{C}^2_{12}$ of the full conformal group
(cf. Appendix A).
Following the same line of argument one concludes that the evolution kernel (Hamiltonian) $\mathbb{H}$
acting on an arbitrary two-particle light-ray operator must be
a certain function of $\mathbb{C}^2_{12}$ alone, which means that any operator from a
particular irreducible component $\mathbb{T}^{(n)}$ is an eigenvector of $\mathbb{H}$
with the same eigenvalue $E_n$. On the other hand,
each $\mathbb{T}^{(n)}$ contains a $SL(2,\mathbb{R})$ invariant subspace
corresponding to quasipartonic operators.
Thus the eigenvalues $E_n$ on the subspaces $\mathbb{T}^{(n)}$, and therefore the operator
$\mathbb{H}$ itself, can be restored from its dependence on $S^2_{12}$
in the quasipartonic sector.

Using this strategy, we derive below a complete set of  $2\to 2$ evolution
kernels for arbitrary operators containing one ``plus'' and one ``minus'' field
in the coordinate-space representation.

\subsection{Fields of the same chirality}
In this section we calculate the two-particle evolution kernels for the tensor product
of two light-ray fields of the same chirality.

With the restriction to collinear twist $E=3$, there are four such operators:
\begin{eqnarray}
\Phi_-(z_1)\otimes \Phi_+(z_2),&& \Phi_+(z_1)\otimes \Phi_-(z_2),
\nonumber\\
\Phi_+(z_1)\otimes D_{-+}\Phi_+(z_2), && D_{-+}\Phi_+(z_1)\otimes \Phi_+(z_2),
\end{eqnarray}
where $\Phi_{\pm}$ are defined in Eq.~(\ref{defPhi}). The first two,
$\Phi_-(z_1)\otimes \Phi_+(z_2)$ and $\Phi_+(z_1)\otimes \Phi_-(z_2)$,
have the same quantum numbers so they can mix under renormalization
and the same is true for the the second pair,
$\Phi_+(z_1)\otimes D_{-+}\Phi_+(z_2)$ and $D_{-+}\Phi_+(z_1)\otimes \Phi_+(z_2)$.
On the contrary, light-ray operators with and without an extra transverse derivative,
e.g. $\Phi_-(z_1)\otimes \Phi_+(z_2)$ and $\Phi_+(z_1)\otimes D_{-+}\Phi_+(z_2)$,
cannot mix because they have different helicity, cf.~Tab.~1.
As a consequence, the evolution kernel has a $2\times2$ matrix form, e.g.
\begin{align}\label{Hmatrix}
\mathbb{H}\begin{pmatrix}\Phi_-\otimes \Phi_+\\
\Phi_+\otimes \Phi_-
\end{pmatrix}=\begin{pmatrix} \mathbb{H}_{11}&\mathbb{H}_{12}\\
\mathbb{H}_{21}&\mathbb{H}_{22}
\end{pmatrix}
\begin{pmatrix}\Phi_-\otimes \Phi_+\\
\Phi_+\otimes \Phi_-
\end{pmatrix}\,
\end{align}
and similar for the other pair.

From the group theory point of view,
the diagonal kernels $\mathbb{H}_{11}$ and $\mathbb{H}_{22} $
are the $SL(2,\mathbb{R})$ invariant operators which map
$T^{j_1}\otimes T^{j_2}\to T^{j_1}\otimes T^{j_2}$. The
off-diagonal kernels intertwine the representations
$T^{j_2}\otimes T^{j_1}$ and  $T^{j_1}\otimes T^{j_2}$,
$$
\mathbb{H}_{12}:T^{j_2}\otimes T^{j_1}\to T^{j_1}\otimes T^{j_2}\qquad
\mathbb{H}_{21}:T^{j_1}\otimes T^{j_2}\to T^{j_2}\otimes T^{j_1}\,.
$$
Our task is to find explicit expressions for these kernels.

In what follows we explain the method in detail on the particular example of the operator
$\psi_-\otimes \psi_+$. The other cases are similar so we only present the results.

As the first step, one has to compare the action of $SO(4,2)$ and $SL(2,\mathbb{R})$ Casimir operators
on the tensor product of two ``plus'' fields. Using Eqs.~(\ref{A:CJ}) and (\ref{A:Ckappa})
from Appendix~\ref{App:A} one finds
\begin{align}\label{C++}
\left[\mathbb{C}^2_{12}+\frac32\right]\,\psi_+(z_1)\otimes\psi_+(z_2)=S_{12}^2\,
\psi_+(z_1)\otimes\psi_+(z_2)=-\partial_1\partial_2\,z_{12}^2\,\psi_+(z_1)\otimes\psi_+(z_2)\,,
\end{align}
where we used the explicit expression for $S_{12}^2$ (\ref{S12})
(for the case $j_1=j_2=1$) to arrive at the last equality.
Note that the action of $SO(4,2)$ and $SL(2,\mathbb{R})$ Casimir operators
on the ``plus'' fields do not coincide, but differ by a constant (\ref{A:Ckappa}).

Thanks to the $SL(2,\mathbb{R})$ invariance it is sufficient to consider the action of the
Casimir operators on functions of two variables that are annihilated by the ``step-down''
operator $(S_1^-+S_2^-)\varphi(z_1,z_2)=0$. This condition is nothing but shift-invariance:
$\varphi(z_1,z_2) = \varphi(z_1-z_2) \equiv \varphi(z_{12})$. It is easy to
see that $\varphi_n(z_1,z_2)=z_{12}^n$ are the eigenfunctions corresponding to the
eigenvalue $(n+2)(n+1)$:
\begin{eqnarray}
  \Big[-\partial_1\partial_2\,z_{12}^2\Big]\, z_{12}^n = j_n(j_n-1)\, z_{12}^n\,,\qquad j_n=n+2\,.
\end{eqnarray}

The second step is to find the explicit expression for the Casimir operator acting on
the $\psi_-\otimes\psi_+$ fields. It has a matrix form
\begin{align}\label{cab}
\left[\mathbb{C}^2_{12}+\frac32\right]
\begin{pmatrix}\psi_-\otimes \psi_+\\\psi_+\otimes \psi_-
\end{pmatrix}=\begin{pmatrix}a&b\\c&d\end{pmatrix}
\begin{pmatrix}
\psi_-\otimes \psi_+\\\psi_+\otimes \psi_-
\end{pmatrix}\,.
\end{align}
The entries $a,b,c,d$ can be found using Eq.~(\ref{A:Jfull}) and explicit
expressions for the conformal generators~(\ref{A:conf}).
This is easy for the diagonal entries $a$ and $d$, because the terms in
the last two lines in Eq.~(\ref{A:Jfull}) do not contribute. One obtains
\begin{align}
a=S_{12}^{2,(j_1=1/2,j_2=1)}+\frac14\,, && d=S_{12}^{2,(j_1=1,j_2=1/2)}+\frac14\,.
\end{align}
The expression for $S_{12}^2$ is again given by~Eq.~(\ref{S12}), but with different values
of the conformal spins as compared to the ``plus--plus'' case. Hereafter
we will not show the values of $j_1,j_2$ explicitly, unless this can lead to a confusion.
It is always implied that parameters that enter the definitions of the operators are
determined by quantum numbers of the fields they act on.

The direct calculation of the off-diagonal elements $b,c$ is rather cumbersome
because in this case all terms in Eq.~(\ref{A:Jfull}) have to be taken into account.
These entries can easily be found, however, with the help of the following trick.

Let us apply the operation $\lambda(\partial/\partial\mu)$ to the both sides of~(\ref{cab}).
Taking into account that by definition $\psi_- =  \mu^\alpha\psi_\alpha$ one obtains
\begin{align}\label{cab-1}
\left[\mathbb{C}^2_{12}+\frac32\right]
\begin{pmatrix}\psi_+\otimes \psi_+\\\psi_+\otimes \psi_+
\end{pmatrix}=\begin{pmatrix}a&b\\c&d\end{pmatrix}
\begin{pmatrix}
\psi_+\otimes \psi_+\\\psi_+\otimes \psi_+
\end{pmatrix}\,.
\end{align}
The l.h.s. of this equality is given by Eq.~(\ref{C++}) so that obviously
\begin{align}
b=S_{12}^{2, (j_1=j_2=1)}-a=\partial_2\, z_{21}\,,&& c=S_{12}^{2, (j_1=j_2=1)}-d=\partial_1\, z_{12}\,.
\end{align}
It is easy to check that the operators $b$ and $c$ intertwine the representations
$T^{1/2}\otimes T^{1}$ and $T^{1}\otimes T^{1/2}$,
$$
b:T^{1}\otimes T^{1/2}\overset{z_{21}}{\longrightarrow}T^{1/2}\otimes T^{0}
\overset{\partial_2}{\longrightarrow}T^{1/2}\otimes T^{1}\,.
$$

Collecting everything, we obtain
\begin{align}
\left[\mathbb{C}^2_{12}+\frac32\right]
\begin{pmatrix}\psi_-\otimes \psi_+\\\psi_+\otimes \psi_-
\end{pmatrix}=\begin{pmatrix}S_{12}^2+1/4&\partial_2z_{21}\\\partial_1z_{12}&S_{12}^2+1/4
\end{pmatrix}
\begin{pmatrix}
\psi_-\otimes \psi_+\\\psi_+\otimes \psi_-
\end{pmatrix}\,.
\end{align}
Remarkably enough, one can represent this operator in the form
\begin{align}
\left[\mathbb{C}^2_{12}+\frac32\right]=\widehat J_{12}\left(\widehat J_{12}-1\right)\,,
\end{align}
where
\begin{align}
\widehat J_{12}=-\begin{pmatrix}0&\partial_2z_{21}\\\partial_1z_{12}&0
\end{pmatrix}\,.
\end{align}
The eigenfunctions of the operator $\widehat J_{12}$ have the form
\begin{align}
\psi_n^\pm(z_1,z_2)=z_{12}^n\,\begin{pmatrix}1\\ \pm 1\end{pmatrix}, && \widehat
J_{12}\psi_n^\pm=\mp(n+1) \psi_n^\pm\,,
\end{align}
so that
\begin{eqnarray}
   \left[\mathbb{C}^2_{12}+\frac32\right]\psi_n^\pm=j_n^\pm(j_n^\pm-1)\psi_n^\pm ,\qquad
  j^+_n=n+2\,, \qquad  j^-_n=(n+1)\,.
\end{eqnarray}

As stated above~\cite{BFKS04},
the Hamiltonian $\mathbb{H}$ must be a function of the full Casimir
operator of the conformal group. This means that they share the same set of eigenfunctions
and the eigenvalues $E_n$ of $\mathbb{H}$ are functions of $j_n$. Moreover, the
expressions for $E_n=E(j_n)$ must be the same for the both collinear twist $E=2$ and $E=3$
sectors:
\begin{eqnarray}
 \mathbb{H}^{E=2} z_{12}^n &=& E(j_n) z_{12}^n\,,
\nonumber\\
 \mathbb{H}^{E=3} \psi_n^\pm &=& E(j_n^\pm)\psi_n^\pm\,,
\label{5:Beisert}
\end{eqnarray}
so that the function $E(j)$ is already known: It is determined by
the Hamiltonian for the quasipartonic operator $\psi_+\otimes \psi_+$, see Eq.~(\ref{4:HqqJ}):
\begin{equation}
 E(j)=2[\psi(j)-\psi(1)-\sigma_q]\,.
\end{equation}
(We omitted here the color structure.)

The remaining steps are purely technical. First, using (\ref{5:Beisert}), we can restore the
action of $\mathbb{H}$ on arbitrary shift-invariant polynomials. It is easy to see that
the dimension of the polynomial is conserved, so we only need to consider functions
of the form
$\begin{pmatrix} \alpha\\ \beta \end{pmatrix}z_{12}^n.$
Using the standard decomposition
 $ \mathbb{H}= E_+ |\psi_n^+\rangle \langle \psi_n^+| + E_-|\psi_n^-\rangle \langle \psi_n^-|$
one gets
\begin{eqnarray}
&&\mathbb{H}\begin{pmatrix} \alpha\\ \beta \end{pmatrix}z_{12}^n \,=\,
\begin{pmatrix} h_{11}(n) & h_{12}(n)\\ h_{21}(n)& h_{22}(n)\end{pmatrix}
\begin{pmatrix} \alpha\\ \beta\end{pmatrix}z_{12}^n\,,
\nonumber\\
&&h_{ik}
\,=\,\frac12\left[E(j_n^+)\begin{pmatrix}1\\1\end{pmatrix}\otimes\begin{pmatrix}1,1\end{pmatrix}+
E(j_n^-)\begin{pmatrix}1\\-1\end{pmatrix}\otimes\begin{pmatrix}1,-1\end{pmatrix}\right]\,,
\end{eqnarray}
so that
\begin{align}
h_{11}(n)=&h_{22}(n)=\frac12\left(E(j^+_n)+E({j_n^-})\right)=
\psi(n+2)+\psi(n+1)-2\psi(1)-2\sigma_q\,,
\notag\\
h_{12}(n)=&h_{21}(n)=\frac12\left(E(j^+_n)-E({j_n^-})\right)=\frac{1}{n+1}\,.
\end{align}
Once the action of $\mathbb{H}$ on shift-invariant polynomials (which correspond to highest weights
of the corresponding representations) is established, one can easily restore the explicit form of
the $SL(2,R)$ invariant operators $\mathbb{H}_{ik}$ in Eq.~(\ref{Hmatrix}) from the requirement that
\begin{eqnarray}
 \begin{pmatrix} \mathbb{H}_{11}&\mathbb{H}_{12}\\\mathbb{H}_{21}&\mathbb{H}_{22} \end{pmatrix}
  \begin{pmatrix} \alpha\\ \beta \end{pmatrix}z_{12}^n &=&
 \begin{pmatrix} h_{11}(n) & h_{12}(n)\\ h_{21}(n)& h_{22}(n)\end{pmatrix}
\begin{pmatrix} \alpha\\ \beta\end{pmatrix}z_{12}^n\,.
\end{eqnarray}
One obtains
\begin{align}
2[\psi(J_{12})-\psi(1)]\begin{pmatrix}\psi_-\otimes \psi_+\\\psi_+\otimes \psi_-
\end{pmatrix}=
\begin{pmatrix} \widehat{\mathcal{H}} &\mathcal{H}^{e,1}_{12}\\
\mathcal{H}^{e,1}_{21} &\widehat{\mathcal{H}}
\end{pmatrix}\begin{pmatrix}\psi_-\otimes \psi_+\\\psi_+\otimes \psi_-
\end{pmatrix}\,
\end{align}
and
\begin{align}
\label{5:psipsi}
\mathbb{H}\begin{pmatrix}\psi_-\otimes \psi_+\\\psi_+\otimes \psi_-
\end{pmatrix}=
-2(t^b\otimes t^b)\begin{pmatrix} \widehat{\mathcal{H}}-2\sigma_q &\mathcal{H}^{e,1}_{12}\\
\mathcal{H}^{e,1}_{21} &\widehat{\mathcal{H}}-2\sigma_q
\end{pmatrix}\begin{pmatrix}\psi_-\otimes \psi_+\\\psi_+\otimes \psi_-
\end{pmatrix},
\end{align}
which is our final result.

For all other cases the calculation is similar so that we only present
the resulting expressions for the $SO(4,2)$ Casimir operator
and the Hamiltonians. Let
\begin{align}\label{5:chiral_doublets}
[X_+]^{j_1j_2}_{i_1 i_2}(z_1,z_2)=\begin{pmatrix}\Phi_-^{j_1}\otimes \Phi_+^{j_2}\\
                                    \Phi_+^{i_1}\otimes \Phi_-^{i_2}
\end{pmatrix},
\qquad
[X_-]^{j_1j_2}_{i_1i_2}(z_1,z_2)=\begin{pmatrix}\Phi_+^{j_1}\otimes [\bar D_{-+}\Phi_+]^{j_2}\\
                                    [\bar D_{+-}\Phi_+]^{i_1}\otimes \Phi_+^{i_2}
\end{pmatrix},
\end{align}
where  $j_1,j_2$ and $i_1=j_1+1/2,i_2=j_2-1/2$ are the conformal spin of the
``upper'' and the ``lower'' pair of fields. Below we assume that $j_1\leq j_2$ so that
in the case of a quark-gluon pair the first field is always the quark and the second one
is the gluon.  The Casimir operator takes the form
\begin{align}\label{CJJ}
\mathbb{C}^2_{12}+\kappa_{12}=\widehat J_{12}\left(\widehat J_{12}-1\right)\,,
\end{align}
where $\kappa_{12}$ is defined in Appendix~\ref{App:A}, Eq.~(\ref{A:Ckappa}) and
has to be calculated for the corresponding pair of ``plus'' fields $\Phi_+\otimes \Phi_+$.
The operator $\widehat J_{12}$ is defined as
\begin{align}
\widehat J_{12}=(i_2-j_1)\mathbb{I}\mp\begin{pmatrix}0& z_{21}\partial_2+2i_2\\
z_{12}\partial_1+2j_1&0\end{pmatrix}.
\end{align}
Here and below the upper sign corresponds to $X_+$ and the lower one to $X_-$.

The Hamiltonians for the ``doublets''
\begin{align}
X_+
=&\Biggl\{\begin{pmatrix}\psi_-\otimes\psi_+\\
               \psi_+\otimes\psi_-
\end{pmatrix},
\begin{pmatrix}\psi_-\otimes\chi_+\\
               \psi_+\otimes\chi_-
\end{pmatrix},
\begin{pmatrix}\chi_-\otimes\psi_+\\
               \chi_+\otimes\psi_-
\end{pmatrix},
\begin{pmatrix}\chi_-\otimes\chi_+\\
               \chi_+\otimes\chi_-
\end{pmatrix},
\begin{pmatrix}f_{+-}\otimes f_{++}\\
                f_{++}\otimes f_{+-}
\end{pmatrix}\Biggr\}, \notag
\\
X_-=&\Biggl\{
\begin{pmatrix}\psi_+\otimes \bar D_{-+}\psi_+\\
               \bar D_{-+}\psi_+\otimes\psi_+
\end{pmatrix},
\begin{pmatrix}\psi_+\otimes \bar D_{-+}\chi_+\\
               \bar D_{-+}\psi_+\otimes\chi_+
\end{pmatrix},  \ldots\,,
\begin{pmatrix}f_{++}\otimes \bar D_{-+}f_{++}\\
                \bar D_{-+} f_{++}\otimes f_{++}
\end{pmatrix}\Biggr\},
\end{align}
(which can be viewed as the higher twist $E=3$ descendants of the  quasipartonic $E=2$
operators of type
\framebox{{\bf A}} and \framebox{\bf E}~)
take the form
\begin{align}
\mathbb{H}X_\pm(z_1,z_2)=-2(t^b\otimes t^b)\begin{pmatrix} \widehat{\mathcal{H}}-\sigma_{X_1X_2}
&\pm\mathcal{H}^{e,1}_{12}\\
\pm\mathcal{H}^{e,1}_{21} &\widehat{\mathcal{H}}-\sigma_{X_1X_2}
\end{pmatrix}X_\pm(z_1,z_2)\,.
\end{align}
Here $\sigma_{X_1 X_2}=\sigma_{X_1}+\sigma_{X_2}$,
$\sigma_X=\{\sigma_q,\sigma_g\}$ and we suppress all color indices (cf. Eq.~(\ref{4:Hqq})).
The Hamiltonian in the antichiral sector is exactly the same.

Next, for the operator pairs
\begin{align}
X_+^{ia}
(z_1,z_2)=&\Biggl\{\begin{pmatrix}\psi_-^i\otimes f^a_{++}\\
               \psi_+^i\otimes f^a_{+-}
\end{pmatrix},
\begin{pmatrix}\chi_-^i\otimes f^a_{++}\\
               \chi_+^i\otimes f^a_{+-}
\end{pmatrix}\Biggr\}
\end{align}
and
\begin{align}
X_-^{ia}(z_1,z_2)=&
\Biggl\{
\begin{pmatrix}\psi_+^i\otimes [\bar D_{-+} f_{++}]^a\\
                [D_{-+}\psi_+]^i\otimes f_{++}^a
\end{pmatrix},
\begin{pmatrix}\chi_+^i\otimes [\bar D_{-+} f_{++}]^a\\
                [D_{-+}\chi_+]^i\otimes f_{++}^a
\end{pmatrix}\Biggr\},
\end{align}
which are the descendants of the quasipartonic operators of type \framebox{{\bf C}},
one gets
\begin{align}
\mathbb{H}X_\pm^{ia}(z_1,z_2)=&-2\Bigg\{(t^b_{ii'}\otimes t^b_{aa'})\begin{pmatrix}
\widehat{\mathcal{H}}-\sigma_{X_1X_2}
&\pm\mathcal{H}^{e,1}_{12}\\
\pm\mathcal{H}^{e,1}_{21} &\widehat{\mathcal{H}}-\sigma_{X_1X_2}
\end{pmatrix}
\notag\\
&\phantom{-2\Biggr\{}\mp (t^{a'}t^a)_{ii'}P_{12}
\begin{pmatrix} \mathcal{H}_{21}^{e,2} &\mp\mathcal{H}^{e,1}_{21}\\
\mp\mathcal{H}^{e,1}_{21} &0
\end{pmatrix}\Biggr\}
\, X_{\pm}^{i'a'}(z_1,z_2)\,.
\end{align}
We remind that all kernels depend on the conformal spins of the fields they
are acting on.

\subsection{Fields of opposite chirality}

The two-particle evolution kernels for the tensor product of two light-ray fields of opposite
chirality can be calculated along the similar lines. A difference is that e.g.
$\bar \Phi_-(z_1)\otimes \Phi_+(z_2)$ cannot mix under renormalization with
$\bar \Phi_+(z_1)\otimes \Phi_-(z_2)$ (since chirality is conserved) and the same is true for the
pair $\bar \Phi_+(z_1)\otimes D_{-+}\Phi_+(z_2)$ and $D_{-+}\bar\Phi_+(z_1)\otimes \Phi_+(z_2)$.
At the same time, mixing of the light-ray operators with and without an extra transverse derivative,
e.g. $\Phi_-(z_1)\otimes \bar \Phi_+(z_2)$ and $\Phi_+(z_1)\otimes D_{-+}\bar \Phi_+(z_2)$,
is allowed.

{}For this reason, instead of (\ref{5:chiral_doublets}), we have to consider the ``doublets''
\begin{align}\label{5:mixed_doublets}
X^{j_1j_2}_{i_1 i_2}(z_1,z_2)=\begin{pmatrix}\Phi_-^{j_1}\otimes \bar\Phi_+^{j_2}\\
                               \Phi_+^{i_1}\otimes \frac12 [D_{-+}\bar \Phi_+]^{i_2}
\end{pmatrix},
\end{align}
where we take $\Phi=\{\psi,\chi,f\}$, $\bar \Phi=\{\bar\psi,\bar\chi,\bar f\}$;
$j_1,j_2$ and $i_1=j_1+1/2$, $i_2=j_2+1/2$ are the corresponding conformal spins.

 The $SO(4,2)$ Casimir operator acting on the the light-ray fields (\ref{5:mixed_doublets})
 can be represented in the form~(\ref{CJJ}) with
\begin{align}
\widehat J=-(j_1+j_2-1/2)\mathbb{I}-\begin{pmatrix}0&z_{21}\\
                                                 F_{12}&0
\end{pmatrix},
\end{align}
where
\begin{equation}\label{S12mixed}
  F_{12} = - \partial_1\partial_2 \,z_{21} + 2(i_1-1)\partial_2
-2(i_2-1)\partial_1\,.
\end{equation}	
The eigenfunctions of $\widehat J$ are
\begin{align}\label{psipm}
\psi_n^+(z_1,z_2)=&z_{21}^{n-1}\begin{pmatrix}z_{21}\\n\end{pmatrix}, \qquad n\geq 0\,,
\\
\psi_n^-(z_1,z_2)=&z_{21}^{n-1}\begin{pmatrix}z_{21}\\ \alpha_n\end{pmatrix},\qquad
n>0\,,
\end{align}
where $\alpha_n=-n-2(j_1+j_2)+1$ and
\begin{align}
\widehat J\,\psi_n^\pm=\mp(n+j_1+j_2-1/2)\psi_n^\pm\,.
\end{align}
Thus, for the Casimir operator~(\ref{CJJ}) one obtains
\begin{align}
[\mathbb{C}_2^{(12)}+\kappa_{12}]\,\psi_n^\pm=j_n^\pm(j_n^\pm-1)\,\psi_n^\pm\,,&&
j_n^\pm=n+j_1+j_2\pm1/2\,,
\end{align}
where, as above, the value of $\kappa_{12}$~(\ref{A:Ckappa}) has to be calculated
using the helicities of the corresponding ``plus'' fields $\Phi_+\otimes \bar\Phi_+$,
cf. Tab.~1.

Using these expressions and proceeding along the same lines as in Sect.~5.1 we can restore
the Hamiltonian in the coordinate space representation.

{}First, consider the ``doublets'' that arise as descendants of the
quasipartonic operators of type \framebox{{\bf B}}:
\begin{align}
X=\Biggl\{\begin{pmatrix}\psi_-\otimes\bar \psi_+\\
                                   \psi_+\otimes \frac12D_{-+}\bar\psi_+\end{pmatrix},
\begin{pmatrix}\chi_-\otimes\bar \psi_+\\
                                   \chi_+\otimes \frac12D_{-+}\bar\psi_+\end{pmatrix},
\begin{pmatrix}\psi_-\otimes\bar \chi_+\\
                                   \psi_+\otimes \frac12D_{-+}\bar\chi_+\end{pmatrix},
\begin{pmatrix}\chi_-\otimes\bar \chi_+\\
                                   \chi_+\otimes \frac12D_{-+}\bar\chi_+\end{pmatrix}
\Biggr\}.
\end{align}
We assume here that the quarks $\psi_-\otimes\bar \psi_+$ are in the flavor-nonsinglet state, so
that there is no mixing with gluons. In this case one finds
\begin{align}\label{Hpsibpsi}
\mathbb{H}\, X(z_1,z_2)=-2(t^b\otimes t^b)\begin{pmatrix}
             \widehat{\mathcal{H}}+ \mathcal{H}^d-2\sigma_q& &z_{21}\mathcal{H}_{12}^+\\
             \dfrac{1}{z_{21}}\Pi_0& &\widehat{\mathcal{H}}-2\mathcal{H}_{12}^+-2\sigma_q
\end{pmatrix}\, X(z_1,z_2)\,,
\end{align}
where $\Pi_0$ is defined by Eq.~(\ref{Pi0}).

Descendants of quasipartonic operators of type \framebox{{\bf D}} fall in two classes
that have to be considered separately.

First, consider the operators
\begin{align}
X^{ia}=\Biggl\{\begin{pmatrix}\psi_-^i\otimes\bar f^a_{++}\\
                                   \psi_+^i\otimes \frac12D_{-+}\bar f^a_{++}\end{pmatrix},
\begin{pmatrix}\chi^i_-\otimes\bar f^a_{++}\\
                                   \chi^i_+\otimes \frac12D_{-+}\bar f^a_{++}\end{pmatrix}\,
\Biggr\}.
\end{align}
The corresponding evolution kernel takes the form
%
%
%
\begin{align}
\mathbb{H}\,X^{ia}(z_1,z_2)=&
-2\Big\{(t^b_{ii'}\otimes t^b_{aa'})\mathbb{H}_1
+(t^{a'}t^{a})_{ii'}\mathbb{H}_2\Big\}
X^{i'a'}(z_1,z_2)\,.
\end{align}
The $2\times2$ matrix kernels
$\mathbb{H}_1$ and $\mathbb{H}_2$ are given by the following expressions:
\begin{equation}
\mathbb{H}_1=
   \begin{pmatrix}
             \widehat{\mathcal{H}}+ \mathcal{H}^d-\sigma_{qg} & & z_{21}\mathcal{H}_{12}^+
   \\
             \dfrac{1}{z_{21}}\Pi_0 & & \widehat{\mathcal{H}}-3\mathcal{H}_{12}^+-\sigma_{qg}
   \end{pmatrix}
,\qquad
\mathbb{H}_2=
   \begin{pmatrix}
             -\mathcal{H}^d & &z_{21} \mathcal{H}_{12}^{-}
   \\
           -\dfrac{2}{z_{21}} \,P_{12}\, \mathcal{H}^{e,2}_{21}\,\Pi_0 & &-3\mathcal{H}_{12}^-
   \end{pmatrix},
\end{equation}
where $\sigma_{qg}=\sigma_q+\sigma_g$ and $\Pi_0$ is defined by Eq.~(\ref{Pi0}).

The second set of operators is
\begin{align}
X^{ai}=\Biggl\{\begin{pmatrix}f_{+-}^a\otimes\bar \psi^i_{+}\\
                                   f_{++}^a\otimes \frac12D_{-+}\bar \psi^i_{+}\end{pmatrix},
\begin{pmatrix}f_{+-}^a\otimes\bar \chi^i_{+}\\
                                   f_{++}^a\otimes \frac12D_{-+}\bar \chi^i_{+}\end{pmatrix}
\Biggr\}.
\end{align}
In this case one obtains
\begin{align}
\mathbb{H}\,X^{ai}(z_1,z_2)=& -2\Big\{(t^b_{aa'}\otimes t^b_{ii'})\mathbb{H}_1
+(t^{a'}t^{a})_{ii'}\mathbb{H}_2\Big\} X^{a'i'}(z_1,z_2)\,
\end{align}
with
\begin{eqnarray}
\mathbb{H}_1&=&
\begin{pmatrix}
             \widehat{\mathcal{H}}-\mathcal{H}^+ +2\mathcal{H}^d-\sigma_{qg}&
z_{21} \left(\mathcal{H}_{12}^+ + 
\widetilde{\mathcal{H}}_{12}^+
\right)\\
             \frac{1}{z_{21}}\Pi_0&
\widehat{\mathcal{H}}-4\mathcal{H}_{12}^+ - 
2\widetilde{\mathcal{H}}_{12}^+
-\sigma_{qg}
\end{pmatrix},
\nonumber\\
\mathbb{H}_2&=&
\begin{pmatrix}
             -2\mathcal{H}^d & 2z_{21} \mathcal{H}_{12}^{-}
\\
           -\frac{2}{z_{21}} \mathcal{H}_{12}^{-}\Pi_0 &-6\mathcal{H}_{12}^-
\end{pmatrix}.
\end{eqnarray}
%
%
%
%

Next, we consider the renormalization of  $f \bar f$ operators which are the descendants
of quasipartonic operators of type $\framebox{\bf F}$. We define
\begin{eqnarray}
G^{ab}&=&\begin{pmatrix}f_{+-}^a\otimes \bar f_{++}^b\\
                           f_{++}^a\otimes \frac12D_{-+}\bar f_{++}^b
\end{pmatrix},
\nonumber\\
\widehat{\mathcal J}^{ab}(z_1,z_2)&=&\begin{pmatrix}
\bar\psi_+(z_1) T^a T^b\psi_-(z_2)-\chi_-(z_2)T^bT^a\bar\chi_+(z_1)\\
\frac12 D_{-+}\bar\psi_+(z_1) T^a
T^b\psi_+(z_2)-\chi_+(z_2)T^bT^a\frac12 D_{-+} \bar\chi_+(z_1)
\end{pmatrix},
\label{5:Gab}
\end{eqnarray}
where in the second operator doublet the sum over flavors is implied.
One obtains
\begin{align}
\mathbb{H}\, G^{ab}(z_1,z_2)=& -2\Big\{
(t^c_{aa'}\otimes t^c_{bb'})\,\mathbb{H}_1+(t^c_{ba'}\otimes
t^c_{ab'})\,\mathbb{H}_2\,
\Big\} G^{a'b'}(z_1,z_2)
\nonumber\\
&+\frac{2i}{z_{12}}\Big\{2\mathbb{H}_3-P_{ab}\Big\}\widetilde{\Pi}_0\,
\widehat{\mathcal J}^{ab}(z_2,z_1)\,.
\end{align}
Here $P_{ab}$ is the operator of permutations in color space.
The kernels $\mathbb{H}_i$ take the form
\begin{eqnarray}
\mathbb{H}_1&=&
\begin{pmatrix}
             \widehat{\mathcal{H}}-2 \mathcal{H}^+ +3\mathcal{H}^d-2\sigma_g& &
z_{21} \left(\mathcal{H}_{12}^+ + 2\widetilde{\mathcal{H}}_{12}^+\right)
\\
             \frac{1}{z_{21}}{\Pi}_0& &
\widehat{\mathcal{H}}-6(\mathcal{H}_{12}^++\widetilde{\mathcal{H}}_{12}^+)-2\sigma_g
\end{pmatrix},
\notag\\
\mathbb{H}_2&=&
3\begin{pmatrix}
             -\mathcal{H}^d & &z_{21} \mathcal{H}_{12}^{-}
\\
           -\dfrac{2}{z_{21}} \mathcal{H}_{12}^{-}\Pi_0 & &-4\mathcal{H}_{12}^-
\end{pmatrix},
\qquad
\mathbb{H}_3\,=\,
         \begin{pmatrix}
0  && -z_{21}\mathcal{H}_{12}^-
\\
\dfrac{1}{z_{21}} P_{12}\,\mathcal{H}^{e,1}_{21} & & 2 \mathcal{H}_{12}^-
          \end{pmatrix}.
\end{eqnarray}
The projection operator $\widetilde{\Pi}_0$ 
is defined as
\begin{align*}
\widetilde{\Pi}_0=\mathbb{I}-\mathbb{P}_0\,,
\end{align*}
where the operator $\mathbb{P}_0$ is the projector to the null subspace of the Casimir
operator, $\mathbb{C}^2_{12}\mathbb{P}_0=0$. It has the form
%
\begin{align}
\mathbb{P}_0=&\begin{pmatrix}
              2\mathcal{H}^d+\dfrac14   \mathcal{P}_1& &-3 z_{21}\mathcal{H}^d\\
             -\dfrac{3}{4z_{21}} \mathcal{P}_1&
&9\mathcal{H}^d\end{pmatrix},
\end{align}
where $\mathcal{P}_1$ is the $SL(2,R)$ projector to the subspace $J=5/2$ on the tensor
product $T^{1/2}\otimes T^{1}$
\begin{align}
[\mathcal{P}_1\varphi](z_1,z_2)=4z_{21}\int_0^1d\alpha\,\alpha^2\bar\alpha\,
[(\partial_2-2\partial_1)\varphi](z_{12}^\alpha,z_{12}^\alpha)\,.
\end{align}
Finally, we present the result for the quark-antiquark pair in the flavor-singlet
state, case \framebox{\bf B}. Let
\begin{eqnarray}
X^{AB}(z_1,z_2)&=&
\Biggl\{
\begin{pmatrix}\psi_-^{A}\otimes \bar\psi_+^B\\
                           \psi_+^{A}\otimes  \frac12 D_{-+}\bar\psi_+^{B}
\end{pmatrix},
\begin{pmatrix}\chi_-^{A}\otimes \bar\chi_+^B\\
                           \chi_+^{A}\otimes  \frac12 D_{-+}\bar\chi_+^{B}
\end{pmatrix}
\Biggr\},
\\
%
\widehat {\mathcal J}^a(z_1,z_2)&=&\begin{pmatrix}\bar \psi_+(z_1) T^a \psi_-(z_2)+
\chi_-(z_2) T^a \bar\chi_+(z_1)\\
                          \frac12 D_{-+}\bar\psi_+(z_1)T^a\psi_+(z_2)+
\chi_+(z_2)T^a\frac12 D_{-+}\bar\chi_+(z_1)
\end{pmatrix},
\end{eqnarray}
where in the first operator $A,B$ are flavor indices of the quark fields and
in the second operator the sum over flavors is implied. One obtains
\begin{eqnarray}
\mathbb{H}\,X^{AB}(z_1,z_2)&=&-2(t^a\otimes t^a)\,\mathbb{H}_1\, X^{AB}(z_1,z_2)
-\delta^{AB}\frac23t^a_{ij}\,\mathbb{P}_0\,\widehat {\mathcal J}^a(z_2,z_1)
\nonumber\\&&{}
-\delta^{AB}\,2iz_{12}(t^a t^b)_{ij}\left[
\mathbb{H}_2+P^{ab}\mathbb{H}_3
\right] \,G^{ab}(z_1,z_2)\,,
\end{eqnarray}
where $G^{ab}(z_1,z_2)$ is defined in Eq.~(\ref{5:Gab}) and
\begin{eqnarray}
\mathbb{H}_1&=&\begin{pmatrix}
             \widehat{\mathcal{H}}+ \mathcal{H}^d-2\sigma_q& &z_{21}\mathcal{H}_{12}^+\\
             \dfrac{1}{z_{21}}\Pi_0&
&\widehat{\mathcal{H}}-2\mathcal{H}_{12}^+-2\sigma_q\end{pmatrix},
\qquad
\mathbb{H}_2=\begin{pmatrix}
\mathcal{H}^+_{12}-\mathcal{H}^d & & -z_{21}\widetilde{\mathcal{H}}_{12}^+
\\
-\dfrac{1}{z_{21}}\mathcal{H}^+_{12}\Pi_0 & &\mathcal{H}^+_{12}+ 3\widetilde{\mathcal{H}}_{12}^+
\end{pmatrix},
\nonumber\\
\mathbb{H}_3&=&\begin{pmatrix}
             \mathcal{H}^d & &-z_{21} \mathcal{H}_{12}^{-}
\\
           \dfrac{2}{z_{21}} \mathcal{H}_{12}^{-}\Pi_0 & &4\mathcal{H}_{12}^-
\end{pmatrix}.
\end{eqnarray}
%


\section{Mixing with Three-Particle Operators}
%
As it was already mentioned in Sect.~\ref{Renorm}, light-ray operators with a
different number of constituents can mix with each other. To the leading order in strong
coupling this mixing has a triangular form~(\ref{BigH}) so that at the level of kernels
we need the $2\to3$ contributions corresponding to the mixing of one "plus" and one
"minus" primary fields with the three-particle quasipartonic operators. It seems at first
sight that the diagonal and off-diagonal blocks in the matrix~(\ref{BigH}) are
independent and a separate calculation is necessary to fix the missing off-diagonal
terms. Surprisingly enough this is not the case. We will show below that all $2\to 3$
kernels are completely determined by the $2\to 2$ kernels in the quasipartonic sector.

In short, the main idea is that the necessary relations are imposed by Lorentz
symmetry. For example, applying the generator of translations in transverse direction
to the light cone to the "plus" quark field one obtains
\begin{equation}\label{6:Pt}
i[{\mathbf P}_{\mu\bar\lambda},\psi_+]=\partial_{\mu\bar\lambda}\psi_+=
2\partial_+\psi_- + igA_{\mu\bar\lambda}\psi_+
+~\text{EOM}\,,
\end{equation}
so that the similar transformation applied to the renormalized quasipartonic light-ray
operator yields
\begin{eqnarray}\label{6:RDD}
\partial_{\mu\bar\lambda}[\psi_+(z_1)\otimes\psi_+(z_2)]_R&=&
[\partial_{\mu\bar\lambda}\psi_+(z_1)\otimes\psi_+(z_2)]_R
+[\psi_+(z_1)\otimes\partial_{\mu\bar\lambda}\psi_+(z_2)]_R
\nonumber\\
&=&2\partial_{z_1}[\psi_-(z_1)\otimes\psi_+(z_2)]_R+
2\partial_{z_2}[\psi_+(z_1)\otimes\psi_-(z_2)]_R
\nonumber\\
&&{}+ig[A_{\mu\bar\lambda}\psi_+(z_1)\otimes\psi_+(z_2)]_R
+ig[\psi_+(z_1)\otimes A_{\mu\bar\lambda}\psi_+(z_2)]_R
\nonumber\\
&&{}+\text{EOM}.
\end{eqnarray}
Eq.~(\ref{6:RDD}) is an operator identity which must be satisfied by the corresponding
operator counterterms order by order in perturbation theory. The operator on the l.h.s.
is quasipartonic and its renormalization is governed by the corresponding BFLK kernel. As
we will see below, application of the transverse derivative to the two-particle
counterterm will generate a sum of contributions of two-particle and three-particle
operators. The operators $\psi_-(z_1)\otimes\psi_+(z_2)$ and
$\psi_+(z_1)\otimes\psi_-(z_2)$ in the second line in Eq.~(\ref{6:RDD}) contain both
two-particle counterterms that we have calculated in the previous Section and the
three-particle counterterms that we do not know so far. Finally, the three-particle
quark-quark-gluon operators in the third line in Eq.~(\ref{6:RDD}) are again
quasipartonic and their renormalization is described in terms of the two-particle BFLK
kernels. Thus Eq.~(\ref{6:RDD}) provides one with a relation for the $2\to 3$ mixing
kernels for the $\psi_-\otimes\psi_+$ light-ray operators and the BFLK kernels. The
question is whether this constraint is sufficient to determine the $2\to 3$ kernels in
principle, and how to solve it in practice. We have found two possibilities to proceed.

The first one is to derive another constraint, applying Lorentz
rotations $M_{\mu\mu}$ instead of $P_{\mu\bar\lambda}$ to the quasipartonic
$\psi_+\otimes\psi_+$ operator. We are able to prove that, taken together, these two
constraints determine the $2\to 3$ kernels uniquely.
An obvious advantage of this approach is that it only relies on exact Lorentz symmetry
and so it may also be applicable beyond one loop. The main disadvantage is that
derivation of the constraint imposed by the $M_{\mu\mu}$ transformation is more
complicated because it also affects the axial gauge fixing term in the QCD Lagrangian.
The corresponding contributions must be taken into account but, at least to the one-loop
accuracy, are relatively simple because of cancellations between different Feynman diagrams
that have a structure typical for a Ward identity.

The second possibility which we eventually found to be the most effective, is to study
properties of the constraint equation corresponding to Eq.~(\ref{6:RDD}) under the
collinear conformal transformations. As it stands, this equation is not $SL(2)$
invariant, but, as we will explain, it can be separated in two $SL(2)$-invariant
equations that are already sufficient to determine the $2\to 3$ kernels of interest.
To the one loop accuracy, both approaches are equivalent and produce identical results.
Some of the results presented below have also been checked by the direct calculation
of relevant Feynman diagrams.

It is worthwhile to note that the same technique can be used to determine the $2\to 2$ kernels
as well.
In this case, however, the approach based on the construction of the $SO(4,2)$ Casimir
operator~\cite{BFKS04}, Sect.~5, is clearly advantageous.

In what follows we explain the details of our approach on the concrete example of
the $\psi_-\otimes \psi_+$ operator and then present the results for all kernels in question.
{}For this example, we consider the following set of operators:
\begin{eqnarray}
\mathcal{O}^{ik}_{+}(z_1,z_2)&=&\psi^i_{+}(z_1)\otimes\psi^k_+(z_2)\,,
\notag\\
\mathcal{O}^{ik}_{1}(z_1,z_2)&=&\psi^i_{-}(z_1)\otimes\psi^k_+(z_2)\,,
\notag\\
\mathcal{O}^{ik}_{2}(z_1,z_2)&=&\psi^i_{+}(z_1)\otimes\psi^k_-(z_2)\,,
\notag\\
\mathcal{O}^{ikb}_f(z_1,z_2,z_3)&=&\psi^i_+(z_1)\otimes \psi^k_+(z_2)\otimes\bar f^b_{++}(z_3)\,,
\end{eqnarray}
where $i,k,b$ are the color indices that in most cases will not be displayed explicitly.
The first operator in this list is quasipartonic. To one-loop accuracy the
renormalized operator $[\mathcal{O}_+]_R$ is written as
\begin{align}\label{RHR}
[\mathcal{O}_+]_R=
   \left(1+\frac{\alpha_s}{4\pi\epsilon}\widetilde{\mathbb{H}}\right)
\mathcal{O}_+\,.
\end{align}
Here we only include the contributions of one-particle irreducible diagrams and ignore the field
renormalization; the  corresponding expression for $\widetilde{\mathbb{H}}$ is obtained from
Eq.~(\ref{4:Hqq}) by throwing away the term $2\sigma_q$, i.e.
$\widetilde{\mathbb{H}}=-2(t^b \otimes t^b) \widehat{\mathcal{H}}$.
Below we also use the notation
$$[O]'_R=[O]_R-O$$ for the divergent (one-loop) contribution
$\sim 1/\epsilon$ so that e.g.
$[\mathcal{O}_+]'_R = \alpha_s/4\pi\epsilon\,\widetilde{\mathbb{H}}\mathcal{O}_+$.

Similar, for the ``plus-minus'' operators $\mathcal{O}_{1,2}$ we define
\begin{align}
\label{Ocount}
[\mathcal{O}_i]'_R(z_1,z_2)=&\frac{\alpha_s}{4\pi\epsilon}\left\{
[\widetilde{\mathbb{H}}^{i\to k} \mathcal{O}_k](z_1,z_2)+[\mathbb{H}^{i\to f}\mathcal{O}_f](z_1,z_2)
\right\}
\end{align}
and our task will be to find explicit expressions for the $2\to 3$
kernels $\mathbb{H}^{1\to f}$ and $\mathbb{H}^{2\to f}$.

\subsection{${\mathbf P}_{\mu\bar\lambda}$ transformation}\label{Ptrafo}
%
An equation for the three-particle counterterms
$\mathbb{H}^{1\to f}$ and $\mathbb{H}^{2\to f}$ can be obtained by the application of the
transverse derivative
$\partial_{\mu\bar\lambda}=\mu^\alpha\partial_{\alpha\dot\alpha}\bar\lambda^{\dot\alpha}$
to the renormalized operator in Eq.~(\ref{RHR}). One obtains
\begin{align}\label{RDD}
\partial_{\mu\bar\lambda}[\mathcal{O}_+]_R=
[\partial_{\mu\bar\lambda}\psi_+\otimes\psi_+]_R+[\psi_+\otimes\partial_{\mu\bar\lambda}\psi_+]_R.
\end{align}
We remind that a light-ray field is defined including the gauge link
$\psi_\pm(z)\equiv [0,z]\psi_\pm(z)$, cf.~(\ref{3:lf}), so that
\begin{equation}
 \partial_{\mu\bar\lambda}\psi_+(z) \equiv
\sigma^\alpha_{\mu\bar\lambda}\frac{\partial}{\partial y^{\alpha}}
[y,zn+y]\psi_+(zn+y){|_{y=0}}\,.
\end{equation}
It is convenient to impose the light-cone gauge condition $A_+=0$ on the background field,
the same as for the quantum field $a_+=0$.
In this case $[y,nz+y]=1$ and $\partial_{\mu\bar\lambda}[y,nz+y] = 0$, so that the gauge links
can be ignored throughout this calculation.

Making use of the Fierz identity for Weyl spinors,  $(ab)(cd)=(ac)(bd)-(ad)(bc)$,
one can rewrite $\partial_{\mu\bar\lambda}\psi_+^i$ as follows
\begin{align}
\partial_{\mu\bar\lambda}\psi_+^i=&[D_{\mu\bar\lambda}\psi_+]^i+
ig t^b_{ii'} A^b_{\mu\bar\lambda}\psi^{i'}_+
=2\partial_+\psi_-^i + ig t^b_{ii'} A^b_{\mu\bar\lambda}\psi^{i'}_+-(\mu\lambda)[\bar\lambda\bar
D\psi]^i\,,
\end{align}
where we also used that $\bar\lambda \bar D\lambda = \lambda D\bar\lambda = 2\partial_+$
thanks to the gauge condition $A_+=0$. The ``plus'' derivative can further be reduced to the
derivative over the  light-cone coordinate:
$\partial_+\psi_-(z) \equiv n^\alpha(\partial/\partial y^\alpha)\psi_-(zn+y)|_{y=0}
= (\partial/\partial z)\psi_-(zn)$.
Thus one obtains, e.g. for the first term in Eq.~(\ref{RDD})
\begin{eqnarray}\label{RWR}
[\partial_{\mu\bar\lambda}\psi_+(z_1)\otimes\psi_+(z_2)]_R &=&
2\partial_{1}[\psi_-(z_1)\otimes\psi_+(z_2)]_R
+ig (t^b\otimes I)
[A^b_{\mu\bar\lambda}(z_1)\psi_+(z_1)\otimes\psi_+(z_2)]_R
\notag\\
&&{}-
(\mu\lambda)[\bar\lambda\bar
D\psi(z_1)\otimes\psi_+(z_2)]_R\,.
\end{eqnarray}
Here and below we do not show color indices for the quarks and adopt a shorthand notation
$(t^b\otimes I) (\psi_+\otimes\psi_+)\equiv (t^b_{ii'}\otimes I_{kk'})
(\psi^{i'}_+\otimes\psi^{k'}_+)$, etc.

Combining Eqs.~(\ref{RWR}),~(\ref{RDD}) and (\ref{RHR}) one derives for the
divergent contribution $\sim 1/\epsilon$:
\begin{eqnarray}\label{WI-1}
\lefteqn{2\partial_{1}[\mathcal{O}_1(z_1,z_2)]'_R+
2\partial_{2}[\mathcal{O}_2(z_1,z_2)]'_R\,=}
\nonumber\\
&=& \partial_{\mu\bar\lambda} [\mathcal{O}_+(z_1,z_2)]'_R
%
+(\mu\lambda)\Big\{[\bar\lambda
\bar D\psi(z_1)\otimes\psi_+(z_2)]'_R+[\psi_+(z_1)\otimes \bar\lambda \bar D\psi(z_2)]'_R\Big\}
\nonumber\\
&-&ig (t^b\otimes I)
[ A^b_{\mu\bar\lambda}(z_1)\psi_+(z_1)\otimes\psi_+(z_2)]'_R
-ig (I\otimes t^b)
[\psi_+(z_1)\otimes\psi_+(z_2) A^b_{\mu\bar\lambda}(z_2)]'_R
\end{eqnarray}
The l.h.s. of the above  equation contains the operators we are interested in. The r.h.s. is a
bit messy so that we still have to do some rewriting. Though we are interested in
contributions of three-particle operators only, for completeness we will keep trace of
all (singular) terms.

Let us start with the second term on the r.h.s. of Eq.~(\ref{WI-1}) which contains the EOM,
$\bar\lambda\bar D\psi(z_1)$. It can be rewritten as follows
\begin{equation}
(\mu\lambda)[\bar\lambda\bar
D\psi(z_1)\otimes\psi_+(z_2)]_R
=(\mu\lambda)[\bar\lambda\bar D\psi(z_1)]_R\otimes\psi_+(z_2)
\end{equation}
and further
\begin{eqnarray}
 (\mu\lambda)[\bar\lambda\bar D\psi]_R &=& Z_-(\mu\lambda)\bar\lambda [\bar D\psi]_0
 = Z_-(2Z_-\partial_+\psi_- - Z_+ D_{\mu\bar\lambda}\psi_+)
\nonumber\\&=&
  2 Z_-(Z_- - Z_+)\partial_+\psi_- + Z_-Z_+ (\mu\lambda) \bar\lambda\bar D\psi\,,
\end{eqnarray}
where $[D\psi]_0$ means that the fields and the coupling constant which enter this expression are
replaced by the bare ones. The factor $Z_+$ and $Z_-$ are the field renormalization constants of
``plus'' and ``minus'' quark field, Eq.~(\ref{Zquark}), and we also used that
$gA_{\mu\bar\lambda}$ (in the covariant derivative) is not renormalized.
We obtain
\begin{eqnarray}\label{EOMZ}
(\mu\lambda)[\bar\lambda\bar D\psi(z_1)\otimes\psi_+(z_2)]'_R &=&
\Big(Z_-Z_+-1\Big)(\mu\lambda) \bar \lambda\bar D\psi(z_1)\otimes\psi_+(z_2)
\notag\\
&&{}+ 2Z_-(Z_-
-Z_+)\partial_{z_1}\psi_-(z_1)\otimes\psi_+(z_2)\,.
\end{eqnarray}
The first term on the r.h.s. of (\ref{EOMZ}) is again EOM, which can safely be omitted, and
the second term gives a contribution to the $2\to2$ kernels but not to the $2\to3$ kernels.

Next, consider the contributions in the third line of Eq.~(\ref{WI-1}).
Since the counterterms all have two-particle structure in one loop,
one can split e.g. the first contribution,
$[A^b_{\mu\bar\lambda}(z_1)\psi_+(z_1)\otimes\psi_+(z_2)]'_R$, in three terms:
\begin{align*}
[\psi_+(z_1)\otimes\psi_+(z_2)]'_R\,
A^b_{\mu\bar\lambda}(z_1) +& [\psi_+(z_1)A^b_{\mu\bar\lambda}(z_1)]'_R\otimes \psi_+(z_2) +
\psi_+(z_1)\otimes[\psi_+(z_2)
A^b_{\mu\bar\lambda}(z_1)]'_R\,.
\end{align*}
The first term it given by Eq.~(\ref{RHR}). The second term involves renormalization
of a local operator. A straightforward calculation gives
\begin{align}\label{af}
[A_{\mu\bar\lambda}^b\psi^i_+]'_R=-\frac1{\epsilon}\frac{ig}{16\pi^2}
t^a_{ii'}\Big\{
-4 D_+\psi^{i'}_- + (\mu\lambda) (\bar\lambda \bar D\psi)^{i'}
\Big\}\,.
\end{align}
Note that the gluon field only enters the covariant derivative, as expected in the
background field formalism. The expression in (\ref{af}) is a one-particle counterterm
which multiplies $ \psi_+(z_2)$, so it contributes to the $2\to2$ kernels but not to
the $2\to3$ ones.
To handle the third term we rewrite it as
\begin{eqnarray}
[\psi_+(z_2)
A^b_{\mu\bar\lambda}(z_1)]'_R&=&[\psi_+(z_2)
\left(A^b_{\mu\bar\lambda}(z_1)-A^b_{\mu\bar\lambda}(z_2)\right)]'_R+
[\psi_+(z_2)A^b_{\mu\bar\lambda}(z_2)]'_R\,
\nonumber\\&=&
-z_{12}(\mu\lambda)\int_0^1\!\!d\tau\,
[\psi_+(z_2)\bar f^b_{++}(z_{12}^\tau)]'_R+[\psi_+(z_2)A^b_{\mu\bar\lambda}(z_2)]'_R\,,
\label{RHS3}
\end{eqnarray}
where we used the identity
\begin{align}
\label{identity}
A^b_{\mu\bar\lambda}(z_1)-A^b_{\mu\bar\lambda}(z_2)=
-z_{12}(\mu\lambda)\int_0^1d\tau\, \bar f^b_{++}(z_{12}^\tau)\,.
\end{align}
The second term in (\ref{RHS3}) is again a one-particle local counterterm, Eq.~(\ref{af}),
so it will not contribute to the $2 \to 3$ kernels, and the first term can be
written in terms of the BFLK kernel of type \framebox{\bf D} in Sect.~\ref{QO},
omitting the field renormalization.

To simplify the following expressions, we introduce the ``averaging'' operator
$\mathbb{S}$ acting on a function of three variables such that
\begin{align}\label{Sdef}
[\mathbb{S}\varphi](z_1,z_2)=\int_{z_2}^{z_1}\!ds\,\varphi(z_1,z_2,s)=
z_{12}\int_0^1d\tau\,\varphi(z_1,z_2,z_{12}^\tau)\,.
\end{align}
In this notation
\begin{eqnarray}\label{PF1}
\psi_+(z_1)\otimes[\psi_+(z_2)A^b_{\mu\bar\lambda}(z_1)]'_R &=&
-(\mu\lambda) \frac{\alpha_s}{4\pi\epsilon}[\mathbb{S}\,\widetilde{\mathbb{H}}_{23}\,
\mathcal{O}_f^b](z_1,z_2) +\ldots
\end{eqnarray}
where the two-particle operator $\widetilde{\mathbb{H}}_{23}$ is given in Eq.~(\ref{4:Hfbq})
omitting the terms in $\sigma_q,\sigma_g$.
The subscripts in $\widetilde{\mathbb{H}}_{23}$ specify that the operator acts on the
coordinates (and color indices)
of the second and the third field in the operator $\mathcal{O}_f(z_1,z_2,z_3)$.
The ellipses stand for the contributions of local two-particle counterterms, Eq.~(\ref{af}).

Adding the second term
$\sim [\psi_+(z_1)\otimes\psi_+(z_2) A^b_{\mu\bar\lambda}(z_2)]'_R$ one obtains for the
full contribution in the last line in  Eq.~(\ref{WI-1}):
\begin{eqnarray}\label{line-3}
&&\hspace*{-1cm}(-ig) \frac{\alpha_s}{4\pi\epsilon}\Big\{
  \Big[(t^b\otimes I) A^b_{\mu\bar\lambda}(z_1)+ (I\otimes t^b ) A^b_{\mu\bar\lambda}(z_2)\Big]\,
 \widetilde{\mathbb{H}}\, \mathcal{O}_+(z_1,z_2)
\nonumber\\
&&\hspace*{0.5cm}{}
- (t^b\otimes I)(\mu\lambda)[\mathbb{S}\,\widetilde{\mathbb{H}}_{23}\,
      \mathcal{O}_f^b](z_1,z_2)
+ (I\otimes t^b)(\mu\lambda)[\mathbb{S}\,\widetilde{\mathbb{H}}_{13}\,
      \mathcal{O}_f^b](z_1,z_2) \Big\}+\ldots
\end{eqnarray}

Last but not least, we have to deal with the first term on the r.h.s. of Eq.~(\ref{WI-1}).
The explicit expression for $\widetilde{\mathbb{H}}\,\mathcal{O}_+$ is given by an
integral over the light-cone positions of the product of fields $\psi_+\otimes\psi_+$,
see Eqs.~(\ref{4:Hqq}) and (\ref{H-list}).
The transverse derivative $\partial_{\mu\bar\lambda}$ of this equation can be worked out
with the help of Eq.~(\ref{RWR}):
\begin{eqnarray}\label{PO1}
\hspace*{-1.0cm}
\partial_{\mu\bar\lambda}\,[\widetilde{\mathbb{H}}\mathcal{O}_{+}](z_1,z_2) &=&
2\,\Big[\widetilde{\mathbb{H}}\Big(\partial_{1}\mathcal{O}_{1}+
\partial_{2}\mathcal{O}_{2}\Big)\Big](z_1,z_2) +
\notag\\
&&\hspace*{-0.1cm}{}+ ig\,  \widetilde{\mathbb{H}}\,
\Big[(t^b\otimes I) A^b_{\mu\bar\lambda}(z_1)+ (I\otimes t^b ) A^b_{\mu\bar\lambda}(z_2)\Big]\,
\mathcal{O}_{+}(z_1,z_2)+\text{EOM}.
\end{eqnarray}
{}Finally, collecting everything and using the definitions of the kernels in Eq.~(\ref{Ocount})
we obtain (omitting EOM contributions)
\begin{eqnarray}
\label{EQ-1}
\sum_{i,k=1}^2\partial_i [\widetilde{\mathbb{H}}^{i\to k} \mathcal{O}_k](z_1,z_2)&=& \sum_{k=1}^2
[(\widetilde{\mathbb{H}}+2\,t^b\otimes t^b)\,\partial_{k}\mathcal{O}_{k}](z_1,z_2)\,,
\\
\label{EQ-2}
\sum_{k=1}^2\partial_k[\mathbb{H}^{k\to f}\mathcal{O}_f](z_1,z_2)&=&
\frac12ig\Big\{[(\widetilde{\mathbb{H}}\Delta-\Delta \widetilde{\mathbb{H}})\mathcal{O}_{+}](z_1,z_2)
\notag\\
&&{}+(\mu\lambda)\Big[\mathbb{S} \Big((t^b\otimes I)\widetilde{\mathbb{H}}_{23}-
(I\otimes t^b)\widetilde{\mathbb{H}}_{13}\Big)\mathcal{O}^b_f\Big](z_1,z_2)
\Big\}\,
\end{eqnarray}
for the contributions of two-particle and three-particle operators, respectively.
Here we introduced the operator $\Delta$ which acts as
\begin{align}\label{OD}
[\Delta\varphi]^{ij}(z_1,z_2)=\Big\{(t^b_{ii'}\otimes I_{jj'}) A^b_{\mu\bar\lambda}(z_1)+
(I_{ii'}\otimes t^b_{jj'}) A^b_{\mu\bar\lambda}(z_2)
\Big\}\,\varphi^{i'j'}(z_1,z_2)\,.
\end{align}
This operator depends explicitly on the gluon field $A$. The commutator
$[\widetilde{\mathbb{H}},\Delta]$ in Eq.~(\ref{EQ-2}) can, however, be rewritten in
terms of the field strength tensor $\bar f_+$ as a consequence of the identity
$[(t^a\otimes I)+(I\otimes t^a), (t^b\otimes t^b)]=0$ which can easily
be verified. Thanks to this identity the commutator $[\widetilde{\mathbb{H}},\Delta]$
would vanish if the gluon fields in the first and the second term in Eq.~(\ref{OD})
were taken at the same space-time point. Hence
$[\widetilde{\mathbb{H}},\Delta] \sim A^b_{\mu\bar\lambda}(z_1)- A^b_{\mu\bar\lambda}(z_2)$
and this difference can be rewritten in terms of $\bar f_+$ using Eq.~(\ref{identity}).

Taking into account the explicit expressions for the BFLK kernels
$\widetilde{\mathbb{H}},~\widetilde{\mathbb{H}}_{23},~\widetilde{\mathbb{H}}_{13}$ one can
bring Eq.~(\ref{EQ-2}) to the following (final) form:
\begin{align}\label{first}
\sum_{k=1}^2[\partial_k\mathbb{H}^{k\to f}\mathcal{O}_f](z_1,z_2)=g(\mu\lambda)
\sum_{i=1}^3 C_i [(\mathbb{S}\mathcal{W}_i-\mathcal{T}_i)\mathcal{O}_f](z_1,z_2)\,,
\end{align}
where $C_i$ are the color structures:
\begin{align}\label{}
C_1=f^{bcd}(t^b\otimes t^c)\,,&&C_2=i(t^b\otimes t^dt^b)\,, && C_3=-i(t^dt^b\otimes t^b)\,,
\end{align}
the operator $\mathbb{S}$ is defined in Eq.~(\ref{Sdef}), $\mathbb{W}_i$ are given
in terms of the invariant kernels as~\footnote{Recall that
the kernels depend implicitly on the conformal spins of the fields they act on.}
\begin{align}
\mathcal{W}_1 =
\widehat{\mathcal{H}}_{23}+\widehat{\mathcal{H}}_{23}-\widehat{\mathcal{H}}_{12}-
2(\mathcal{H}_{23}^+ + \mathcal{H}_{13}^+)\,,
&&
\mathcal{W}_2=2\mathcal{H}_{23}^{-}\,,
&&
\mathcal{W}_3=2\mathcal{H}_{13}^{-}
\end{align}
and
\begin{align}\label{}
\mathcal{T}_1=\mathcal{V}_{13}+\mathcal{V}_{23}\,, &&
\mathcal{T}_2=\mathcal{V}_{23}\,, &&
\mathcal{T}_2=\mathcal{V}_{13}\,,
\end{align}

with
\begin{align}\label{}
[\mathcal{V}_{13}\varphi](z_1,z_2)=&z_{12}\int_0^1d\alpha\int_{\bar\alpha}^1d\beta\,\frac{\bar
\alpha}{\alpha}\,
\varphi(z_{12}^\alpha,z_2,z_{21}^\beta)\,,
\notag\\
[\mathcal{V}_{23}\varphi](z_1,z_2)=&z_{12}\int_0^1d\alpha\int_{\bar\alpha}^1d\beta\,\frac{\bar
\alpha}{\alpha}\,
\varphi(z_1,z_{21}^\alpha,z_{12}^\beta)\,.
\end{align}
It remains  to solve this equation and find the explicit expressions for the $2\to 3$ kernels
$\mathbb{H}^{1,2\to f}$.

\subsection{$SL(2,\mathbb{R})$ decomposition}
%

Taking matrix elements of the operator $\mathcal{O}_f$ over suitable states one can view
Eq.~(\ref{first})
as an operator identity
\begin{align}\label{First}
\sum_{k=1}^2\partial_k\mathbb{H}^{k\to f}=g(\mu\lambda)
\sum_{i=1}^3 C_i (\mathbb{S}\mathcal{W}_i-\mathcal{T}_i)\,
\end{align}
on the space of functions $\varphi^{ija}(z_1,z_2,z_3) \mapsto \varphi^{ij}(z_1,z_2)$.
We will show that the kernels $\mathbb{H}^{2\to f}$ are completely determined by this equation.
To this end we have to examine the properties of Eq.~(\ref{First}) under the
$SL(2,\mathbb{R})$ conformal transformations.

The conformal invariance of one-loop QCD evolution equations implies that
the kernels of interest, $\mathbb{H}^{1\to f}$ and $\mathbb{H}^{2\to f}$, are
$SL(2,\mathbb{R})$--invariant operators mapping
\begin{align}\label{intertwine1}
\mathbb{H}^{1\to f}:{}&T^{1}\otimes T^{1}\otimes T^{3/2} \mapsto T^{1/2}\otimes T^{1},
\notag\\
\mathbb{H}^{2\to f}:{}&T^{1}\otimes T^{1}\otimes T^{3/2} \mapsto T^{1}\otimes T^{1/2}.
\end{align}
The precise statement is that the operators
$\mathbb{H}^{k\to f}$ intertwine $SL(2,\mathbb{R})$ generators in the corresponding representations:
Let
\begin{equation}
\vec{S}_{12}^{(j_1,j_2)}=\vec{S}_{1}^{j_1}+\vec{S}_{2}^{j_2}\,,
\qquad\vec{S}_{123}^{(j_1,j_2,j_3)}=\vec{S}_{1}^{j_1}+\vec{S}_{2}^{j_2}+\vec{S}_{3}^{j_3}\,
\end{equation}
be the generators acting on the two-particle and three-particle states with
the given conformal spins, respectively. For example, the first equation in
(\ref{intertwine1}) means that
\begin{equation}
 \mathbb{H}^{1\to f} \vec{S}_{123}^{(1,1,3/2)} = \vec{S}_{12}^{(1/2,1)}\,  \mathbb{H}^{1\to f}
\label{intertwine2}
\end{equation}
and similar for the second operator.

In turn, for the operators $\mathcal{W}_i, \mathbb{S},\mathcal{T}_i$ appearing
on the r.h.s. of  Eq.~(\ref{First}) one obtains
\begin{align}\label{}
\mathcal{W}_i:{}&T^{1}\otimes T^{1}\otimes T^{3/2} \mapsto T^{1}\otimes T^{1}\otimes T^{3/2},
\notag\\
\mathbb{S}:{}&T^{1}\otimes T^{1}\otimes T^{1} \mapsto T^{1}\otimes T^{1},
\notag\\
\mathcal{T}_i:{}&T^{1}\otimes T^{1}\otimes T^{1} \mapsto T^{1}\otimes T^{1}\,,
\end{align}
where the first equation follows from the definition of $\mathcal{W}_i$ in terms
of invariant kernels introduced in Sect.~4, and the last two can be checked by
explicit calculation.

Let us apply Eq.~(\ref{First}) to the step-up operator $S_{123}^{+(1,1,3/2)}$.
On the r.h.s., taking into account that $S_{123}^{+(1,1,3/2)}=S_{123}^{+(1,1,1)}+z_3$ one finds
\begin{align}\label{6:rhs}
(\mathbb{S}\mathcal{W}_i-\mathcal{T}_i)S_{123}^{+(1,1,3/2)}=
S_{12}^{+(1,1)}(\mathbb{S}\mathcal{W}_i-\mathcal{T}_i)+
(\mathbb{S}z_3\mathcal{W}_i-\mathcal{T}_i z_3)\,.
\end{align}
Similarly for the l.h.s., using $S_{12}^{+(1/2,1)}=S_{12}^{+(0,1)}+z_1$ and
$\partial_z T^{0}=T^{1}\partial_z$ one derives
\begin{align}\label{6:lhs}
\sum_{k=1}^2\partial_k\mathbb{H}^{k\to f}\,S_{123}^{+(1,1,3/2)}
=S_{12}^{+(1,1)}\,\sum_{k=1}^2\partial_k\mathbb{H}^{k\to f}+
\sum_{k=1}^2\partial_k\, z_k\,\mathbb{H}^{k\to f}\,.
\end{align}
Thanks to Eq.~(\ref{First}) the terms in (\ref{6:rhs}), (\ref{6:lhs})
multiplying $S_{12}^{+(1,1)}$ are equal; hence the remaining terms have to be equal as well:
\begin{align}\label{Second}
\sum_{k=1}^2\partial_k\, z_k\,\mathbb{H}^{k\to f}=
g(\mu\lambda)\sum_{i=1}^3 C_i(\mathbb{S}z_3\mathcal{W}_i-\mathcal{T}_i z_3)\,.
\end{align}
The two equations in (\ref{First}) and (\ref{Second}) can be rewritten equivalently as
\begin{align}\label{dva}
\partial_1z_{12}\mathbb{H}^{1\to f}+\mathbb{H}^{2\to f}=&g(\mu\lambda)\,
\sum_{i=1}^3 C_i(\mathbb{S}\,(z_3-z_2)\mathbb{W}_i-\mathcal{T}_iz_3+z_2\mathcal{T}_i)
\equiv\sum_{i=1}^3 C_i \mathcal{A}_i\,,
\notag\\
\partial_2z_{21}\mathbb{H}^{2\to f}+\mathbb{H}^{1\to f}=&
g(\mu\lambda)\,\sum_{i=1}^3 C_i(\mathbb{S}\,(z_3-z_1)\mathbb{W}_i-\mathcal{T}_iz_3+z_1\mathcal{T}_i)
\equiv \sum_{i=1}^3 {C}_i \mathcal{B}_i\,.
\end{align}
These equations are already $SL(2,\mathbb{R})$ invariant and correspond to the mappings
$T^{1}\otimes T^{1}\otimes T^{3/2} \mapsto T^{1}\otimes T^{1/2}$ and
$T^{1}\otimes T^{1}\otimes T^{3/2} \mapsto T^{1/2}\otimes T^{1}$, respectively.
The expressions on the r.h.s. can be simplified.
After a straightforward calculation one derives the following expressions
%
\begin{align}\label{}
[\mathcal{A}_1\varphi](z_1,z_2)&=z_{12}^2\left(\int_0^1d\beta \bar\beta\,\varphi(z_1,z_2,z_{12}^\beta)-
\int_0^1d\alpha\int_0^{\bar\alpha}d\beta \,\beta\, \varphi(z_1,z_{21}^\alpha,z_{12}^\beta)\right)\,,
\notag
\\
[\mathcal{A}_2\varphi](z_1,z_2)&=z_{12}^2\int_0^1d\alpha\int_{\bar \alpha}^1d\beta
\frac{\bar\alpha\bar\beta}{\alpha}
\,\varphi(z_1,z_{21}^\alpha,z_{12}^\beta)\,,
\notag
\\
[\mathcal{A}_3\varphi](z_1,z_2)&=z_{12}^2
\int_0^1d\alpha\int_{\bar \alpha}^1d\beta \frac{\bar\alpha}{\alpha^2}(\bar\alpha-\beta)\,
\varphi(z_{12}^\alpha,z_2,z_{21}^\beta)\,,
\notag\\
[\mathcal{B}_1\varphi](z_1,z_2)&=-z_{12}^2\left(\int_0^1d\beta \bar\beta\,\varphi(z_1,z_2,z_{21}^\beta)-
\int_0^1d\alpha\int_0^{\bar\alpha}d\beta \,\beta \,\varphi(z_{12}^\alpha,z_2,z_{21}^\beta)\right)\,,
\notag
\\
[\mathcal{B}_2\varphi](z_1,z_2)&=-z_{12}^2
\int_0^1d\alpha\int_{\bar \alpha}^1d\beta \frac{\bar\alpha}{\alpha^2}(\bar\alpha-\beta)\,
\varphi(z_{21}^\alpha,z_{2},z_{12}^\beta)\,,
\notag
\\
[\mathcal{B}_3\varphi](z_1,z_2)&=-z_{12}^2\int_0^1d\alpha\int_{\bar \alpha}^1d\beta
\frac{\bar\alpha\bar\beta}{\alpha}\,
\varphi(z_1,z_{12}^\alpha,z_{21}^\beta)\,.
\end{align}

As the last step, we can prove that the two equations in (\ref{dva}) or, equivalently,
Eqs.~(\ref{First}) and (\ref{Second}) determine the kernels  $\mathbb{H}^{k\to f}$ uniquely.
Using the first equation in (\ref{dva}) to eliminate $\mathbb{H}^{2\to f}$ one obtains
for $\mathbb{H}^{1\to f}$
\begin{align}\label{HW}
(\partial_2 z_{21}\partial_1 z_{12}-1)\mathbb{H}^{1\to f}\,=\,
g(\mu\lambda)\sum_{i=1}^3C_i\left(\partial_2 z_{21} \mathcal{A}_i-\mathcal{B}_i\right)\,.
\end{align}
The differential operator on the l.h.s. of this equation is nothing else as the Casimir
operator for the conformal spins $(j_1,j_2)=(1/2,1)$:\,
$(\partial_2 z_{21}\partial_1 z_{12}-1)=S_{12}^2-3/4$. This operator has zero modes
$(S_{12}^2-3/4)\psi_0=0$. However, functions of the form $\psi=\mathbb{H}^{1\to f}\varphi$
do not belong to this subspace~\footnote{
Indeed, Eq.~(\ref{intertwine2}) implies that $\mathbb{H}^{1\to f}$ maps the
eigenspace of the Casimir operator $S_{123}^2$ onto the eigenspace (with the same
eigenvalue) of the operator $S_{12}^2$.
Since, the minimal eigenvalue of $(S^2)_{123}^{(1,1,3/2)}$ is $35/4$, one concludes that
if $\psi=\mathbb{H}^{1\to f}\varphi$ then $((S^2)_{12}^{(1/2,1)}-3/4)\psi\neq 0$.}.
Therefore, this operator can be inverted and $\mathbb{H}^{1\to f}$ restored as
%
\begin{align}\label{}
\mathbb{H}^{1\to f}=g(\mu\lambda)\sum_{i=1}^3C_i
\mathbb{R}\left(\partial_2 z_{21} \mathcal{A}_i-\mathcal{B}_i\right)\,,
\end{align}
where the $\mathbb{R}$ is the inverse operator to $(S_{12}^2-3/4)$ on the subspace
that excludes zero modes, $(S_{12}^2-3/4)\psi\neq 0$:
\begin{align}\label{}
[\mathbb{R}\psi](z_1,z_2)=\int_0^1d\alpha\int_0^{\bar\alpha}d\beta\,
\frac{\bar\beta}{\bar\alpha-\beta}
\,\psi(z_{12}^\alpha,z_{21}^\beta)\,.
\end{align}
In practice, this last transformation is not necessary as the solution to Eqs.~(\ref{dva})
can easily be guessed. The final result is presented below in 
Eq.~(\ref{7:first}) in Sect.~\ref{listH}.
It coincides with the corresponding expression in Ref.~\cite{Braun:2008ia} obtained by a direct
calculation.

As the final remark, we want to note that the case of quark-quark operators with the same
chirality considered here proves to be the most complicated in this respect. In most
other cases solving the similar pair of equations is straightforward since, as it turns
out, one (or both) of the $2\to3$ kernels enters without a derivative.

\subsection{${\mathbf M}_{\mu\mu}$ transformation}\label{Mtrafo}
%

Equation~(\ref{Second}) that we derived using conformal $SL(2,\mathbb{R})$ transformations can
also be obtained in a different way, by applying the generator of Lorentz rotations
${\mathbf M}_{\mu\mu}$ to the quasipartonic operator $\mathcal{O}_+$. This alternative
derivation may be interesting as it shows that Lorentz symmetry alone is sufficient to
restore the ``nondiagonal'' $2\to 3$ operator mixing, so in what follows we outline the
main steps.

The general strategy is similar to the case of the ${\mathbf P}_{\mu\bar \lambda}$ transformation
considered in Sect.~\ref{Ptrafo}. Using the definition in (\ref{A:conf}) one obtains
\begin{equation}
   i [{\mathbf M}_{\mu\mu},\psi_+(z)] = -(\mu\lambda) \delta_M \psi_+(z)\,,
\qquad \delta_M \,=\, \frac12 z \partial_{\mu\bar\lambda} +
      \mu\frac{\partial}{\partial\lambda}\,.
\end{equation}
The first difference to the ${\mathbf P}_{\mu\bar \lambda}$ case is that contributions from the
rotation of the gauge links in the definitions of light-ray operators are nonzero and have
to be taken into account. Explicit calculation yields
\begin{equation}
  \delta_M [0,z]\psi_+(z) = (z\partial_z+1)\psi_-(z)
   -\frac{ig}{2}z^2 (\mu\lambda)\int_0^1\!du\,u \bar f_{++}(uz)\psi_+(z)\,.
\end{equation}
A bigger problem is that $\delta_M$ is not a symmetry transformation of the QCD Lagrangian
because of the gauge fixing term
\begin{equation}
   i\int d^4 x\,  \mathcal{L}_{gauge} = -\frac{i}{2\xi}\int d^4 x\,(n_\mu a^\mu(x))^2\,.
\label{gaugefix}
\end{equation}

 Applying the Lorentz rotation to the path integral, one obtains for a generic composite
 operator $O$
\begin{equation}
    {}[\delta_M O]'_R = \delta_M [O]'_R + [(\delta_M S)O]'_R
\end{equation}
i.e. there is an extra term due to the variation of the QCD action~\footnote{Strictly
speaking there are also terms due to breaking of Lorentz symmetry of the counterterms to
$\mathcal{L}_{QCD}$ itself, i.e. field and coupling renormalization. Such corrections
generally correspond to one-particle-reducible self-energy insertions and are of no
relevance here.}
\begin{eqnarray}
 \delta_M  S =  -\frac{i}{\xi}\int d^4x\,a_+(x)\delta_M a_+(x)\,.
\label{delta_S}
\end{eqnarray}

In our case one obtains
\begin{eqnarray*}
\lefteqn{
  \delta_M\mathcal{O}_+(z_1,z_2) \,=\,
  (z_1\partial_1+1)\mathcal{O}_1(z_1,z_2)+ (z_2\partial_2+1)\mathcal{O}_2(z_1,z_2)
   -}
  \nonumber\\&&{}
  -\frac{ig}{2}(\mu\lambda)\int_0^1\tau d\tau
  \Big \{z_1^2(t^b\otimes I)\mathcal{O}_f^b(z_1,z_2,\tau z_1)
   +z_2^2(I\otimes t^b)\mathcal{O}_f^b(z_1,z_2,\tau z_2)
  \Big \}
  +\text{EOM}
\end{eqnarray*}
so that
\begin{eqnarray}
\lefteqn{(z_1\partial_1+1)[\mathcal{O}_1]'_R(z_1,z_2)
            + (z_2\partial_2+1)[\mathcal{O}_2]'_R(z_1,z_2)\,=\,}
\nonumber\\&=&
  -\frac{ig}{2}(\mu\lambda)\int_0^1\tau d\tau
  \Big \{z_1^2(t^b\otimes I)[\mathcal{O}_f^b]'_R(z_1,z_2,\tau z_1)
  +z_2^2(I\otimes t^b)[\mathcal{O}_f^b]'_R(z_1,z_2,\tau z_2)\Big\}
\nonumber\\&&{}+ \delta_M[\mathcal{O}_+]'_R(z_1,z_2)
- [\mathcal{O}_+ (\delta_M  S)]'_R
+\ldots
\label{WI-22}
\end{eqnarray}
cf.~Eq.~(\ref{WI-1}). Note the l.h.s. of this equation is the same as in
Eq.~(\ref{Second}). The r.h.s. is similar to the r.h.s of Eq.~(\ref{WI-1})
except for the additional term $[\mathcal{O}_+ (\delta_M  S)]'_R$. Let us
consider this contribution more closely.

The gluon propagator corresponding to (\ref{gaugefix}) is in general (in vector notation)
\begin{equation}
  G_{\mu\nu}(k) = \frac{i}{k^2}\left[-g_{\mu\nu} + \frac{k_\mu n_\nu+n_\mu k_\nu}{kn} -
   \frac{k_\mu k_\nu}{(kn)^2} (n^2+\xi k^2)\right]
\label{axial_prop}
\end{equation}
and the light-cone gauge corresponds to taking limits $n^2\to 0$ and $\xi\to 0$.

The Lorentz rotation of the action (\ref{delta_S}) is formally $\mathcal{O}(1/\xi)$.
However, since $n^\mu G_{\mu\nu} = -i\xi k_\nu/(kn)$ (\ref{axial_prop}), each
insertion of $a_+(x)$ is effectively  $\mathcal{O}(\xi)$, so that
terms in $\delta_M S$ containing $a_+$ more than once can safely be neglected.
Hence it is sufficient to keep the first term only in the expression for the transformed field
\begin{equation}
  \delta_M a_+(x) = \frac12 a_{\mu\bar\lambda}(x)+\frac{1}{2(\mu\lambda)}(\mu x\bar\partial\mu)a_+(x)
\end{equation}
and consider zero-momentum insertions of the operator
\begin{equation}
  \begin{minipage}{3.2cm}
{\epsfxsize3.2cm\epsfbox{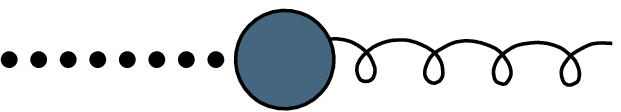}}
  \end{minipage}
 \,=\, \frac{1}{2\xi} a_+(x) a_{\mu\bar\lambda}(x)\,.
\end{equation}

\begin{figure}[t]
\centerline{\epsfysize2.9cm\epsfbox{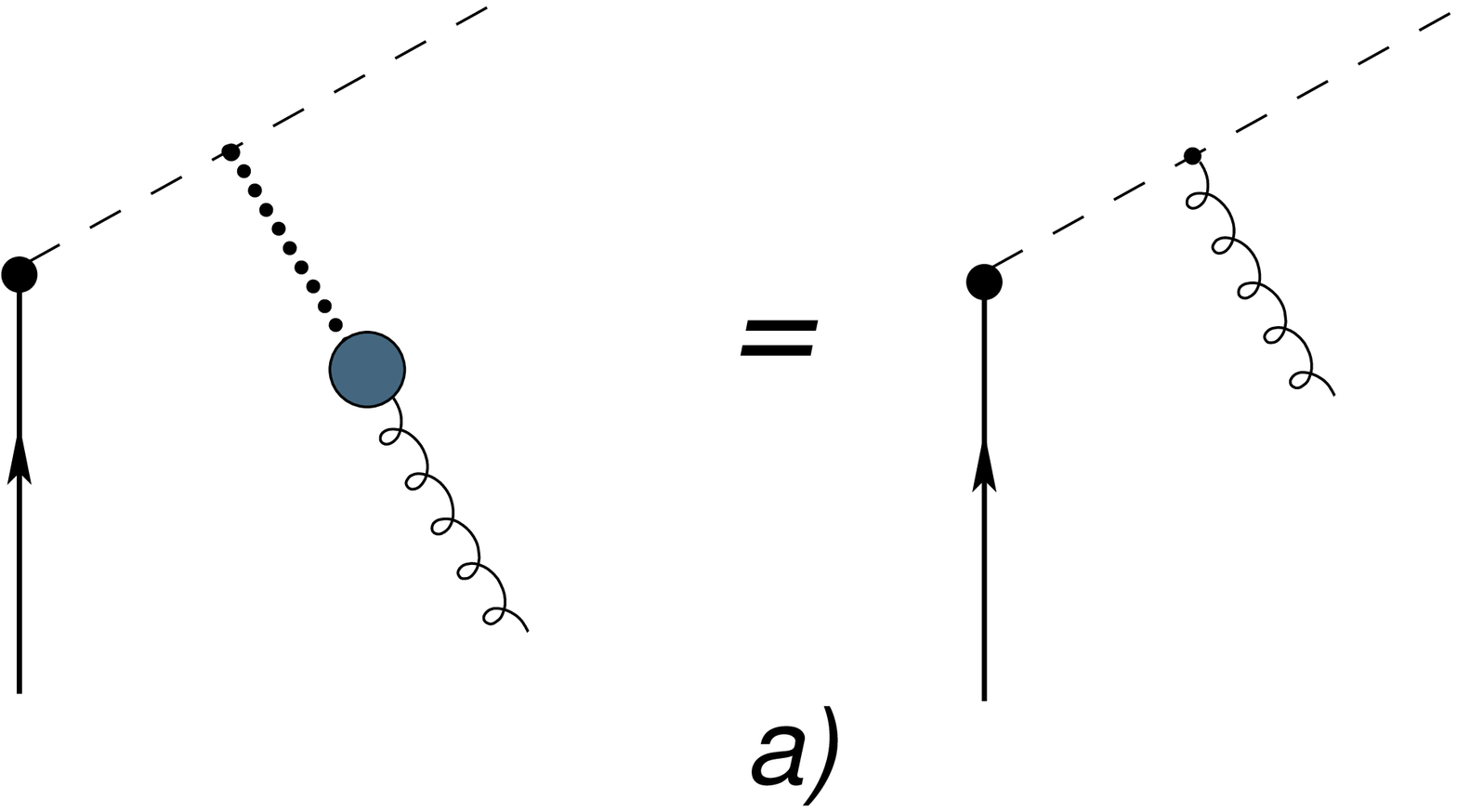}\qquad\qquad\qquad
            \epsfysize3.0cm\epsfbox{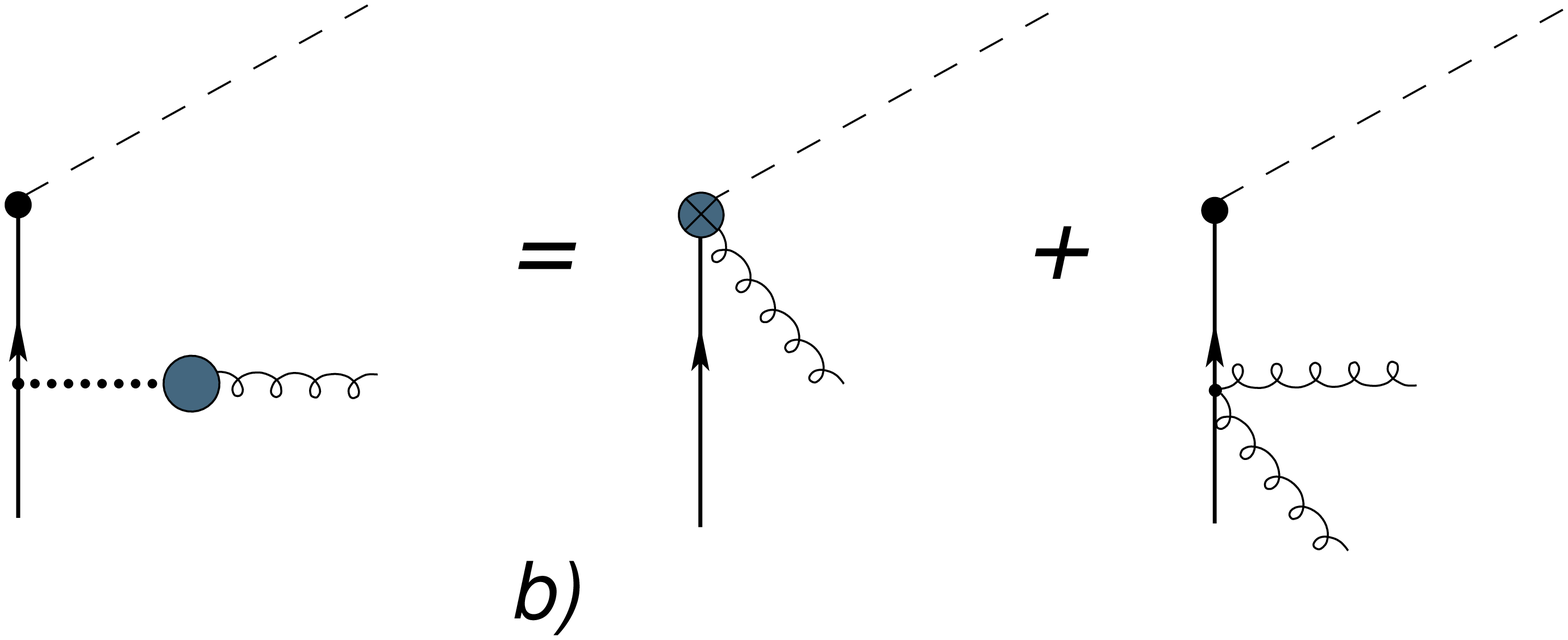}
            }
\caption{\small{$\delta_M S$ insertions, see text}}
\label{fig:insert}
\end{figure}

A detailed analysis shows that the effect of such insertions reduces to the following
transformation of the light-ray fields:
\begin{eqnarray}
(\delta_M S) [0,z] \psi_+(z) &=& \frac{ig}{2}\left[
 z \int_0^1 du\, a_{\mu\bar\lambda}(uz) - \left(\frac{1}{\partial_+}
a_{\mu\bar\lambda}(z)\right)+\ldots
\right]\psi_+(z)\,.
\label{ward}
\end{eqnarray}
This equation is illustrated in Fig.~\ref{fig:insert}. The first term on the r.h.s.
comes from the gauge link, Fig.~\ref{fig:insert}a. Note that
$\wick{1}{<1 {a}_{+}(x) >1 {a}_{+}}(y) = -i\xi \delta^{(4)}(x-y)$ so that the insertion of
$\delta_M S$ simply replaces $a_+(uz)$ by the physical transverse gluon field
$a_{\mu\bar\lambda}(uz)$ at the same position.
The second term on the r.h.s. of (\ref{ward}) corresponds to the first
contribution in Fig~\ref{fig:insert}b, which originates from the contraction of the
quark propagator. There is also another term that involves the classical background
field $A_{\mu\bar\lambda}$ and arises when one completes the covariant derivative to obtain EOM.
Such contributions cancel, however, in the sum of all one-loop Feynman diagrams so we do not
show this term in Eq.~(\ref{ward}) explicitly.
We have checked that all other insertions of $\delta_M S$ (and also the same insertions
as in  Fig.~\ref{fig:insert} but with the contraction of the transverse field
$a_{\mu\bar\lambda}(x)$ in the effective vertex instead of $a_+(x)$) cancel as well.
The structure of these cancellations strongly suggests that they can be put in a form
of a certain Ward identity which we did not work out in the operator form, however.

Using (\ref{identity}) the contributions $\sim a_{\mu\bar\lambda}$ in Eq.~(\ref{ward})
can be rewritten in terms of the field strength tensor $\bar f_{++}$ so that the
last term  in Eq.~(\ref{WI-22}), $[\mathcal{O}_+ (\delta_M  S)]'_R$,
reduces to the sum of contributions of the corresponding BFLK kernels.
The rest of the calculation is straightforward and follows closely the calculation
described in Sect.~\ref{Ptrafo} so we skip the details. The result reproduces
Eq.~(\ref{Second}).

\section{Results for the  $2\to 3$ Kernels}\label{listH}
In this Section we present a complete list of the $2\to 3$ kernels. Altogether,
there exist 72 different twist-3 pairs of the primary fields,
$\{\psi_-\otimes\psi_+,\, \psi_-\otimes\chi_+,\,\psi_+\otimes D_{-+}\bar\chi_+,\ldots\}$.
Many of them are related to each other by parity $P$ and charge conjugation $C$
symmetries, however, so that there are only $16$ independent pairs.
Eight of them can be chosen as the descendants of the quasipartonic
$X_+\otimes X_+$ operators with both fields having the same chirality
and the other eight as the descendants of quasipartonic operators containing primary fields
of opposite chirality.
In each case, the renormalized operator $[X]_R(z_1,z_2)$ in the light-cone gauge
can be written to the one-loop accuracy as
\begin{align}\label{7:ren1}
 [X]_R(z_1,z_2)=X_B(z_1,z_2)+\frac{\alpha_s}{4\pi\epsilon}[\mathbb{H}^{(2\to 2)} X](z_1,z_2)
+\frac{\alpha_s}{4\pi\epsilon}[\mathbb{H}^{(2\to3)} Y](z_1,z_2)\,.
\end{align}
The results for the $2\to 2$ Hamiltonians, $\mathbb{H}^{(2\to 2)}$, are collected in Sect.~\ref{NQO}.
The expressions for the 16 independent three-particle counterterms
$[\mathbb{H}^{(2\to3)} Y](z_1,z_2)$, where
$Y(z_1,z_2,z_3)= X^{(1)}_+(z_1)\otimes X^{(2)}_+(z_2) \otimes X^{(3)}_+(z_3)$  is a
 quasipartonic operator with proper quantum numbers, will be given in what follows.

The representation in (\ref{7:ren1}) is somewhat schematic since several three-particle
operators $Y_1$, $Y_2$,\ldots can contribute to the r.h.s. and all have to be taken into account,
which makes the corresponding expressions quite cumbersome.
{}For example, the operator $f_{+-}\otimes\psi_+$ can mix  with
the operator $f_{++}\otimes\psi_+\otimes\bar f_{++}$ and also with a three-quark operator
$\psi_+\otimes \psi_+\otimes \bar\psi_+$, etc, so that, effectively,
the gluon pair $f_{++}\otimes \bar f_{++}$ is replaced by the quark-antiquark
pair $\psi_+\otimes\bar\psi_+$. Such quark-antiquark pairs always appear 
in special combinations
\begin{align}\label{}
J^a(z_1,z_2)&=\bar\psi_+(z_1) T^a\psi_+(z_2)+\chi_+(z_1)T^a\bar\chi_+(z_2)=
\bar q(z_1) T^a\gamma_+ q(z_2)\,,
&&J^a(z)\equiv J^{a}(z,z)\,,
\notag\\
J^{ab}(z_1,z_2)&=\bar\psi_+(z_1)T^a T^b\psi_+(z_2)-\chi_+(z_2) T^bT^a\bar\chi_+(z_1)\,,
\notag\\
J^{abc}(z_1,z_2)&=\bar\psi_+(z_1)T^a T^b T^c\psi_+(z_2)+\chi_+(z_2) T^cT^bT^a\bar\chi_+(z_1)\,,
\end{align}
where the sum over flavors is implied, $\bar\psi_+  T^a\psi_+ \equiv \sum_A 
\bar\psi_+^A  T^a\psi_+^A$, etc. The operators $J^a, J^{ab}, J^{abc}$ are of course related to each
other. However, imposing such relations complicates the expressions considerably, 
so we do not attempt this.

As mentioned above, sixteen independent $2\to3$ RG kernels will be presented explicitly. 
The remaining $56 = 72-16$ ones can be restored as follows. 

{}First, the left-handed and right-handed quark spinors appearing in 
(\ref{7:ren1}) can be interchanged freely: 
 $\psi_{\pm}\leftrightarrow \chi_{\pm}$ and $\bar\psi_{\pm}\leftrightarrow\bar\chi_{\pm}$
(on both sides of Eq.~(\ref{7:ren1}) simultaneously, but separately for each flavor).
 The form of the kernels is not affected by such transformation so that,
 for example, the four kernels 
\begin{eqnarray*}
&& \psi_-\otimes \psi_+ \to  \psi_-\otimes \psi_+ \otimes \bar f_{++}\,, 
 \quad  
\chi_-\otimes \psi_+  \to  \chi_-\otimes \psi_+ \otimes \bar f_{++}\,, 
\\
&& \psi_-\otimes \chi_+  \to  \psi_-\otimes \chi_+ \otimes \bar f_{++}\,, 
\quad 
\chi_-\otimes \chi_+  \to \chi_-\otimes \chi_+ \otimes \bar f_{++}\, 
\end{eqnarray*}
are equal to each other~\footnote{
The annihilation-type contributions $\sim \delta_{AB}$ in Eq.~(\ref{7:seventeen}) 
are only present for $\bar\psi^A_+\psi^B_-, \,\, \bar\chi^A_+\chi^B_-$
operators and have to be discarded for $\bar\chi^A_+\psi^B_-,\, \bar\psi^A_+\chi^B_-$.}.
Making use of such simple substitution rules one ends up with 36 kernels:
$16=4\times 4$ quark--quark, $16=2\times 8$ quark--gluon and $4$ gluon--gluon ones.

The remaining $36=72-36$ kernels can be obtained by
applying hermitian conjugation to the corresponding quantum operators. To this end,
the color factors like $i(t^dt^c)\otimes t^c$ should be treated as $c-$numbers, i.e.
\begin{align*}
\Big(i(t^dt^c)_{ii'}\otimes t^c_{jj'}\psi_-^{i'}(z_1) \psi_+^{j'}(z_2)\Big)^\dagger&=
\Big(i(t^dt^c)_{ii'}\otimes t^c_{jj'}\Big)^* (\psi_+^{j'}(z_2))^\dagger (\psi_-^{i'}(z_1))^\dagger
\\
&=i(t^dt^c)_{ii'}\otimes t^c_{jj'}\bar\psi_+^{j'}(z_2)\bar\psi_-^{i'}(z_1)\,.
\end{align*}
Here in the first line all $SU(N)$ generators are taken in the
quark representation, $t^a=T^a$, whereas in  the second line 
$t^a=(-T^a)^T$ are the generators in the antiquark representation, cf. Eq.~(\ref{3:generators}).

We remind that under hermitian conjugation
\begin{align}\label{}
\psi_\pm^\dagger=\bar\psi_\pm\,,&&  \chi_\pm^\dagger=\bar\chi_\pm\,,&&
f_{++}^\dagger=\bar f_{++}\,,&& f_{+-}^\dagger=\bar f_{+-}\,, && (D_{-+})^*=D_{+-}\,
\end{align}
and in addition one has to replace $(\mu\lambda) \to (\bar\lambda\bar\mu)$. 
One also finds easily that
\begin{equation}
(J^{a}(z_1,z_2))^\dagger=J^{a}(z_2,z_1)\,,\quad
(J^{ab}(z_1,z_2))^\dagger=J^{ba}(z_2,z_1)\,,\quad
(J^{abc}(z_1,z_2))^\dagger=J^{cba}(z_2,z_1)\,.
\end{equation}
\subsection{$\psi_-\psi_+$ or $\frac12 D_{-+}\bar\psi_+\bar\psi_+$}
%
\begin{eqnarray*}
&&\boxed{
\begin{array}{cl}
  X^{ij}(z_1,z_2)=\psi_-^i(z_1)\psi_+^j(z_2)\,, 
& \multirow{3}{*}{$~Y^{ijd}(z_1,z_2,z_3)=g(\mu\lambda)\psi_+^i(z_1)\psi_+^j(z_2)\bar f_{++}^d(z_3)$}
\\
\text{or} &
\\
 X^{ij}(z_1,z_2)=\frac12 D_{-+}\bar\psi^i_+(z_1)\bar\psi^j_+(z_2) &
\end{array}
}
\end{eqnarray*} 

The three-particle counterterm in both cases takes the form   
\begin{equation}
[X^{ij}]'_R=\frac{\alpha_s}{4\pi\epsilon} \sum^3_{k=1} (C_k)^{ij}_{i'j'd} \mathcal{H}_k Y^{i'j'd}\,, 
\end{equation}
where $C_k$ are the color structures
\begin{equation}
(C_1)^{ij}_{i'j'd} \,=\, f^{dbc}t^b_{ii'} t^c_{jj'}\,,\quad
(C_2)^{ij}_{i'j'd} \,=\, i(t^dt^b)_{ii'} t^b_{jj'}\,,\quad
(C_3)^{ij}_{i'j'd} \,=\, i\, t^b_{ii'} (t^d t^b)_{jj'}\,.
\label{7:c1-c3}
\end{equation}
The $SL(2,\mathbb{R})$ invariant kernels for the first case, $X^{ij}=\psi_-^i\psi_+^j$, 
are given by:
\begin{align}\label{7:first}
[\mathcal{H}_1\varphi](z_1,z_2)&=z_{12}^2\int_0^1d\alpha\int_0^{\bar\alpha}d\beta \beta\,
                                  \varphi(z_{12}^\alpha,z_2,z_{21}^\beta)\,,
\notag\\
[\mathcal{H}_2\varphi](z_1,z_2)&=z_{12}^2
     \int_0^1d\alpha\int_{\bar \alpha}^1d\beta \frac{\bar\alpha\bar\beta}{\alpha}\,
                   \varphi(z_{12}^\alpha,z_2,z_{21}^\beta)\,.
\end{align}
(the third color structure, $C_3$, does not contribute).

{}For the second case, $X^{ij}=\frac12 D_{-+}\bar\psi^i_+\bar\psi^j_+$,
one obtains
\begin{align}\label{7:second}
[\mathcal{H}_1\varphi](z_1,z_2)&=z_{12}\Biggl\{
\int_0^1d\beta\,\bar\beta\,\varphi(z_1,z_{2},z_{12}^\beta)+
\int_0^1d\alpha\int_{\bar\alpha}^1d\beta\,\bar\beta\,
\varphi(z_1,z_{12}^\alpha,z_{21}^\beta)
\notag\\
&\phantom{=z_{12}\Biggl\{}
+\int_0^1 d\alpha\int_{\bar\alpha}^1d\beta\frac{\bar\beta^2}{\beta}\,
\varphi(z_{21}^\alpha,z_{2},z_{12}^\beta)
\Biggr\}\,,
\notag\\
[\mathcal{H}_2\varphi](z_1,z_2)&=z_{12}
\int_0^1d\alpha\int_{\bar\alpha}^1d\beta\,\frac{\bar\alpha\beta}{\alpha}
\left(2-\frac{\bar\alpha\bar\beta}{\alpha\beta}\right)\varphi(z_{12}^\alpha,z_2,z_{21}^\beta)\,,
\notag\\
[\mathcal{H}_3\varphi](z_1,z_2)&=-z_{12}
 \int_0^1d\alpha\int_{\bar\alpha}^1d\beta
\,\beta\,\varphi(z_1,z_{21}^\alpha,z_{12}^\beta)\,.
\end{align}

\subsection{$\frac12 D_{-+}\bar f_{++}\bar f_{++}$}
%
\begin{eqnarray*}
&&\boxed{
\begin{array}{cl}
  X^{ab}(z_1,z_2)=\frac12 D_{-+}\bar f^{a}_{++}(z_1)\bar f_{++}^b(z_1)\,,
& ~Y^{abd}(z_1,z_2,z_3)=g(\mu\lambda)\bar f^{a}_{++}(z_1)\bar f_{++}^b(z_1)\bar f_{++}^d(z_3)
\end{array}
}
\end{eqnarray*} 
The three-particle counterterm takes the form  
\begin{equation}
[X^{ab}]'_R=\frac{\alpha_s}{4\pi\epsilon} \sum^3_{k=1} (C_k)^{ab}_{a'b'd} \mathcal{H}_k Y^{a'b'd}\,,
\end{equation}
where the color structures $C_k$ are the same as in Eq.~(\ref{7:c1-c3}) (but with all generators
in the adjoint representation, cf. (\ref{3:generators})). 
The $SL(2,\mathbb{R})$ invariant kernels are:
\begin{align}\label{dbfbf}
[\mathcal{H}_1\varphi](z_1,z_2)&=z_{12}\Biggl\{
\int_0^1d\beta\,\bar\beta\,\varphi(z_1,z_2,z_{12}^\beta)
+\int_0^1d\alpha\int_{\bar\alpha}^1d\beta\frac{\alpha\bar\beta^2}{\beta}
\left(2-\frac{\bar\alpha\bar\beta}{\alpha\beta}\right)\varphi(z_{21}^\alpha,z_2,z_{12}^\beta)
\notag\\
&\phantom{=z_{12}\Biggl\{\int}
+\int_0^1d\alpha\int_{\bar\alpha}^1d\beta\,\alpha\bar\beta
\left(2-\frac{\bar\alpha\bar\beta}{\alpha\beta}\right)\varphi(z_1,z_{12}^\alpha,z_{21}^\beta)
\Biggr\}\,,
\notag\\
[\mathcal{H}_2\varphi](z_1,z_2)&=z_{12}
\int_0^1d\alpha\int_{\bar\alpha}^1d\beta\frac{\bar\alpha^2\beta}{\alpha}
\varphi(z_{12}^\alpha,z_2,z_{21}^\beta)\,,
\notag\\
[\mathcal{H}_3\varphi](z_1,z_2)&=-z_{12}
\int_0^1d\alpha\int_{\bar\alpha}^1d\beta\frac{\bar\alpha^2\bar\beta}{\alpha}
\varphi(z_1,z_{21}^\alpha,z_{12}^\beta)\,.
\end{align}

\subsection{$\frac12D_{-+}\bar\psi_+\bar f_{++}$ or $\bar\psi_+\frac12D_{-+}\bar f_{++}$}
%
\begin{eqnarray*}
&&\boxed{
\begin{array}{cl}
   X^{ia}(z_1,z_2) = \frac12D_{-+}\bar\psi_+^{i}(z_1)\bar f_{++}^a(z_2)
& \multirow{3}{*}{$~Y^{iad}(z_1,z_2,z_3)=g(\mu\lambda)\bar\psi_+^{i}(z_1)\bar
f_{++}^a(z_2)\bar f^d_{++}(z_3)$} 
\\
\text{or} &
\\
 X^{ia}(z_1,z_2)=\bar\psi_+^{i}(z_1)\frac12D_{-+}\bar f_{++}^a(z_2) &
\end{array}
}
\end{eqnarray*} 
The three-particle counterterm takes the form  
\begin{equation}
[X^{ia}]'_R=\frac{\alpha_s}{4\pi\epsilon} \sum^5_{k=1} (C_k)^{ia}_{i'a'd} \mathcal{H}_k Y^{i'a'd}\,.
\end{equation}
The first three color structures $C_1,C_2,C_3$ are the same as in Eq.~(\ref{7:c1-c3}) 
(with the generators in the appropriate representation), and there are 
two new structures:
\begin{equation}\label{7:c4-c5}
(C_4)^{ia}_{i'a'd}=i(t^dt^{a'}t^a)_{ii'}\,,\qquad (C_5)^{ia}_{i'a'd}=i(t^{a'} t^d t^a)_{ii'}\,,
\end{equation}
The invariant kernels for the first case, $X^{ia}=\frac12D_{-+}\bar\psi_+^{i}\bar f_{++}^a$, 
are given by:
\begin{align}\label{7:six}
[\mathcal{H}_1\varphi](z_1,z_2)&=z_{12}\Biggl\{
\int_0^1d\beta\bar\beta\,\varphi(z_{1},z_2,z_{12}^\beta)
+\int_0^1d\alpha\int_{\bar\alpha}^1d\beta\,\frac{\bar\beta^2}{\beta\phantom{{}^2}}
\,\varphi(z_{21}^\alpha,z_2,z_{12}^\beta)
\notag\\
&
\phantom{=z_{12}\Biggl\{}+\int_0^1d\alpha\int_{\bar\alpha}^1d\beta\,\alpha\bar\beta\,
\left(2-\frac{\bar\alpha\bar\beta}{\alpha\beta}\right)
\,\varphi(z_{1},z_{12}^\alpha,z_{21}^\beta)
\Biggr\}\,,
\notag\\
[\mathcal{H}_2\varphi](z_1,z_2)&=z_{12}
\int_0^1d\alpha\int_{\bar\alpha}^1d\beta\frac{\bar\alpha\beta}{\alpha}
\left(2-\frac{\bar\alpha\bar\beta}{\alpha\beta}\right)
\varphi(z_{12}^\alpha,z_2,z_{21}^\beta)\,,
\notag\\
[\mathcal{H}_3\varphi](z_1,z_2)&=-z_{12}
\int_0^1d\alpha\int_{\bar\alpha}^1d\beta\frac{\bar\alpha^2\bar\beta}{\alpha}\,
\varphi(z_{1},z_{21}^\alpha,z_{12}^\beta)\,,
\notag\\
[\mathcal{H}_4\varphi](z_1,z_2)&=-z_{12}
\int_0^1d\alpha\int_{\bar\alpha}^1d\beta\,\bar\alpha\beta\,\varphi(z_2,z_{12}^\alpha,z_{21}^\beta)\,,
\notag\\
[\mathcal{H}_5\varphi](z_1,z_2)&=-z_{12}
\int_0^1d\alpha\int_0^{\bar\alpha}d\beta\,\bar\alpha\beta\,\varphi(z_2,z_{12}^\alpha,z_{21}^\beta)\,.
\end{align}
The invariant kernels for the second case, $X^{ia}=\bar\psi_+^{i}(z_1)\frac12D_{-+}\bar f_{++}^a(z_2)$, 
are:
\begin{align}\label{7:seven}
[\mathcal{H}_1\varphi](z_1,z_2)&=z_{12}\Biggl\{
\int_0^1d\beta\bar\beta\,\varphi(z_{1},z_2,z_{12}^\beta)
+\int_0^1d\alpha\int_0^{\bar\alpha}d\beta\,\beta
\,\varphi(z_{12}^\alpha,z_2,z_{21}^\beta)
\notag\\
&
\phantom{=-z_{12}\Biggl\{}+\int_0^1d\alpha\int_{\bar\alpha}^1d\beta\,\frac{\alpha\bar\beta^2}{\beta}
\left(2-\frac{\bar\alpha\bar\beta}{\alpha\beta}\right)
\,\varphi(z_{1},z_{12}^\alpha,z_{21}^\beta)
\Biggr\}\,,
\notag\\
[\mathcal{H}_2\varphi](z_1,z_2)&=z_{12}\int_0^1d\alpha\int_{\bar\alpha}^1d\beta\,\beta\,
                                   \varphi(z_{12}^\alpha,z_2,z_{21}^\beta)\,,
\notag\\
[\mathcal{H}_3\varphi](z_1,z_2)&=-z_{12}\int_0^1d\alpha\int_{\bar\alpha}^1d\beta\,
                   \frac{\bar\alpha^2\beta}{\alpha}
                    \varphi(z_1,z_{21}^\alpha,z_{12}^\beta)\,,
\notag\\
[\mathcal{H}_4\varphi](z_1,z_2)&=-z_{12}
         \int_0^1d\alpha\int_{\bar\alpha}^1d\beta\,\bar\alpha\bar\beta\,
                   \varphi(z_2,z_{12}^\alpha,z_{21}^\beta)\,,
\notag\\
[\mathcal{H}_5\varphi](z_1,z_2)&=-z_{12}
               \int_0^1d\alpha\int_0^{\bar\alpha}d\beta\,\bar\alpha \bar \beta\,
                \varphi(z_2,z_{12}^\alpha,z_{21}^\beta)\,.
\end{align}
%

\subsection{$\bar f_{++}\psi_{-}$ or $\frac12D_{-+}\bar f_{++}\psi_{+}$}
%
\begin{eqnarray*}
&&\boxed{
\begin{array}{cl}
  X^{ai}(z_1,z_2)=\bar f_{++}^{a}(z_1)\psi_{-}^i(z_2)\,, 
& \multirow{3}{*}{$~Y^{aid}(z_1,z_2,z_3)=g(\mu\lambda)\bar
f_{++}^{a}(z_1)\psi_{+}^i(z_2)\bar f^d_{++}(z_3)$} 
\\
\text{or} &
\\
 X^{ai}(z_1,z_2)= \frac12D_{-+}\bar f_{++}^{a}(z_1)\psi_{+}^i(z_2) &
\end{array}
}
\end{eqnarray*} 

The three-particle counterterm takes the form  
\begin{equation}
[X^{ai}]'_R=\frac{\alpha_s}{4\pi\epsilon} \sum^6_{k=1} (C_k)^{ai}_{a'i'd} \mathcal{H}_k Y^{a'i'd}\,.
\end{equation}
The color structures $C_1,C_2,C_3$ are specified in Eq.~(\ref{7:c1-c3})
(with the generators in the appropriate representation), $C_4,C_5$ are given in
Eq.~(\ref{7:c4-c5}), and there is one new structure:
\begin{equation}\label{7:c6}
 (C_6)^{ai}_{a'i'd}=i(t^{a'}  t^a t^d)_{ii'}\,.
\end{equation}
The invariant kernels for the first case, $X^{ai}=\bar f_{++}^{a}\psi_{-}^i$
are given by:
\begin{align}\label{7:eight}
[\mathcal{H}_1\varphi](z_1,z_2)&=-z_{12}^2\Biggl\{
\int_0^1d\alpha\int_{\bar\alpha}^1d\beta\,\bar\beta\,
\varphi(z_1,z_{12}^\alpha,z_{21}^\beta)
\notag\\
&\phantom{=-z_{12}\Biggl\{}+
\int_0^1d\alpha\int_0^{\bar\alpha}d\beta\int_{\beta}^{\bar\alpha}d\gamma\,
\bar\alpha\gamma\left(2-\frac{\alpha\gamma}{\bar\alpha\bar\gamma}\right)
\varphi(z_{12}^\alpha,z_{21}^\beta,z_{21}^\gamma)
\Biggr\}\,,
\notag\\
[\mathcal{H}_2\varphi](z_1,z_2)&=-z_{12}^2
\int_0^1d\alpha\int_0^{\bar\alpha}d\beta\int_{\bar\alpha }^1 d\gamma \,
 \frac{\bar\alpha^2\bar\gamma}{\alpha}
\varphi(z_{12}^\alpha,z_{21}^\beta,z_{21}^\gamma)\,,
\notag\\
[\mathcal{H}_3\varphi](z_1,z_2)&=z_{12}^2\Biggl\{
\int_0^1d\alpha\int_{\bar\alpha}^1
d\beta\frac{\bar\alpha\bar\beta}{\alpha}\varphi(z_1,z_{21}^\alpha,z_{12}^\beta)
\\
&+
\int_0^1d\alpha\int_0^{\bar\alpha}d\beta\int_0^{\beta}d\gamma\,
\bar\alpha\gamma\,\left(2-\frac{\alpha\gamma}{\bar\alpha\bar\gamma}\right)
\varphi(z_{12}^\alpha,z_{21}^\beta,z_{21}^\gamma)\Biggr\}\,,
\notag\\
[\mathcal{H}_4\varphi](z_1,z_2)&=z_{12}^2
\int_0^1d\alpha\int_{\bar\alpha}^1d\beta\int_0^{\bar\alpha}d\gamma\,
\bar\alpha\gamma\left(2-\frac{\alpha\gamma}{\bar\alpha\bar\gamma}\right)\,
\varphi(z_{12}^\alpha,z_{21}^\beta,z_{21}^\gamma)\,,
\notag\\
[\mathcal{H}_5\varphi](z_1,z_2)&=z_{12}^2
\int_0^1d\alpha\int_{\bar\alpha}^1d\beta\int_{\bar\alpha}^1d\gamma\,
\frac{\bar\alpha^2\bar\gamma}{\alpha}\,
\varphi(z_{12}^\alpha,z_{21}^\beta,z_{21}^\gamma)\,,
\notag\\
[\mathcal{H}_6\varphi](z_1,z_2)&=-z_{12}^2
\int_0^1d\alpha\int_{\bar\alpha}^1d\beta\int_{\beta}^1d\gamma\,
\frac{\bar\alpha^2\bar\gamma}{\alpha}\,
\varphi(z_{12}^\alpha,z_{21}^\beta,z_{21}^\gamma)\,.
\end{align}
For the second case,  $X^{ai} = \frac12D_{-+}\bar f_{++}^{a}\psi_{+}^i$
one obtains
\begin{align}\label{7:neun}
[\mathcal{H}_1\varphi](z_1,z_2)&=z_{12}\Biggl\{
\int_0^1d\beta\,\bar\beta\, \varphi(z_1,z_2,z_{12}^\beta)
+\int_0^1d\alpha\int_{\bar\alpha}^1d\beta\,\frac{\alpha\bar\beta^2}{\beta}
\left(2-\frac{\bar\alpha\bar\beta}{\alpha\beta}\right)
\varphi(z_{21}^\alpha,z_2,z_{12}^\beta)
\notag\\
&\phantom{=z_{12}^2\Biggl\{}+
\int_0^1d\alpha\int_0^{\bar\alpha}d\beta\int^{\bar\alpha}_{\beta} d\gamma\,
\bar\alpha\gamma\,\left(4-\frac{\alpha\gamma}{\bar\alpha\bar\gamma}\right)
\varphi(z_{12}^\alpha,z_{21}^\beta,z_{21}^\gamma)\Biggr\}\,,
\notag\\
[\mathcal{H}_2\varphi](z_1,z_2)&=z_{12}\Biggl\{
\int_0^1d\alpha \int_{\bar\alpha}^1d\beta\,\frac{\bar\alpha^2\beta}{\alpha}
\varphi(z_{12}^\alpha,z_{2},z_{21}^\beta)
\notag\\
&\phantom{=z_{12}^2\Biggl\{}
+\int_0^1d\alpha\int_0^{\bar\alpha}d\beta\int_{\bar\alpha}^1 d\gamma \,
    \bar\alpha\gamma \left(2+\frac{\bar\alpha\bar\gamma}{\alpha\gamma}\right)
\varphi(z_{12}^\alpha,z_{21}^\beta,z_{21}^\gamma)\Biggr\}\,,
\notag\\
[\mathcal{H}_3\varphi](z_1,z_2)&=-z_{12}
\int_0^1d\alpha\int_0^{\bar\alpha}d\beta\int_0^{\beta}d\gamma\,
\bar\alpha\gamma\left(4-\frac{\alpha\gamma}{\bar\alpha\bar\gamma}\right)\,
\varphi(z_{12}^\alpha,z_{21}^\beta,z_{21}^\gamma)\,,
\notag\\
[\mathcal{H}_4\varphi](z_1,z_2)&=-z_{12}
\int_0^1d\alpha\int_{\bar\alpha}^1d\beta\int_0^{\bar\alpha}d\gamma\,
\bar\alpha\gamma\left(4-\frac{\alpha\gamma}{\bar\alpha\bar\gamma}\right)\,
\varphi(z_{12}^\alpha,z_{21}^\beta,z_{21}^\gamma)\,,
\notag\\
[\mathcal{H}_5\varphi](z_1,z_2)&=-z_{12}
\int_0^1d\alpha\int_{\bar\alpha}^1d\beta\int_{\bar\alpha}^1d\gamma\,\bar\alpha\gamma\,
\left(2+\frac{\bar\alpha\bar\gamma}{\alpha\gamma}\right)\,
\varphi(z_{12}^\alpha,z_{21}^\beta,z_{21}^\gamma)\,,
\notag\\
[\mathcal{H}_6\varphi](z_1,z_2)&=z_{12}
\int_0^1d\alpha\int_{\bar\alpha}^1d\beta\int_{\beta}^1d\gamma\,\bar\alpha\gamma\,
\left(2+\frac{\bar\alpha\bar\gamma}{\alpha\gamma}\right)\,
\varphi(z_{12}^\alpha,z_{21}^\beta,z_{21}^\gamma)\,.
\end{align}
%

\subsection{$f_{+-}\psi_{+}$ or $f_{++}\psi_{-}$}
%
\begin{eqnarray*}
&&\boxed{
\begin{array}{cl}
 X^{ai}(z_1,z_2)= f_{+-}^{a}(z_1)\psi_{+}^i(z_2)~~{} 
 & {}~Y^{aid}(z_1,z_2,z_3)=g(\mu\lambda)f_{++}^{a}(z_1)\psi_{+}^i(z_2)\bar f^d_{++}(z_3)
\\
 \text{or}
 & {}~~~~~~\,J^{ai}(z_1,z_2)=g(\mu\lambda)J^a(z_1)\,\psi_+^i(z_2)
\\
 X^{ai}(z_1,z_2)= f_{++}^{a}(z_1)\psi_{-}^i(z_2)~~{}
 & {}~Z^{abi}(z_1,z_2,z_3)=g(\mu\lambda)J^{ab}(z_1,z_2)\, \psi_+^i(z_3)
\end{array}
}
\end{eqnarray*} 
The three-particle counterterm for the first case, $X^{ai}= f_{+-}^{a}\psi_{+}^i$,
takes the form  
\begin{align}\label{7:ten}
[X^{ai}]'_R=\frac{\alpha_s}{4\pi\epsilon}\left(\sum_{k=1}^2 (C_k)^{ai}_{a'i'd}
\mathcal{H}_k Y^{a'i'd} +(t^b_{aa'}\otimes t^b_{ii'})\widetilde{\mathcal{H}}_1 J^{a'i'}+
t^b_{ii'}
\widetilde{\mathcal{H}}_2 Z^{abi'}\right)\,,
\end{align}
where the color structures $C_1,C_2$ are specified in Eq.~(\ref{7:c1-c3}).
The invariant kernels are
\begin{align}
[\mathcal{H}_1\varphi](z_1,z_2)&=
z_{12}^2\int_0^1d\alpha\int_0^{\bar\alpha}d\beta\,\bar\alpha\beta\,
\varphi(z_{12}^\alpha,z_{2},z_{21}^\beta)\,,
\notag\\
[\mathcal{H}_2\varphi](z_1,z_2)&=z_{12}^2\int_0^1d\alpha\int_{\bar\alpha}^1d\beta\,
\frac{\bar\alpha^2\bar\beta}{\alpha}\varphi(z_{12}^\alpha,z_2,z_{21}^\beta)\,,
\notag\\
[\widetilde{\mathcal{H}}_1\varphi](z_1,z_2)&=z_{12}\int_0^1d\alpha\,\bar\alpha^2\,
\varphi(z_{12}^\alpha,z_2)\,,
\notag\\
 [\widetilde{\mathcal{H}}_2\varphi](z_1,z_2)&=-z_{12}\int_0^1d\alpha\int_{\bar\alpha}^1d\beta
\frac{\bar\beta}{\beta}\, \varphi(z_{21}^\alpha,z_{12}^\beta,z_2)\,.
\end{align}
{}For the second operator,  $X^{ai} = f_{++}^{a}\psi_{-}^i$, we obtain
\begin{align}\label{7:eleven}
[X^{ai}]'_R=\frac{\alpha_s}{4\pi\epsilon}\left(\sum_{k=1}^5 (C_k)^{ai}_{a'i'd}
\mathcal{H}_k Y^{a'i'd} +(t^{a'}t^a)_{ii'}\widetilde{\mathcal{H}}_1 J^{a'i'}+
t^b_{ii'}
\widetilde{\mathcal{H}}_2 Z^{bai'}\right)\,,
\end{align}
where the color structures $C_k$ are defined in Eqs.~(\ref{7:c1-c3}),(\ref{7:c4-c5}).
The invariant kernels are
\begin{align}\label{}
[\mathcal{H}_1\varphi](z_1,z_2)&=
-z_{12}^2\int_0^1d\alpha\int_0^{\bar\alpha}d\beta\,\beta\,
\varphi(z_1,z_{21}^\alpha,z_{12}^\beta)\,,
\notag\\
[\mathcal{H}_3\varphi](z_1,z_2)&=z_{12}^2\int_0^1d\alpha\int_{\bar\alpha}^1d\beta\,
\frac{\bar\alpha\bar\beta}{\alpha}\varphi(z_1,z_{21}^\alpha,z_{12}^\beta)\,,
\notag\\
[\mathcal{H}_4\varphi](z_1,z_2)&=-z_{12}^2
\int_0^1d\alpha\int_0^{\bar \alpha}d\beta\,\alpha\beta\,\varphi(z_{12}^\alpha,z_1,z_{21}^
\beta)\,,
\notag\\
[\mathcal{H}_5\varphi](z_1,z_2)&=-z_{12}^2
\int_0^1d\alpha\int_0^{\bar \alpha}d\beta\,\alpha\beta\,\varphi(z_{21}^\alpha,z_1,z_{12}^\beta)\,,
\end{align}
(the second color structure does not contribute) and
\begin{align}\label{}
[\widetilde{\mathcal{H}}_1\varphi](z_1,z_2)&=z_{12}\int_0^1d\alpha\,\alpha\bar\alpha\,
\varphi(z_{12}^\alpha,z_1)\,,
\notag\\
 [\widetilde{\mathcal{H}}_2\varphi](z_1,z_2)&=z_{12}\int_0^1d\alpha\int_0^{\bar\alpha}d\beta
\, \varphi(z_{12}^\alpha,z_1,z_{21}^\beta)\,.
\end{align}

\subsection{$\bar\psi_+f_{+-}$ or $\frac12 D_{-+}\bar\psi_+f_{+-}$}
%
\begin{eqnarray*}
&&\boxed{
\begin{array}{cl}
 X^{ia}(z_1,z_2)= \bar\psi_+^i(z_1)f_{+-}^a(z_2)~~{} 
 & {}~Y^{iad}(z_1,z_2,z_3)=g(\mu\lambda)\bar\psi_+^i(z_1)f_{++}^a(z_2)\bar f^d_{++}(z_3)
\\
 \text{or}
 & {}~~~~~~\,J^{ia}(z_1,z_2)=g(\mu\lambda)\bar\psi_+^i(z_1)\otimes J^a(z_2) 
\\
 X^{ia}(z_1,z_2)=\frac12 D_{-+}\bar\psi_+^i(z_1)f_{+-}^a(z_2)~~{}
 & {}~Z^{iab}(z_1,z_2,z_3)=g(\mu\lambda)\bar\psi_+^i(z_1)\otimes J^{ab}(z_2,z_3) 
\end{array}
}
\end{eqnarray*} 
The three-particle counterterm for the first case,  $X^{ia}= \bar\psi_+^if_{+-}^a$,
takes the form 
\begin{align}\label{7:twelve}
[X^{ia}]'_R=\frac{\alpha_s}{4\pi\epsilon}\left(\sum_{k=1}^6 (C_k)^{ai}_{a'i'd}
\mathcal{H}_k Y^{a'i'd} \!+(t^b_{ii'}\otimes t^b_{aa'})\widetilde{\mathcal{H}}_1 J^{i'a'}
\!+(t^{a'}t^a)_{ii'}\widetilde{\mathcal{H}}_2 J^{i'a'}
\!+t^b_{ii'}\widetilde{\mathcal{H}}_3 Z^{i'ab}\right),
\end{align}
with the color structures $C_k$ as defined in Eqs.~(\ref{7:c1-c3}),(\ref{7:c4-c5}),(\ref{7:c6}).
The invariant kernels are
\begin{align}\label{}
[\mathcal{H}_1\varphi](z_1,z_2)&=-z_{12}^2\Biggl\{
\int_0^1d\alpha\int_0^{\bar\alpha}d\beta\,\bar\alpha\beta\,\varphi(z_1,z_{21}^\alpha,z_{12}^\beta)
\notag\\
&\phantom{-z_{12}^2\Biggl\{}
+\int_0^1d\alpha\int_0^{\bar\alpha}d\beta\int_{\beta}^{\bar\alpha}d\gamma\,
(\bar\beta\gamma+\beta\bar\gamma)\,\varphi(z_{12}^\alpha,z_{21}^\beta,z_{21}^\gamma)
\Biggr\}\,,
\notag\\
[\mathcal{H}_2\varphi](z_1,z_2)&=-z_{12}^2
\int_0^1d\alpha\int_0^{\bar\alpha}d\beta\int_{\bar\alpha}^1 d\gamma\,
(\bar\beta\gamma+\beta\bar\gamma)\,
\varphi(z_{12}^\alpha,z_{21}^\beta,z_{21}^\gamma)\,,
\notag\\
[\mathcal{H}_3\varphi](z_1,z_2)&=z_{12}^2\Biggl\{
\int_0^1d\alpha\int_{\bar\alpha}^1d\beta\,
\frac{\bar\alpha^2\bar\beta}{\alpha}\,\varphi(z_1,z_{21}^\alpha,z_{12}^\beta)
\notag\\
&\phantom{=z_{12}^2\Biggl\{}
+2\int_0^1d\alpha\int_0^{\bar\alpha}d\beta\int_0^\beta d\gamma\,\bar\beta\gamma\,
\varphi(z_{12}^\alpha,z_{21}^\beta,z_{21}^\gamma)\Biggr\}\,,
\notag\\
[\mathcal{H}_4\varphi](z_1,z_2)&=-z_{12}^2
\int_0^1d\alpha\int_{\bar\alpha}^1d\beta\,
\int_{\beta}^1d\gamma
\left(\bar\beta\gamma+{\beta\bar\gamma}\right)\,
\varphi(z_{12}^\alpha,z_{21}^\beta,z_{21}^\gamma)\,,
\notag\\
[\mathcal{H}_5\varphi](z_1,z_2)&=-2z_{12}^2
\int_0^1d\alpha\int_{\bar\alpha}^1d\beta\,
\int_0^{\beta}d\gamma
\bar\beta\gamma\,
\varphi(z_{12}^\alpha,z_{21}^\beta,z_{21}^\gamma)\,,
\notag\\
[\mathcal{H}_6\varphi](z_1,z_2)&=2z_{12}^2
\int_0^1d\alpha\int_{\bar\alpha}^1d\beta\,
\int_0^{\bar\alpha}d\gamma
\bar\beta\gamma\,
\varphi(z_{12}^\alpha,z_{21}^\beta,z_{21}^\gamma)\,
\end{align}
and
 \begin{align}
[\widetilde{\mathcal{H}}_1\varphi](z_1,z_2)&=-z_{12}\Biggl\{\int_0^1d\beta\,
\bar\beta^2\,\varphi(z_{1}, z_{21}^\beta)+2\int_0^1d\alpha\int_0^{\bar\alpha}d\beta\,
\beta\bar\beta\,\varphi(z_{12}^\alpha, z_{21}^\beta)
\Biggl\}
\,,
\notag\\
[\widetilde{\mathcal{H}}_2\varphi](z_1,z_2)&=-2z_{12}\int_0^1d\alpha\int_{\bar\alpha}^1d\beta\,
\beta\bar\beta\,\varphi(z_{12}^\alpha, z_{21}^\beta)\,,
\notag\\
[\widetilde{\mathcal{H}}_3\varphi](z_1,z_2)&
=z_{12}\Biggl\{\int_0^1\!d\alpha\int_{\bar\alpha}^1\!d\beta\,
\frac{\bar\beta}{\beta}\varphi(z_1,z_{12}^\alpha,z_{21}^\beta)+
\int_0^1\!d\alpha\int_0^{\bar\alpha}\!d\beta\int_{\beta}^{\bar\alpha}\!d\gamma
\,\varphi(z_{12}^\alpha, z_{21}^\beta,z_{21}^\gamma)
\Biggr\}\,.
\end{align}
{}For the second operator, $X^{ia}=\frac12 D_{-+}\bar\psi_+^if_{+-}^a$, we obtain 
\begin{eqnarray}\label{7:thirteen}
[X^{ia}]'_R&=&\frac{\alpha_s}{4\pi\epsilon}\Biggl(\sum_{k=1}^6 (C_k)^{ai}_{a'i'd}
\mathcal{H}_k Y^{a'i'd} +(t^b_{ii'}\otimes t^b_{aa'})\widetilde{\mathcal{H}}_1 J^{i'a'}
+(t^{a'}t^a)_{ii'}\widetilde{\mathcal{H}}_2 J^{i'a'}
\nonumber\\
&&{}\hspace*{1cm}+
t^b_{ii'} \widetilde{\mathcal{H}}_3 Z^{i'ab}
+ t^b_{ii'} \widetilde{\mathcal{H}}_4 Z^{i'ba}\Biggr),
\end{eqnarray}
with the same color structures $C_k$  as in the first case.
The invariant kernels are
\begin{align}\label{}
[\mathcal{H}_1\varphi](z_1,z_2)&=z_{12}\Biggl\{\int_0^1d\beta\,\bar\beta\,\varphi(z_1,z_2,z_{12}^\beta)+
\int_0^1d\alpha\int_{\bar\alpha}^1d\beta\frac{\bar\beta^2}{\beta}\varphi(z_{21}^\alpha,z_2,z_{12}^\beta)
\notag\\
&\phantom{=z_{12}\Biggl\{}
+2\int_0^1d\alpha\int_0^{\bar\alpha}d\beta\int_{\beta}^{\bar\alpha}d\gamma\,
(2\bar\beta\gamma+\beta\bar\gamma)\,\varphi(z_{12}^\alpha,z_{21}^\beta,z_{21}^\gamma)\Biggr\}\,,
\notag\\
[\mathcal{H}_2\varphi](z_1,z_2)&=z_{12}\Biggl\{\int_0^1d\alpha\int_{\bar\alpha}^1d\beta
\,\frac{\bar\alpha\beta}{\alpha}\left(2-\frac{\bar\alpha\bar\beta}{\alpha\beta}\right)\,
\varphi(z_{12}^\alpha,z_2,z_{21}^\beta)
\notag\\
&\phantom{=z_{12}\Biggl\{}
+2\int_0^1 d\alpha \int_0^{\bar\alpha} d\beta\int_{\bar\alpha}^1 d\gamma\,
(2\bar\beta\gamma+\beta\bar\gamma)\,
\varphi(z_{12}^\alpha,z_{21}^\beta,z_{21}^\gamma)
\Biggr\}
\notag\\
[\mathcal{H}_3\varphi](z_1,z_2)&=-6z_{12}
\int_0^1d\alpha\int_0^{\bar\alpha}d\beta\int_0^\beta d\gamma\,\bar\beta\gamma\,
\varphi(z_{12}^\alpha,z_{21}^\beta,z_{21}^\gamma)\,,
\notag\\
[\mathcal{H}_4\varphi](z_1,z_2)&=2z_{12}
\int_0^1d\alpha\int_{\bar\alpha}^1d\beta\,
\int_{\beta}^1d\gamma
\left(2\bar\beta\gamma+{\beta\bar\gamma}\right)\,
\varphi(z_{12}^\alpha,z_{21}^\beta,z_{21}^\gamma)\,,
\notag\\
[\mathcal{H}_5\varphi](z_1,z_2)&=6z_{12}
\int_0^1d\alpha\int_{\bar\alpha}^1d\beta\,
\int_0^{\beta}d\gamma
\bar\beta\gamma\,
\varphi(z_{12}^\alpha,z_{21}^\beta,z_{21}^\gamma)\,,
\notag\\
[\mathcal{H}_6\varphi](z_1,z_2)&=-6z_{12}
\int_0^1d\alpha\int_{\bar\alpha}^1d\beta\,
\int_0^{\bar\alpha}d\gamma
\bar\beta\gamma\,
\varphi(z_{12}^\alpha,z_{21}^\beta,z_{21}^\gamma)\,,
\end{align}
and
\begin{align}\label{}
[\widetilde{\mathcal{H}}_1\varphi](z_1,z_2)&=6\int_0^1d\alpha\int_{0}^{\bar\alpha}d\beta\,
\beta\bar\beta\,\varphi(z_{12}^\alpha,z_{21}^\beta)\,,
\notag\\
[\widetilde{\mathcal{H}}_2\varphi](z_1,z_2)&=6\int_0^1d\alpha\int_{\bar\alpha}^1d\beta\,
\beta\bar\beta\,\varphi(z_{12}^\alpha, z_{21}^\beta)\,,
\notag\\
[\widetilde{\mathcal{H}}_3\varphi](z_1,z_2)&=
-2\int_0^1d\alpha\int_0^{\bar\alpha}d\beta\int_{\beta}^{\bar\alpha}d\gamma
\,\varphi(z_{12}^\alpha, z_{21}^\beta,z_{21}^\gamma)\,,
\notag\\
[\widetilde{\mathcal{H}}_4\varphi](z_1,z_2)&=-\int_0^1d\alpha\,\varphi(z_1,z_{12}^\alpha,z_2)\,.
\end{align}

\subsection{$f_{+-} f_{++}$}
%
\begin{eqnarray*}
&&\boxed{
\begin{array}{cl}
 \multirow{4}{*}{$X^{ab}(z_1,z_2)=f_{+-}^a(z_1) f_{++}^b(z_2)~~{}$}
 & {}~Y^{abd}(z_1,z_2,z_3)=g(\mu\lambda)f_{++}^a(z_1) f_{++}^b(z_2) \bar f_{++}^d(z_3)
\\
 & {}~~~~~~G^{ab}(z_1,z_2)=g(\mu\lambda)J^a(z_1) f_{++}^b(z_2)
\\
 & {}~~E^{ab}(z_1,z_2,z_3)=g(\mu\lambda) f_{++}^{c}(z_1) J^{acb}(z_2,z_3)
\\
 & {}~Z^{abc}(z_1,z_2,z_3)=g(\mu\lambda) J^{ab}(z_1,z_2)f_{++}^{c}(z_3)
\end{array}
}
\end{eqnarray*} 
The three-particle counterterm takes the form 
\begin{eqnarray}\label{7:fourteen}
{}[X^{ab}]'_R&=&\frac{\alpha_s}{4\pi\epsilon}\Bigg(\sum_{k=1}^2 (C_k)^{ab}_{a'b'd}
\mathcal{H}_k Y^{a'b'd} +(t^c_{aa'} \otimes t^c_{bb'})\widetilde{\mathcal{H}}_1 G^{a'b'}
+\widetilde{\mathcal{H}}_2 E^{ab}
\notag\\ &&{}
+t^c_{bb'} \widetilde{\mathcal{H}}_3 Z^{acb'} 
+t^c_{ab'} \widetilde{\mathcal{H}}_4 Z^{cbb'}\Bigg),
\end{eqnarray}
with the color structures $C_1,C_2$ as defined in Eq.~(\ref{7:c1-c3}).
The invariant kernels are in this case
\begin{align}\label{}
[\mathcal{H}_1\varphi](z_1,z_2)&=z_{12}^2\int_0^1d\alpha\int_{0}^{\bar\alpha}d\beta\,
\bar\alpha\beta\, \varphi(z_{12}^\alpha,z_2,z_{21}^\beta)\,,
\notag\\
[\mathcal{H}_2\varphi](z_1,z_2)&=z_{12}^2\int_0^1d\alpha\int_{\bar\alpha}^1d\beta\,
\frac{\bar\alpha^2\bar\beta}{\alpha}\,\varphi(z_{12}^\alpha,z_2,z_{21}^\beta)\,,
\end{align}
and
\begin{align}
[\widetilde{\mathcal{H}}_1\varphi](z_1,z_2)&=z_{12}\int_0^1d\alpha\,
\bar\alpha^2\,\varphi(z_{12}^\alpha,z_{2})\,,
\notag\\
[\widetilde{\mathcal{H}}_2\varphi](z_1,z_2)&=z_{12}\int_0^1d\alpha\int_{\bar\alpha}^1d\beta\,
\bar\alpha\,\varphi(z_{12}^\alpha,z_{21}^\beta,z_2)\,,
\notag\\
[\widetilde{\mathcal{H}}_3\varphi](z_1,z_2)&=-z_{12}\int_0^1d\alpha\int_{\bar\alpha}^1d\beta\,
\frac{\bar\beta}{\beta}\,\varphi(z_{21}^\alpha,z_{12}^\beta,z_2)\,,
\notag\\
[\widetilde{\mathcal{H}}_4\varphi](z_1,z_2)&=-z_{12}\int_0^1d\alpha\int_0^{\bar\alpha}d\beta\,
{\bar\beta}\,\varphi(z_{21}^\alpha,z_2,z_{12}^\beta)\,.
\end{align}

\subsection{$\bar f_{++} f_{+-}$ or $\frac12  D_{-+}\bar f_{++} f_{++}$}
%
\begin{eqnarray*}
&&\boxed{
\begin{array}{cl}
 & {}~Y^{abd}(z_1,z_2,z_3)=g(\mu\lambda)\bar f_{++}^a(z_1) f_{++}^b(z_2) \bar f_{++}^d(z_3)
\\
   X^{ab}(z_1,z_2)=\bar f_{++}^a(z_1) f_{+-}^b(z_2)~~{}
 & {}~~~~~~G^{ab}(z_1,z_2)=g(\mu\lambda)\bar f_{++}^a(z_1)\, J^b(z_2)
\\
\text{or}
\\
X^{ab}(z_1,z_2)=\frac12  D_{-+}\bar f_{++}^a(z_1) f_{++}^b(z_2)
 & {}~~E^{ab}(z_1,z_2,z_3)=g(\mu\lambda) \bar f_{++}^{c}(z_1) J^{bca}(z_2,z_3)
\\
 & {}~Z^{abc}(z_1,z_2,z_3)=g(\mu\lambda)\bar f_{++}^{a}(z_3) J^{bc}(z_2,z_3)
\end{array}
}
\end{eqnarray*} 
The three-particle counterterm for the first case, $X^{ab}=\bar f_{++}^a f_{+-}^b$, 
takes the form 
\begin{align}\label{7:fifteen}
[X^{ab}]'_R&=\frac{\alpha_s}{4\pi\epsilon}\Bigg(\sum_{k=1}^6 (C_k)^{ab}_{a'b'd}
\mathcal{H}_k Y^{a'b'd} +(t^c_{aa'}\otimes t^c_{bb'})\left(\widetilde{\mathcal{H}}_1 G^{a'b'}+
\widetilde{\mathcal{H}}_2 G^{b'a'}
\right)
\notag\\
&
+\widetilde{\mathcal{H}}_3 E^{ab}+
t^c_{aa'} \widetilde {\mathcal{H}}_4 Z^{a'bc}
+t^c_{ba'} \widetilde {\mathcal{H}}_5 Z^{a'ca} \Bigg).
\end{align}
The color structures $C_1,C_2,C_3$ are defined in Eq.~(\ref{7:c1-c3}) and $C_4,C_5,C_6$ are
given by 
\begin{align}\label{7:c4-c6:new}
(C_4)^{ab}_{a'b'd}=f^{d ec} t^{e}_{ba'}\, t^c_{ab'}\,, &&
(C_5)^{ab}_{a'b'd}= i(t^dt^c)_{ba'}\, t^c_{ab'}\,, &&
(C_6)^{ab}_{a'b'd}= i(t^c)_{ba'}\, (t^dt^c)_{ab'}\,.
\end{align}
Note that these structures are different from those in Eqs.~(\ref{7:c4-c5}) and (\ref{7:c6}).
The invariant kernels are 
\begin{align}\label{}
[\mathcal{H}_1\varphi](z_1,z_2)&=-z_{12}^2\Biggl\{
\int_0^1d\alpha\int_0^{\bar\alpha}d\beta\,\bar\alpha\beta\,\varphi(z_1,z_{21}^\alpha,z_{12}^\beta)
\notag\\
&\phantom{=-z_{12}^2\Biggl\{}+
\int_0^1d\alpha\int_0^{\bar\alpha}d\beta\int_{\beta}^{\bar\alpha}d\gamma\,
\bar\alpha\bar\beta\gamma\left(2-\frac{\alpha\gamma}{\bar\alpha\bar\gamma}+
2\frac{\beta\bar\gamma}{\bar\beta\gamma}\right)\varphi(z_{12}^\alpha,z_{21}^\beta,z_{21}^\gamma)
\Biggr\}\,,
\notag\\
[\mathcal{H}_2\varphi](z_1,z_2)&=-z_{12}^2
\int_0^1d\alpha\int_0^{\bar\alpha}d\beta\int_{\bar\alpha}^1d\gamma\,\bar\alpha\beta\bar\gamma
\left(2+\frac{\bar\alpha\bar\beta}{\alpha\beta}\right)
\varphi(z_{12}^\alpha,z_{21}^\beta,z_{21}^\gamma)\,,
\notag\\
[\mathcal{H}_3\varphi](z_1,z_2)&=z_{12}^2\Biggl\{
\int_0^1d\alpha\int_{\bar\alpha}^1d\beta\,\frac{\bar\alpha^2\bar\beta}{\alpha}
\varphi(z_{1},z_{21}^\alpha,z_{12}^\beta)
\notag\\
&\hspace*{-0.65cm}
+\int_0^1d\alpha\int_0^{\bar\alpha}d\beta\int_0^{\beta}d\gamma\,\gamma
\left(2\,\bar\alpha\bar\beta\left(2-\frac{\alpha\gamma}{\bar\alpha\bar\gamma}\right)
+\alpha\beta\left(2-\frac{\bar\beta\gamma}{\beta\bar\gamma}\right)\right)
\varphi(z_{12}^\alpha,z_{21}^\beta,z_{21}^\gamma)
\Biggr\}\,,
\notag\\
%
[{\mathcal{H}}_4\varphi](z_1,z_2)&=-3z_{12}^2
\int_0^1d\alpha\int_{\bar\alpha}^1d\beta\int_{\beta}^{\bar\alpha}d\gamma\,
\frac{\bar\alpha^2\bar\beta\bar\gamma}{\alpha}\,\varphi(z_{12}^\alpha,z_{21}^\beta,z_{21}^\gamma)\,,
\notag\\
[{\mathcal{H}}_5\varphi](z_1,z_2)&=\phantom{-}
3z_{12}^2\int_0^1d\alpha\int_{\bar\alpha}^1d\beta\int_0^{\bar\alpha}d\gamma\,
\bar\alpha\bar\beta\gamma\left(2-\frac{\alpha\gamma}{\bar\alpha\bar\gamma}\right)
\varphi(z_{12}^\alpha,z_{21}^\beta,z_{21}^\gamma)\,,
\notag\\
[{\mathcal{H}}_6\varphi](z_1,z_2)&=
-3z_{12}^2\int_0^1d\alpha\int_{\bar\alpha}^1d\beta\int_{\beta}^1d\gamma\,
\frac{\bar\alpha^2\bar\beta\bar\gamma}{\alpha}\,\varphi(z_{12}^\alpha,z_{21}^\beta,z_{21}^\gamma)\,
\end{align}
and
\begin{align}\label{}
[\widetilde{\mathcal{H}}_1\varphi](z_1,z_2)&=-{z_{12}}\biggl\{
\int_0^1d\alpha\,\alpha^2 \varphi(z_1,z_{12}^\alpha)+
2\int_0^1d\alpha\int_0^{\bar\alpha}d\beta\,\bar\alpha\beta\bar\beta
\left(2+\frac{\alpha\beta}{\bar\alpha\bar\beta}\right)
\varphi(z_{12}^\alpha,z_{21}^\beta)
\biggr\}\,,
\notag\\
[\widetilde{\mathcal{H}}_2\varphi](z_1,z_2)&=-6{z_{12}}
\int_0^1d\alpha\int_{0}^{\bar\alpha}d\beta\,\alpha\beta\bar\beta\,
\varphi(z_{21}^\alpha,z_{12}^\beta)\,,
\notag\\
[\widetilde{\mathcal{H}}_3\varphi](z_1,z_2)&=2{z_{12}}
\int_0^1d\alpha\int_0^{\bar\alpha}d\beta\int_{\bar\alpha}^1d\gamma\,\bar\alpha\,
\varphi(z_{12}^\alpha,z_{21}^\beta,z_{21}^\gamma)\,,
\notag\\
[\widetilde{\mathcal{H}}_4\varphi](z_1,z_2)&={z_{12}}\biggl\{
\int_0^1\!d\alpha\int_{\bar\alpha}^1\!\!d\beta\,\frac{\bar\beta}{\beta}\,
\varphi(z_1,z_{12}^\alpha,z_{21}^\beta)+
2\!\int_0^1\!d\alpha\int_0^{\bar\alpha}\!\!d\beta\int_{\beta}^{\bar\alpha}\!\!d\gamma\,
\bar\alpha
\,\varphi(z_{12}^\alpha,z_{21}^\beta,z_{21}^\gamma)\biggr\}
\notag\\
[\widetilde{\mathcal{H}}_5\varphi](z_1,z_2)&=-2{z_{12}}
\int_{0}^1d\alpha\int_{\bar\alpha}^1d\beta\int_{\beta}^1d\gamma\bar\alpha
\,\varphi(z_{12}^\alpha,z_{21}^\beta,z_{21}^\gamma)\,.
\end{align}

{}For the second operator, $X^{ab}=\frac12  D_{-+}\bar f_{++}^a f_{++}^b$, we obtain
\begin{align}\label{7:sixteen}
[X^{ab}]'_R&=\frac{\alpha_s}{4\pi\epsilon}\Bigg(\sum_{k=1}^6 (C_k)^{ab}_{a'b'd}
\mathcal{H}_k Y^{a'b'd} +(t^c_{aa'}\otimes t^c_{bb'})\left(\widetilde{\mathcal{H}}_1 G^{a'b'}+
\widetilde{\mathcal{H}}_2 G^{b'a'}
\right)
\notag\\
&
+\widetilde{\mathcal{H}}_3 E^{ab} +\widetilde{\mathcal{H}}_4 E^{ba}
+ t^c_{aa'}\left(\widetilde {\mathcal{H}}_5 Z^{a'cb}+ \widetilde {\mathcal{H}}_6 Z^{a'bc}\right)
+ t^c_{ba'} \widetilde {\mathcal{H}}_7 Z^{a'ca}\Bigg).
\end{align}
The color structures in this expression, $C_1,C_2,C_3$ and $C_4,C_5,C_6$, are defined in
Eqs.~(\ref{7:c1-c3}) 
and (\ref{7:c4-c6:new}), respectively. The invariant kernels are given by
\begin{align}\label{}
[\mathcal{H}_1\varphi](z_1,z_2)&=z_{12}\Biggl\{
\int_0^1d\beta\bar\beta\varphi(z_1,z_2,z_{12}^\beta)
%
%
+\int_0^1d\alpha\int_{\bar\alpha}^1d\beta\,\frac{\alpha\bar\beta^2}{\beta}
\left(2-\frac{\bar\alpha\bar\beta}{\alpha\beta}\right)
\varphi(z_{21}^\alpha,z_2,z_{12}^\beta)
\notag\\
%
%
%
&
+2\int_0^1d\alpha\int_{0}^{\bar\alpha}d\beta\int_\beta^{\bar\alpha}d\gamma\,\bar\alpha\bar\beta\gamma\,
\left(4+\frac{\alpha\beta}{\bar\alpha\bar\beta}-\frac{\alpha\gamma}{\bar\alpha\bar\gamma}+2
\frac{\beta\bar\gamma}{\bar\beta\gamma}\right)\varphi(z_{12}^\alpha,z_{21}^\beta,z_{21}^\gamma)
\Biggr\}\,,
\notag
\\
[\mathcal{H}_2\varphi](z_1,z_2)&=z_{12}\Biggl\{\int_0^1d\alpha\int_{\bar\alpha}^1d\beta
\frac{\bar\alpha^2\beta}{\alpha}\,\varphi(z_{12}^\alpha,z_2,z_{21}^\beta)
\notag
\\
&
\hspace*{-0.8cm}
+2\int_0^1d\alpha\int_0^{\bar\alpha}d\beta\int_{\bar\alpha}^1d\gamma\,\bar\alpha\left(
\bar\beta\gamma\left(2+\frac{\alpha\beta}{\bar\alpha\bar\beta}\right)+
\beta\bar\gamma\left(2+\frac{\bar\alpha\bar\beta}{\alpha\beta}\right)\right)
\varphi(z_{12}^\alpha,z_{21}^\beta,z_{21}^\gamma)\Biggr\}\,,
\notag
\\
[\mathcal{H}_3\varphi](z_1,z_2)&=-6z_{12}
\int_0^1d\alpha\int_0^{\bar\alpha}d\beta\int_0^{\beta}d\gamma\,\bar\alpha\bar\beta\gamma
\left(2+\frac{\alpha\beta}{\bar\alpha\bar\beta}-\frac{\alpha\gamma}{\bar\alpha\bar\gamma}\right)
\varphi(z_{12}^\alpha,z_{21}^\beta,z_{21}^\gamma)\,,
\end{align}
\begin{align}\label{}
[{\mathcal{H}}_4\varphi](z_1,z_2)&=\phantom{-}6z_{12}
\int_0^1d\alpha\int_{\bar\alpha}^1d\beta\int_{\beta}^{\bar\alpha} d\gamma\,
\bar\alpha\bar\beta\gamma\left(1+\frac{\bar\alpha\bar\gamma}{\alpha\gamma}\right)
\varphi(z_{12}^\alpha,z_{21}^\beta,z_{21}^\gamma)\,,
\notag\\
[{\mathcal{H}}_5\varphi](z_1,z_2)&=-6z_{12}
\int_0^1d\alpha\int_{\bar\alpha}^1d\beta\int_0^{\bar\alpha}d\gamma\,
\bar\alpha\bar\beta\gamma\left(3-\frac{\alpha\gamma}{\bar\alpha\bar\gamma}\right)
\varphi(z_{12}^\alpha,z_{21}^\beta,z_{21}^\gamma)\,,
\notag\\
[{\mathcal{H}}_6\varphi](z_1,z_2)&=
\phantom{-}6z_{12}\int_0^1d\alpha\int_{\bar\alpha}^1d\beta\int_{\beta}^1d\gamma\,
\bar\alpha\bar\beta\gamma\left(1+\frac{\bar\alpha\bar\gamma}{\alpha\gamma}\right)
\varphi(z_{12}^\alpha,z_{21}^\beta,z_{21}^\gamma)\,.
\end{align}
and
\begin{align}\label{}
[\widetilde{\mathcal{H}}_1\varphi](z_1,z_2)&=
6\int_0^1d\alpha\int_0^{\bar\alpha}d\beta\,\bar\alpha\beta\bar\beta
\left(3+\frac{\alpha\beta}{\bar\alpha\bar\beta}\right)
\varphi(z_{12}^\alpha,z_{21}^\beta)\,,
\notag\\
[\widetilde{\mathcal{H}}_2\varphi](z_1,z_2)&=24
\int_0^1d\alpha\int_{0}^{\bar\alpha}d\beta\,\alpha\beta\bar\beta\,
\varphi(z_{21}^\alpha,z_{12}^\beta)\,,
\notag\\
[\widetilde{\mathcal{H}}_3\varphi](z_1,z_2)&=-6
\int_0^1d\alpha\int_0^{\bar\alpha}d\beta\int_{\bar\alpha}^1d\gamma\,\bar\alpha\,
\varphi(z_{12}^\alpha,z_{21}^\beta,z_{21}^\gamma)\,,
\notag\\
[\widetilde{\mathcal{H}}_4\varphi](z_1,z_2)&=
-\int_0^1d\alpha \,\bar\alpha\, \varphi(z_{12}^\alpha,z_1,z_2)\,,
\notag\\
[\widetilde{\mathcal{H}}_5\varphi](z_1,z_2)&=-\int_0^1d\alpha\,\varphi(z_1,z_{12}^\alpha,z_2)\,,
\\
[\widetilde{\mathcal{H}}_6\varphi](z_1,z_2)&=
-6\int_0^1d\alpha\int_0^{\bar\alpha}d\beta\int_{\beta}^{\bar\alpha}d\gamma\,\bar\alpha\,
\varphi(z_{12}^\alpha,z_{21}^\beta,z_{21}^\gamma)\,,
\notag\\
[\widetilde{\mathcal{H}}_7\varphi](z_1,z_2)&=
6\int_0^1d\alpha\int_{\bar\alpha}^1d\beta\int_{\beta}^1d\gamma\bar\alpha
\varphi(z_{12}^\alpha,z_{21}^\beta,z_{21}^\gamma)\,.
\end{align}

\subsection{$\bar\psi_+\psi_{-}$ or $ \frac12 D_{-+}\bar\psi_+\psi_{+}$}
%
\begin{eqnarray*}
&&\boxed{
\begin{array}{cl}
 X^{ij}_{AB}(z_1,z_2)=\bar\psi^{i,A}_+(z_1)\psi_{-}^{j,B}(z_2)~~{}
 & {}\!Y^{ijd}_{AB}(z_1,z_2,z_3)=g(\mu\lambda)\bar\psi^{i,A}_+(z_1)\psi_{+}^{j,B}(z_2)\bar f_{++}^d(z_3)
\\
\text{or}
 & {}~~~\,W^{ab}(z_1,z_2) = g(\mu\lambda){J}^{a}(z_1)\,\bar f^b_{++}(z_2)
\\
X^{ij}_{AB}(z_1,z_2)= \frac12 D_{-+}\bar\psi^{i,A}_+(z_1)\psi_{+}^{j,B}(z_2)~~{}
 & {}~~\,\,G^{abc}(z_1,z_2)=g(\mu\lambda)\bar f^a_{++}(z_1)f^b_{++}(z_2)\bar f^c_{++}(z_3)
\end{array}
}
\end{eqnarray*} 
Note that in this case we display the flavor indices $A,B$ of the quark fields.

The three-particle counterterms for the both operators, 
$X^{ij}_{AB}=\bar\psi^{i,A}_+\psi_{-}^{j,B}$ and 
$X^{ij}_{AB}= \frac12 D_{-+}\bar\psi^{i,A}_+\psi_{+}^{j,B}$,
have the same structure:
\begin{eqnarray}\label{7:seventeen}
[X^{ij}_{AB}]'_R&=&\frac{\alpha_s}{4\pi\epsilon}
\Bigg[\sum_{k=1}^3 (C_k)^{ij}_{i'j'd}\mathcal{H}_k Y_{AB}^{i'j'd}
\nonumber \\
&&{}+
\delta_{AB}\left( 
  i(t^{a}t^b)_{ji} \widetilde{\mathcal{H}}_1 W^{ab}
+ i(t^{b}t^a)_{ji} \widetilde{\mathcal{H}}_2 W^{ab}
+ \sum_{k=1}^6 (L_k)^{ij}_{abc} \bar{\mathcal{H}}_k G^{abc}
\right)\Bigg].
\end{eqnarray}
The color structures $C_k$ are defined in Eq.~(\ref{7:c1-c3}) and 
there are six new structures $(L_k)^{ij}_{abc}$:
\begin{align}\label{7:l1-l6}
L_1=(t^{a}t^b t^c)_{ji}\,,&& L_2=(t^{c}t^a t^b)_{ji}\,,&& L_3=(t^{a}t^c t^b)_{ji}\,,
\notag\\
L_4=(t^{c}t^b t^a)_{ji}\,,&& L_5=(t^{b}t^a t^c)_{ji}\,,&& L_6=(t^{b}t^c t^a)_{ji}\,.
\end{align}
The invariant kernels for the first operator, $X^{ij}_{AB}=\bar\psi^{i,A}_+\psi_{-}^{j,B}$,
are given by
\begin{align}\label{bchipsi}
[\mathcal{H}_1\varphi](z_1,z_2)&=-z^2_{12}\Biggl\{
\int_0^1d\alpha\int_{\bar\alpha}^1d\beta \,\bar\beta\,\varphi(z_1,z_{12}^\alpha,z_{21}^\beta)
\notag\\
&\phantom{=-z^2_{12}\Biggl\{}+
\int_0^1d\alpha\int_0^{\bar\alpha}d\beta\int_{\beta}^{\bar\alpha}d\gamma\,\gamma\,
\varphi(z_{12}^\alpha,z_{21}^\beta,z_{21}^\gamma)\Biggr\}\,,
\notag\\
[\mathcal{H}_2\varphi](z_1,z_2)&=-z^2_{12}
\int_0^1d\alpha\int_0^{\bar\alpha}d\beta\int_{\bar\alpha}^1 d\gamma\,  \gamma\,
\varphi(z_{12}^\alpha,z_{21}^\beta,z_{21}^\gamma)\,,
\notag\\
[\mathcal{H}_3\varphi](z_1,z_2)&=z^2_{12}
\Biggl\{\int_0^1d\alpha\int_{\bar\alpha}^1\!d\beta \frac{\bar\alpha\bar\beta}{\alpha}
\varphi(z_1,z_{21}^\alpha,z_{12}^\beta)
\notag\\
&\phantom{=z^2_{12}\Biggl\{}
+\int_0^1d\alpha\int_0^{\bar\alpha}\! d\beta\int_0^{\beta}\! d\gamma \gamma
\varphi(z_{12}^\alpha,z_{21}^\beta,z_{21}^\gamma)
\Biggr\}\,,
\end{align}
\begin{align}\label{}
[\widetilde{\mathcal{H}}_1\varphi](z_1,z_2)&=z_{12}^2
\int_0^1d\alpha\int_{\bar\alpha}^1d\beta\,\bar\alpha^2\bar\beta\,\varphi(z_{12}^\alpha,z_{21}^\beta)\,,
\notag\\
[\widetilde{\mathcal{H}}_2\varphi](z_1,z_2)&=z_{12}^2
\int_0^1d\alpha\int_{\bar\alpha}^1d\beta\,\alpha\bar\alpha\bar\beta
\left(2-\frac{\bar\alpha\bar\beta}{\alpha\beta}\right)\varphi(z_{21}^\alpha,z_{12}^\beta)\,.
\end{align}
and
\begin{align}\label{}
[\bar{\mathcal{H}}_1\varphi](z_1,z_2)&=z_{12}^3
\int_0^1d\alpha\int_{\bar\alpha}^1d\beta\int_{\beta}^1d\gamma
\,\frac{\bar\alpha^2\bar\beta\bar\gamma}{\alpha}\,\varphi(z_{12}^\alpha,z_{21}^\beta,z_{21}^\gamma)\,,
\notag\\
[\bar{\mathcal{H}}_2\varphi](z_1,z_2)&=z_{12}^3
\int_0^1d\alpha\int_{\bar\alpha}^1d\beta\int_{0}^{\bar\alpha}\bar\alpha\bar\beta\gamma
\left(2-\frac{\alpha\gamma}{\bar\alpha\bar\gamma}\right)
\varphi(z_{12}^\alpha,z_{21}^\beta,z_{21}^\gamma)\,,
\notag\\
[\bar{\mathcal{H}}_3\varphi](z_1,z_2)&=z_{12}^3
\int_0^1d\alpha\int_{\bar\alpha}^1d\beta\int_{\bar\alpha}^\beta d\gamma
\,\frac{\bar\alpha^2\bar\beta\bar\gamma}{\alpha}\,\varphi(z_{12}^\alpha,z_{21}^\beta,z_{21}^\gamma)\,,
\notag\\
[\bar{\mathcal{H}}_4\varphi](z_1,z_2)&=z_{12}^3
\int_0^1d\alpha\int_0^{\bar\alpha}d\beta\int_0^\beta d\gamma\,\bar\alpha\bar\beta\gamma
\left(1+\frac{\alpha\beta}{\bar\alpha\bar\beta}\left(1-\frac{\bar\beta\gamma}{\beta\bar\gamma}\right)
\right)\,
\varphi(z_{12}^\alpha,z_{21}^\beta,z_{21}^\gamma)\,,
\notag\\
[\bar{\mathcal{H}}_5\varphi](z_1,z_2)&=z_{12}^3
\int_0^1d\alpha\int_0^{\bar\alpha}d\beta\int_{\bar\alpha}^1d\gamma\,\bar\alpha\beta\bar\gamma\,
\varphi(z_{12}^\alpha,z_{21}^\beta,z_{21}^\gamma)\,,
\notag\\
[\bar{\mathcal{H}}_6\varphi](z_1,z_2)&=z_{12}^3
\int_0^1d\alpha\int_0^{\bar\alpha}d\beta\int_\beta^{\bar\alpha}d\gamma\,\bar\alpha\beta\bar\gamma\,
\varphi(z_{12}^\alpha,z_{21}^\beta,z_{21}^\gamma)\,.
\end{align}

For the second operator, 
$X^{ij}_{AB}= \frac12 D_{-+}\bar\psi^{i,A}_+\psi_{+}^{j,B}$,
we obtain
\begin{align}\label{dbchipsi}
[\mathcal{H}_1\varphi](z_1,z_2)&=z_{12}\Biggl\{
\int_0^1 d\beta\,\bar\beta\,\varphi(z_1,z_{2},z_{12}^\beta)+
\int_0^1 d\alpha\int_{\bar\alpha}^1d\beta\,\frac{\bar\beta^2}{\beta}\,
\varphi(z_{21}^\alpha,z_{2},z_{12}^\beta)
\notag\\
&\phantom{=z_{12}\Biggl\{}
+2\int_0^1d\alpha\int_0^{\bar\alpha}d\beta\int_{\beta}^{\bar\alpha}d\gamma\,\gamma\,
\varphi(z_{12}^\alpha,z_{21}^\beta,z_{21}^\gamma)
\Biggl\}\,,
\notag\\
[\mathcal{H}_2\varphi](z_1,z_2)&=z_{12}\Biggl\{
\int_0^1d\alpha\int_{\bar\alpha}^1d\beta\frac{\bar\alpha\beta}{\alpha}
\left(2-\frac{\bar\alpha\bar\beta}{\alpha\beta}\right)
\varphi(z_{12}^\alpha,z_{2},z_{21}^\beta)
\notag\\
&
\phantom{=z_{12}\Biggl\{}
+2\int_0^1d\alpha\int_0^{\bar\alpha}d\beta\int_{\bar\alpha}^1 d\gamma\,\gamma\,
\varphi(z_{12}^\alpha,z_{21}^\beta,z_{21}^\gamma)
\Biggr\}\,,
\notag\\
[\mathcal{H}_3\varphi](z_1,z_2)&=2\,z_{12}
\int_0^1d\alpha\int_0^{\bar\alpha}\! d\beta\int_0^{\beta}\! d\gamma \gamma
\varphi(z_{12}^\alpha,z_{21}^\beta,z_{21}^\gamma)\,,
\end{align}
\begin{align}\label{}
[\widetilde{\mathcal{H}}_1\varphi](z_1,z_2)&=-z_{12}
\int_0^1d\alpha\int_{\bar\alpha}^1d\beta\,\alpha\bar\alpha\beta\,
\left(2+\frac{\bar\alpha\bar\beta}{\alpha\beta}\right)\varphi(z_{12}^\alpha,z_{21}^\beta)\,,
\notag\\
[\widetilde{\mathcal{H}}_2\varphi](z_1,z_2)&=-z_{12}
\int_0^1d\alpha\int_{\bar\alpha}^1d\beta\,\alpha\bar\alpha\bar\beta
\left(4-\frac{\bar\alpha\bar\beta}{\alpha\beta}\right)\varphi(z_{21}^\alpha,z_{12}^\beta)\,
\end{align}
and
\begin{align}\label{}
[\bar{\mathcal{H}}_1\varphi](z_1,z_2)&=-z_{12}^2
\int_0^1d\alpha\int_{\bar\alpha}^1d\beta\int_{\beta}^1d\gamma
\,\bar\alpha\bar\beta\gamma\left(2+\frac{\bar\alpha\bar\gamma}{\alpha\gamma}\right)
\varphi(z_{12}^\alpha,z_{21}^\beta,z_{21}^\gamma)\,,
\notag\\
[\bar{\mathcal{H}}_2\varphi](z_1,z_2)&=-z_{12}^2
\int_0^1d\alpha\int_{\bar\alpha}^1d\beta\int_{0}^{\bar\alpha}d\gamma\,\bar\alpha\bar\beta\gamma
\left(4-\frac{\alpha\gamma}{\bar\alpha\bar\gamma}\right)
\varphi(z_{12}^\alpha,z_{21}^\beta,z_{21}^\gamma)\,,
\notag\\
[\bar{\mathcal{H}}_3\varphi](z_1,z_2)&=-z_{12}^2
\int_0^1d\alpha\int_{\bar\alpha}^1d\beta\int_{\bar\alpha}^{\beta}d\gamma
\,\bar\alpha\bar\beta\gamma\left(2+\frac{\bar\alpha\bar\gamma}{\alpha\gamma}\right)
\varphi(z_{12}^\alpha,z_{21}^\beta,z_{21}^\gamma)\,,
\notag\\
[\bar{\mathcal{H}}_4\varphi](z_1,z_2)&=-z_{12}^2
\int_0^1d\alpha\int_0^{\bar\alpha}d\beta\int_0^\beta d\gamma\,\bar\alpha\bar\beta\gamma
\left(2+\frac{\alpha\beta}{\bar\alpha\bar\beta}\left(2-\frac{\bar\beta\gamma}{\beta\bar\gamma}\right)
\right)\,
\varphi(z_{12}^\alpha,z_{21}^\beta,z_{21}^\gamma)\,,
\notag\\
[\bar{\mathcal{H}}_5\varphi](z_1,z_2)&=-z_{12}^2
\int_0^1d\alpha\int_0^{\bar\alpha}d\beta\int_{\bar\alpha}^1d\gamma\,\bar\alpha\bar\beta\gamma\,
\left(1+\frac{\alpha\beta}{\bar\alpha\bar\beta}+\frac{\beta\bar\gamma}{\bar\beta\gamma}\right)\,
\varphi(z_{12}^\alpha,z_{21}^\beta,z_{21}^\gamma)\,,
\notag\\
[\bar{\mathcal{H}}_6\varphi](z_1,z_2)&=-z_{12}^2
\int_0^1d\alpha\int_0^{\bar\alpha}d\beta\int_\beta^{\bar\alpha}d\gamma\,\bar\alpha\bar\beta\gamma\,
\left(1+\frac{\alpha\beta}{\bar\alpha\bar\beta}+\frac{\beta\bar\gamma}{\bar\beta\gamma}\right)\,
\varphi(z_{12}^\alpha,z_{21}^\beta,z_{21}^\gamma)\,.
\end{align}

\section{Conclusions}

Extending the work by Bukhvostov, Frolov, Lipatov and Kuraev (BFLK) 
\cite{Bukhvostov:1985rn}, in this paper we derive a complete set of two-particle 
$2\to2$ RG kernels for collinear twist $E=3$ operators and also the $2\to 3$ mixing kernels 
into three-particle quasipartonic operators, cf. Fig.~1.  
The kernels are written for the renormalization of light-ray
operators built of chiral fields in a particular basis such that the conformal
symmetry is manifest. They serve as building blocks in the RG equations 
for arbitrary gauge-invariant twist-four operators in Quantum Chromodynamics.  
The results can easily be recast in momentum fraction space,
in the form of evolution equations for higher-twist generalized parton distributions
(cf.~\cite{Belitsky:2000vx,Belitsky:2005bu}). 
Specific applications will be considered elsewhere. 

To this end we suggest a new technique which is based on using mostly algebraic methods 
and bypasses calculation of Feynman diagrams. The main idea is that RG equations for 
operators containing field components of different collinear 
twist are related by Poincar{\e} symmetry. In this way the $SL(2,\mathbb{R})$ invariance of 
BFLK kernels for quasipartonic operators becomes extended to the full conformal
group $SO(4,2)$. As the result, the $2\to 2$ RG kernels involving ``minus'' quark 
and/or gluon fields can be obtained from the results of Ref.~\cite{Bukhvostov:1985rn}
by a simple replacement of the two-particle quadratic $SL(2,\mathbb{R})$ Casimir operator
by the $SO(4,2)$ one, cf. Ref.~\cite{BFKS04}. The way to explicit expressions for 
$2\to 3$ kernels is somewhat longer: They can be obtained by applying translations 
and Lorentz rotations in the transverse plane to the renormalized leading twist 
operators. We found this technique to be very efficient. It also offers a new 
insight in the algebraic structure of operator renormalization in gauge theories and 
may be useful in a more general context. It would be interesting to investigate  
whether the same methods can be used beyond the one-loop approximation.

\section*{Acknowledgements}

This work was supported by the German Research Foundation (DFG),
grants 92090175 and 9209282, grant RNP 2.1.1/1575 and by the RFFI grants 07-02-92166, 09-01-93108.


\appendix

\section*{Appendices}

\section{The Conformal Group}\label{App:A}

Generators of the conformal group act on the fundamental fields in
the spinor representation in the following way \cite{MS69}:
%
\begin{eqnarray}\label{A:conf}
i[{\mathbf P}_{\alpha\dot\alpha},\Phi(x)]&=&\partial_{\alpha\dot\alpha} \Phi(x)
\,\equiv\, iP_{\alpha\dot\alpha}\Phi(x)\,,
\nonumber\\
i[{\mathbf D},\Phi(x)]&=&\frac12\left(x_{\alpha\dot\alpha}\partial^{\alpha\dot\alpha}+2\twist+
\xi^\alpha\frac{\partial}{\partial\xi^\alpha}
+\bar\xi_{\dot\alpha}\frac{\partial}{\partial\bar\xi_{\dot\alpha}}\right)\Phi(x)
\,\equiv\, iD\,\Phi(x) \,,
\nonumber\\
i[{\mathbf M}_{\alpha\beta},\Phi(x)]&=&\frac14\left(x_{\alpha\dot\gamma}
{\partial_{\beta}}^{\dot\gamma}+
x_{\beta\dot\gamma} {\partial_{\alpha}}^{\dot\gamma}
-2\xi_{\alpha}\frac{\partial}{\partial{\xi^\beta}}-2\xi_{\beta}\frac{\partial}{\partial{\xi^\alpha}}
\right)\Phi(x)
\,\equiv\, iM_{\alpha\beta} \Phi(x)\,,
\\
i[ \bar{\mathbf  M}_{\dot\alpha\dot\beta},\Phi(x)]&=&\frac14\left(x_{\gamma\dot\alpha}
{\partial^\gamma}_{\dot\beta}
+
x_{\gamma\dot\beta}{\partial^\gamma}_{\dot\alpha}
-
2\bar\xi_{\dot\alpha}\frac{\partial}{\partial{\bar\xi^{\dot\beta}}}-
2\bar\xi_{\dot\beta}\frac{\partial}{\partial{\bar\xi^{\dot\alpha}}}
\right)\Phi(x)
\,\equiv\, i\bar M_{\dot\alpha\dot\beta} \Phi(x)\,,
\nonumber\\
i[{\mathbf K}_{\alpha\dot\alpha},\Phi(x)]&=&
\left(x_{\alpha\dot\gamma} x_{\gamma\dot\alpha}\,\partial^{\gamma\dot\gamma}+
2\twist x_{\alpha\dot\alpha}+2\xi_\alpha {\bar x_{\dot\alpha}}^{\phantom{\alpha}\beta}
\frac{\partial}{\partial\xi^\beta}
+
2\bar\xi_{\dot\alpha} x_{\alpha\dot\beta}\frac{\partial}{\partial\bar\xi_{\dot\beta}}\right)\Phi(x)
\,\equiv\, iK_{\alpha\dot\alpha} \Phi(x)
\,.
\nonumber
\end{eqnarray}
%
Here $\Phi=(\Phi_\xi,\bar \Phi_\xi)$ with $\Phi_\xi=\{\psi_\xi,\chi_\xi, f_{\xi\xi}\}$,
$\bar \Phi_{\xi}=\{\bar\psi_\xi,\bar \chi_\xi, \bar f_{\xi\xi}\}$ where $\xi$ is an
auxiliary spinor. The generators $P_{\alpha\dot\alpha}$, $D$, $M_{\alpha\beta}$ ($\bar
M_{\dot\alpha\dot\beta}$)
and $K_{\alpha\dot\alpha}$ correspond to translations, dilatation, Lorentz rotations and special
conformal transformations, respectively.

The chiral fields $\Phi_\xi$ transform according to the $(s,0)$ representation of the Lorentz
group
$$
\frac12M^2\Phi_\xi^{(s 0)}=s(s+1)\Phi_\xi^{(s 0)}\,,\qquad
\frac12\bar M^2 \Phi_\xi^{(s 0)}= 0\,,
$$
where $M^2\equiv M_{\alpha\beta}M^{\alpha\beta}$, $\bar M^2 \equiv \bar
M_{\dot\alpha\dot\beta}\bar M^{\dot\alpha\dot\beta}$
and $s=1/2$ or $s=1$ for quarks and gluons. In turn, the antichiral fields $\bar \Phi_\xi$
transform according to the $(0,\bar s)$ representation
$$
\frac12M^2\bar\Phi_\xi^{ 0 \bar s)}= 0\,,\qquad
\frac12\bar M^2 \bar \Phi_\xi^{(0 \bar s)}= \bar s(\bar s+1) \bar \Phi_\xi^{(0 \bar s)}\,.
$$
Finally $\twist=1$ is the {\em geometric}\, twist~\cite{Gross:1971wn} of the field $\Phi$:
It is defined as  $\twist = \ell^{\rm can}-s-\bar s$ where $\ell^{\rm can}$ is the canonical dimension.

The commutation relations for the generators read
\begin{eqnarray}
[P^{\alpha\dot\alpha},D] &= &-iP^{\alpha\dot\alpha}\,,\hspace*{3.5cm}
{}[K^{\alpha\dot\alpha},D]\,=\,iK^{\alpha\dot\alpha}\,,
\nonumber
\\
{}[P_{\alpha\dot\alpha},M_{\beta\gamma}]&= &\frac{i}2\left[\epsilon_{\alpha\beta}P_{\gamma\dot\alpha}
+\epsilon_{\alpha\gamma} P_{\beta\dot\alpha}
\right]\,,\qquad
{}[P_{\alpha\dot\alpha},\bar M_{\dot\beta\dot\gamma}]\,=\,-
\frac{i}2\left[\epsilon_{\dot\alpha\dot\beta}P_{\alpha\dot\gamma}
+\epsilon_{\dot\alpha\dot\gamma} P_{\alpha\dot\beta}
\right]\,,
\nonumber
\\ \notag
{}[M_{\alpha\beta}, M_{\gamma\delta}]&=&-\frac{i}2\left(
\epsilon_{\gamma\beta}
M_{\alpha\delta}+\epsilon_{\delta\beta}M_{\alpha\gamma}+\epsilon_{\gamma\alpha}M_{\beta\delta}
+\epsilon_{\delta\alpha}M_{\beta\gamma}
\right)\,,
\\ \notag
{}[\bar M_{\dot\alpha\dot\beta}, \bar M_{\dot\gamma\dot\delta}]&=&-\frac{i}2\left(
\epsilon_{\dot\beta\dot\gamma}
\bar M_{\dot\alpha\dot\delta}+\epsilon_{\dot\beta\dot\delta}\bar M_{\dot\alpha\dot\gamma}
+\epsilon_{\dot\alpha\dot\gamma}\bar M_{\dot\beta\dot\delta}
+\epsilon_{\dot\alpha\dot\delta}\bar M_{\dot\beta\dot\gamma}
\right)\,,
\\
{}[P^{\beta\dot\beta},
K_{\alpha\dot\alpha}]&=&-4i
\left(\delta^{\beta}_{\alpha}\,\delta^{\dot\beta}_{\dot\alpha}\, D+
\delta^{\dot\beta}_{\dot\alpha}\, {M_{\alpha}}^{\beta}
+\delta^{\beta}_{\alpha}\,{\bar M_{\dot\alpha}}^{\phantom{\dot\alpha}\dot\beta}
 \right)\,.
\end{eqnarray}
The expression for the quadratic $SO(4,2)$ Casimir operator
$\mathbb{C}^2$ reads:
\begin{align}
\mathbb{C}^2=&\frac18\Big(K^{\alpha\dot\alpha}
P_{\alpha\dot\alpha}+P_{\alpha\dot\alpha}K^{\alpha\dot\alpha}\Big)
-\frac12D^2+\frac12\Big(
M_{\alpha\beta}M^{\alpha\beta}+
\bar M_{\dot\alpha\dot\beta}\bar M^{\dot\alpha\dot\beta}\Big)\,.
\end{align}
One finds
\begin{align}
\mathbb{C}^2\,\Phi_\xi^{(s 0)}=
\frac32(s^2-1)\,\Phi_\xi^{(s 0)}\,, &&
\mathbb{C}^2 \, \bar\Phi_\xi^{(0\bar s)}=
\frac32(\bar s^2-1)\, \bar\Phi_\xi^{(0\bar s)}\,.
\end{align}
The generators of the collinear $SL(2,\mathbb{R})$ subgroup are defined as
\begin{eqnarray}
&&S_+\,=\,\frac{i}2(\mu\, K\,\bar\mu)\,, \qquad S_-\,=\,-\frac{i}2(\lambda\, P\,\bar\lambda)\,,
\nonumber\\ &&
S_0\,=\,\frac{i}2\left((\mu\lambda)(\bar\lambda\bar\mu) D
 -(\bar\lambda\bar\mu)\mu^\alpha\lambda^\beta M_{\alpha\beta}
-(\mu\lambda)\bar M_{\dot\alpha\dot\beta}\bar\mu^{\dot\alpha}\bar\lambda^{\dot\beta}\right)\,,
\end{eqnarray}
In order to shorten the following expressions,
without loss of generality one can assume \cite{Braun:2008ia}
\begin{eqnarray}
  \lambda^{\alpha}&=&(1,0)\,, \qquad  \lambda_\alpha\,=\,(0,1)\,,
\nonumber \\
 \mu^\alpha&=&(0,1)\,,\qquad  \mu_\alpha\,=\,(-1,0) \,,
\label{n-choice}
\end{eqnarray}
so that $(\mu\lambda)=(\bar\lambda\bar\mu) = 1$. Using this convention the same generators can
be written as
\begin{align}
S_+=\frac{i}{2}K_{2\dot 2}\,,&& S_-=-\frac{i}2 P_{1\dot 1}\,,&&
S_0=\frac{i}{2}\Big(D-M_{12}-\bar M_{\dot 1\dot 2}\Big)\,.
\end{align}
The $SL(2,\mathbb{R})$ generators commute with the operators of the collinear twist
$E$ and chirality $H$ defined as
\begin{align}
E=i\big(D+M_{12}+\bar M_{\dot 1\dot 2}\big)\,, &&
H=i\big(\bar M_{\dot 1\dot 2}-M_{12}\big)\,.
\end{align}
For our purposes it is useful to separate the contribution of the light-cone subgroup
to the expression for the two-particle Casimir operator $\mathbb{C}^2_{12}$:
\begin{eqnarray}\label{A:Jfull}
\mathbb{C}_{12}^2-\mathbb{C}^2_{1}-\mathbb{C}^2_{2}&=&
(\vec{S}_1+\vec{S}_2)^2-\vec{S}_1^2-\vec{S}_2^2
+\frac12 E_1 E_2+H_1H_2
\nonumber\\
&+& \frac14\left(K_1^{1\dot 2} P_{2,1\dot 2}+K_1^{2\dot1} P_{2,2\dot1}+K_2^{1\dot 2}
P_{1,1\dot2}+
K_2^{2\dot 1}
P_{1,2\dot 1}+K_{1}^{2\dot 2} P_{2,2\dot 2}+K_{2}^{2\dot 2} P_{1,2\dot 2}\right)
\nonumber\\
&+& \Big(M_{1}^{2 2} M_{2}^{1 1}+
M_{1}^{11} M_{2}^{22}+\bar M_{1}^{\dot2\dot2} \bar M_{2}^{\dot1\dot1}+\bar M_{1}^{\dot1\dot1} \bar
M_{2}^{\dot2\dot2}\Big)\,,
\end{eqnarray}
where e.g. $P_{2,1\dot 2}$ stands for the $(1\dot 2)$ component of the translation
operator acting on the second field, etc. The terms in the last two lines
all vanish if applied to the tensor product of two ``plus'' light-ray fields,
so that one finds
\begin{align}\label{A:CJ}
\mathbb{C}_{12}^2\,\Phi^{(1)}_+\otimes
\Phi^{(2)}_+=\Big(S_{12}^2-\kappa_{12}\Big) \Phi^{(1)}_+\otimes \Phi^{(2)}_+\,.
\end{align}
Here $S_{12}^2=(\vec{S}_1+\vec{S}_2)^2$ is the two-particle Casimir operator of the
collinear $SL(2,\mathbb{R})$ subgroup and $\kappa_{12}$ is a constant:
\begin{align}\label{A:Ckappa}
\kappa_{12}=\left[2-\frac12\left(h_1+h_2\right)^2\right]\,,
\end{align}
where $h_1, h_2$ are the helicities of the fields:
$H_k\,\Phi^{(k)}_+=h_k\,\Phi^{(k)}_+ $.

\section{Invariant kernels}\label{App:B}
In this appendix we formulate a simple rule how to construct $SL(2,\mathbb{R})$ 
invariant kernels.

{}First, note that if a function $\varphi(z_1,z_2)$ transforms according to the 
representation $T^{j_1}\otimes T^{j_2}$ of the $SL(2,\mathbb{R})$ group, i.e.
\begin{align}\label{}
\varphi(z_1,z_2)\to \varphi'(z_1,z_2)=
\frac{1}{(cz_1+d)^{2j_1}(cz_2+d)^{2j_2}}\,\varphi\left(\frac{az_1+b}{cz_1+d},\frac{az_2+b}{cz_2+d}
\right),
\end{align}
then the function $f(z_1,z_2)=(z_1-z_2)^{2n}\varphi(z_1,z_2)$, $n$ being integer or half-integer,
transforms according to the representation $T^{j_1-n}\otimes T^{j_2-n}$. 
An invariant operator $\mathcal{H}$ which maps $T^{j_1}\otimes T^{j_2}\to T^{i_1}\otimes T^{i_2}$
exists if and only if $j_1+j_2=i_1+i_2+2m$, where $m$ is, again, integer or half-integer. 
It is clear that invariant kernels for $m\not=0$, $\mathcal{H}^{(m)}$, can be
represented in the form $\mathcal{H}^{(m)}=(z_1-z_2)^{-2m}\,\mathcal{H}$, where $\mathcal{H}$ is
an invariant kernel for the case  $j_1+j_2=i_1+i_2$. It is, therefore, sufficient to
consider this case only.

To write down the most general expression for an
invariant operator $\mathcal{H}:T^{j_1}\otimes T^{j_2}\mapsto T^{i_1}\otimes T^{i_2}$
acting on the space of functions of two variables $\varphi(w_1,w_2)\mapsto \varphi(z_1,z_2)$,
consider the following diagram:
\begin{figure}[h]
\psfrag{a}[cc][cc][0.9]{$\alpha_{11},p_{11}$}
\psfrag{b}[cc][cc][0.9]{$\alpha_{22},p_{22}$}
\psfrag{c}[lc][cc][0.9]{$\alpha_{21},p_{21}$}
\psfrag{d}[lc][cc][0.9]{$\alpha_{12},p_{12}$}
\psfrag{z1}[cc][cc][0.9]{$z_1$}
\psfrag{z2}[cc][cc][0.9]{$z_2$}
\psfrag{w1}[cc][cc][0.9]{$w_1$}
\psfrag{w2}[cc][cc][0.9]{$w_2$}
\centerline{\includegraphics[width=3.5cm]{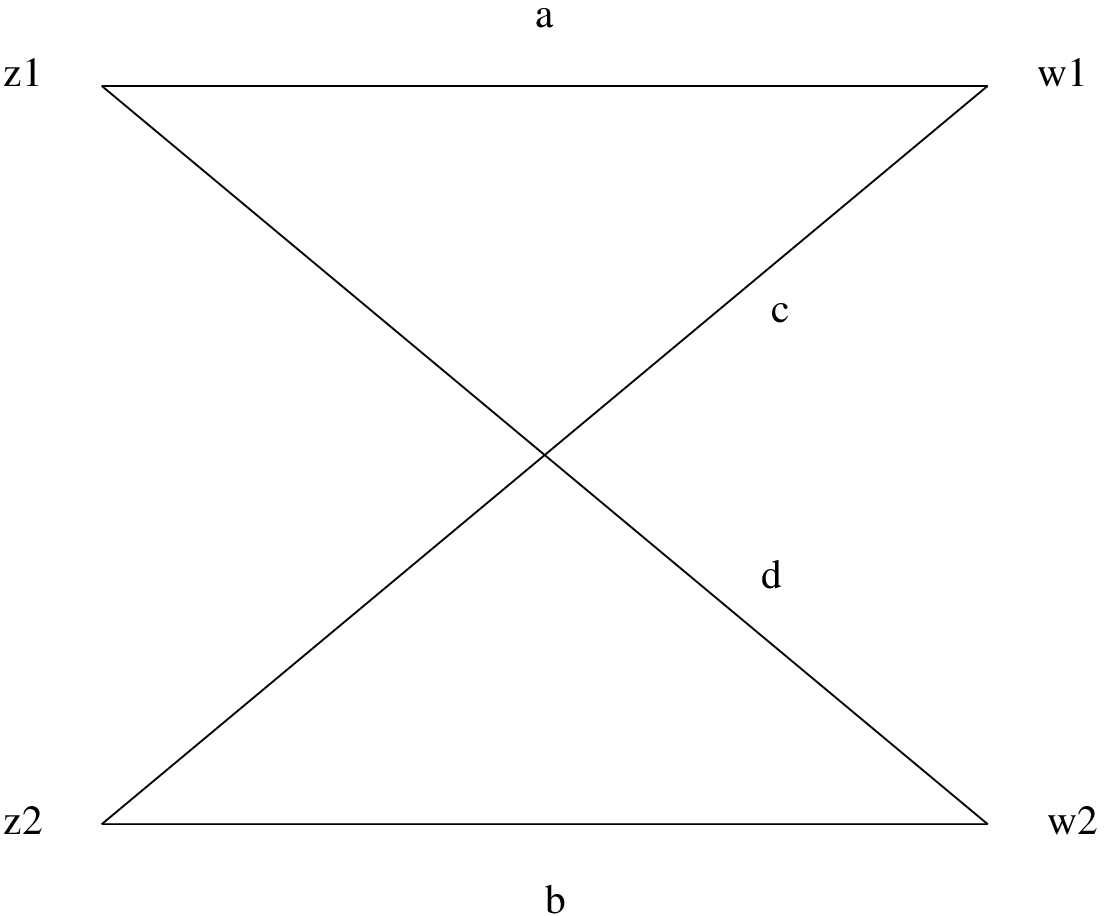}                        }
\end{figure}
In this diagram the points $w_1,w_2$ are connected with $z_1,z_2$ by four lines.
Each line, e.g. connecting the points $z_i$ and $w_k$, carries a pair of indices, 
$(\alpha_{ik},p_{ik})$, $i,k=1,2$.
The indices on  different lines are subject to the constraints:
\begin{align}\label{XY}
\alpha_{1k}+\alpha_{2k}=1\,, &&p_{1k}+p_{2k}=2j_k\,,&&p_{k1}+p_{k2}=2i_k\,, && k=1,2\,.
\end{align}
An invariant kernel has the form~\cite{Belitsky:2004sc}
\begin{align}\label{Xform}
[\mathcal{H}\varphi](z_1,z_2)=\int_0^1d\alpha_{21}\int_{0}^1 d\alpha_{12} \prod^2_{i,k=1} 
\alpha_{ik}^{p_{ik}-1}\, 
\omega\left(\frac{\alpha_{21}\alpha_{12}}{\alpha_{11}\alpha_{22}}\right)\,
\varphi(w_1(\alpha),w_2(\alpha))\,,
\end{align}
where 
$$
w_k(\alpha)=\sum_{m=1}^2 z_m \alpha_{mk}
$$
and $\omega(x)$ is an arbitrary function.

The choice of $p$-parameters which satisfy Eq.~(\ref{XY}) on is not unique, 
but the ratio of the integration weights corresponding to two different solutions 
is a power of the invariant (anharmonic) ratio~$(\alpha_{21}\alpha_{12})/(\alpha_{11}\alpha_{22})$, 
i.e. a different choice results in the redefinition of the function $\omega$.

Changing the notation to 
\begin{align*}
\alpha_{11}=1-\alpha\,,&&
\alpha_{21}=\alpha,&& \alpha_{12}=\beta,&& \alpha_{22}=1-\beta\,,
&& p_{11}=\hat a,&& p_{21}=a,&& p_{22}=\hat b,&& p_{12}=b 
\end{align*}	
Eq.~(\ref{Xform}) takes a more familiar form
\begin{align}\label{t22inv}
[\mathcal{H}\varphi](z_1,z_2)=\int_0^1\!d\alpha\int_0^1\!d\beta\,
\bar\alpha^{\hat a-1}\alpha^{a-1}\bar\beta^{\hat b-1}\beta^{b-1}\, \omega\left(
\frac{\alpha\beta}{\bar\alpha\bar\beta}
\right)\,\varphi(z_{12}^\alpha,z_{21}^\beta)\,.
\end{align}

Let us compare this general expression with the kernels given in~(\ref{H-list}).
Notice that the integrations in (\ref{H-list}) do not go over the unit square, 
$0<\alpha,\beta<1$ as in (\ref{t22inv}), but over the simplex either $0<\alpha+\beta<1$ 
or $1<\alpha+\beta$. These integration regions are not arbitrary, since the only
way to obtain them is to choose the function $\omega(x)$ proportional to the 
theta-function of the anharmonic ratio. In particular, the choice $\omega(x)=\theta(1-x)\, 
\tilde \omega(x)$ results in the integration region $0<\alpha+\beta<1$. 
The kernels in~(\ref{H-list}) that contain one-dimensional integrals arise when $w(x)$ 
is proportional to a $\delta-$function. For example, $\mathcal{H}^d$ corresponds to the choice
$\omega(x)=\delta(1-x)$.  The `exchange' kernels $\mathcal{H}_{12}^{e,(k)}$ correspond
to the solutions with
\begin{align*}	
\hat a=2j_1-k-1\,, && a=k+1\,, &&\hat b=2j_2-1\,,&& b=1
\end{align*}	
and $\omega(x)=\delta(x)$.

More complicated, e.g. $2\to 3$ kernels $\mathcal{H}: T^{j_1}\otimes T^{j_2}\otimes
T^{j_3}\mapsto T^{i_1}\otimes T^{i_2}$, can be constructed similarly. 
It is again sufficient to consider the case $j_1+j_2+j_3=i_1+i_2$. 
To this end, draw a diagram with  $w_1,w_2,w_3$ points  connected
to each of the $z_1,z_2$ points:    
\begin{figure}[ht]
\psfrag{x11}[rc][cc][0.9]{$\alpha_{11},p_{11}$}
\psfrag{x22}[lc][cc][0.9]{$\alpha_{22},p_{22}$}
\psfrag{x21}[lc][cc][0.9]{$\alpha_{21},p_{21}$}
\psfrag{x12}[lc][cc][0.9]{$\alpha_{12},p_{12}$}
\psfrag{x23}[rc][cc][0.9]{$\alpha_{23},p_{23}$}
\psfrag{x13}[lc][cc][0.9]{$\alpha_{13},p_{13}$}
\psfrag{z1}[cc][cc][0.9]{$z_1$}
\psfrag{z2}[cc][cc][0.9]{$z_2$}
\psfrag{w1}[cc][cc][0.9]{$w_1$}
\psfrag{w2}[cc][cc][0.9]{$w_2$}
\psfrag{w3}[cc][cc][0.9]{$w_3$}
\centerline{\includegraphics[width=3.5cm]{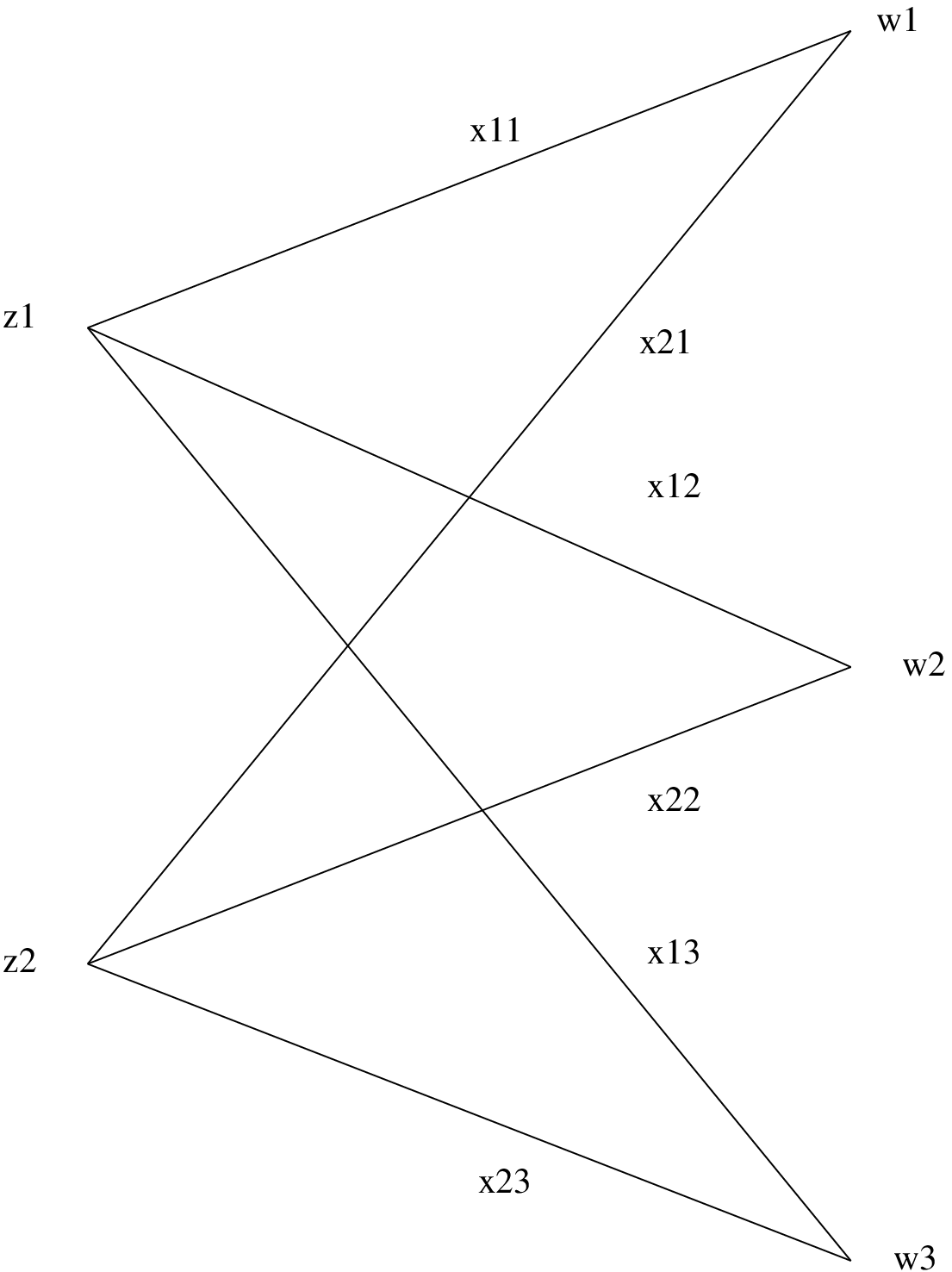}                        }
\end{figure}

Similar to the $2\to 2$ case we  furnish each line with a pair
of indices which have to satisfy the following equations:
\begin{align*}
&\alpha_{1k}+\alpha_{2k}=1\,, &&k=1,2,3\\
&p_{1k}+p_{2k}=2j_k\,, &&k=1,2,3 \\
& p_{m1}+p_{m2}+p_{m3}=2i_m\,, && m=1,2\,.
\end{align*}
The most general invariant kernel $\mathcal{H}$ has the form
\begin{align*}
[\mathcal{H}\varphi](z_1,z_2)=\prod_{k=1}^3\int_0^1d\alpha_{1k} \prod_{i=1}^2
\prod_{m=1}^3 \alpha_{im}^{p_{im}-1}\, 
\omega\left(\frac{\alpha_{21}\alpha_{12}}{\alpha_{11}\alpha_{22}},\frac{\alpha_{21}\alpha_{13}}{\alpha_{11}\alpha_{23}}\right)\,
\varphi(w_1(\alpha),w_2(\alpha), w_3(\alpha))\,,
\end{align*}
where $w_k(\alpha)=\sum_{m=1}^2 z_m \alpha_{mk}$ and $\omega(x,y)$ is an arbitrary 
function of two variables.

\end{document}